# Absolute energy level positions in tin and lead based halide perovskites

*Shuxia Tao[1,\*], Ines Schmidt[2], Geert Brocks[1,3], Junke Jiang[1], Ionut Tranca[4], Klaus Meerholz[2], and Selina Olthof[2,\*]*

1 Center for Computational Energy Research, Department of Applied Physics, Eindhoven University of Technology, P.O. Box 513, 5600MB, Eindhoven, The Netherlands
2 Department of Chemistry, University of Cologne, Luxemburger Straße 116, Cologne 50939, Germany
3 Computational Materials Science, Faculty of Science and Technology and MESA+ Institute for Nanotechnology, University of Twente, P.O. Box 217, 7500 AE Enschede, The Netherlands
4 Energy Technology, Department of Mechanical Engineering, Eindhoven University of Technology, 5600 MB Eindhoven, The Netherlands

Metal-halide perovskites are promising materials for future optoelectronic applications. One intriguing property, important for many applications, is the tunability of the band gap via compositional engineering. While experimental reports on changes in absorption or photoluminescence show rather good agreement for wide variety of compounds, the physical origins of these changes, namely the variations in valence band and conduction band positions, are not well characterized. Knowledge of these band positions is of importance for optimizing the energy level alignment with charge extraction layers in optoelectronic devices. Here, we determine ionization energy and electron affinity values of all primary tin and lead based perovskites using photoelectron spectroscopy data, supported by first-principles calculations. Through analysis of the chemical bonding, we characterize the key energy levels and elucidate their trends via a tight-binding analysis. We demonstrate that energy level variations in perovskites are primarily determined by the relative positions of the atomic energy levels of metal cations and halide anions. Secondary changes in the perovskite energy levels result from the cation-anion interaction strength, which depends on the volume and structural distortions of the perovskite lattices. These results mark a significant step towards understanding the electronic structure of this material class and provides the basis for rational design rules regarding the energetics in perovskite optoelectronics.

Metal halide perovskites are solution-processable semiconducting materials with a general formula $AMX_3$, where A are monovalent cations ($Cs^+$, $MA^+ = (CH_3NH_3)^+$, or $FA^+ = (CH(NH_2)_2)^+$), M are metal cations ($Pb^{2+}$ or $Sn^{2+}$), and X are halide anions ($I^-$ or $Br^-$ or $Cl^-$). This material class has received enormous attention in the scientific community recently due to breakthroughs in perovskite-based optoelectronic applications, mainly in photovoltaics[1-5], but also in photodetectors[6], light emission[7-9], and lasing[10]. Intriguingly, by exchanging or mixing different A, M, and/or X ions, it is possible to tune the optical gap of these semiconductors, which is exploited, e.g., to optimize the overlap with the solar spectrum in tandem solar cells[11], or to tune the emission wavelength of LEDs[12]. These changes in band gap are well characterized for a large number of primary $AMX_3$ compounds, as well as for more complex perovskite mixtures, in experimental [13-15] as well as computational[16-25] studies. However, two fundamental questions have not been resolved yet: (i) what is the underlying physical origin of the changes in the band gaps and (ii) how do the absolute positions of the valence band maximum (VBM) and conduction band minimum (CBM) change as a function of the composition of the perovskites? The answers to these questions are not only fundamentally highly interesting, but are also needed to develop strategies for tailoring desired optoelectronic properties and to optimally match perovskite energy levels to contacts and extraction layers for efficient charge transport through a device.

The challenges in answering these questions originate from the complex interplay of a few subtle yet correlated factors when combining different A, M, and X, such as the type and the size of ions, the crystal structure, and the degree of distortion with respect to the ideal perovskite structure[18,23]. Experimental studies, mainly based on photoelectron spectroscopy, have been limited to a small subset of $ABX_3$ compounds and suffer from significant variations in reported energy level values, which are due to variations brought on by preparation conditions[26-28], as well as by different data evaluation protocols[29]. Moreover, unlike in all-organic semiconductors, it is non-trivial to determine the energy onsets of the valence and conduction bands in metal halide perovskites experimentally using direct and inverse photoelectron spectroscopy due to a low density of states (DOS) at the band edges[30]. To reliably learn something about energy level trends in these systems, comparative studies are needed for subsets of compounds; these are however scarce and mostly limited to halide variations in $MAPbX_3$ perovskites, see e.g. Refs. [14,30-32]. Computational studies are also insightful for identifying trends in band gaps[16,18-20,23], key characteristics of DOSs[20-23], and band structures[24,25]. However, predictions of the absolute energy levels and their trends are challenging to make, due to intrinsic approximations in the methods, variations in the choice of structural models[17], and the need to model the crystal terminations (the surfaces).

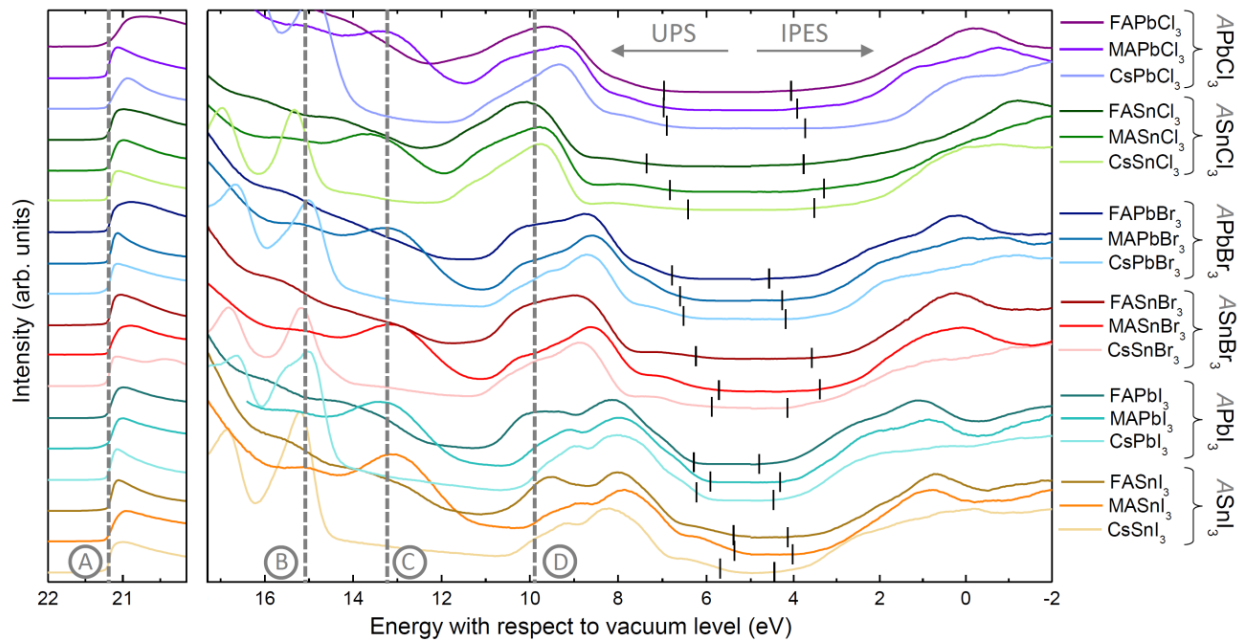

*Figure 1: Combined UPS and IPES measurements of 18 metal halide perovskite systems; for IPES, only the smoothed data trend is shown (raw data are included in the Supplementary Figure S6). For better comparability, the curves are offset vertically and the high energy cutoffs are aligned at the excitation energy of 21.22 eV, marked by line A; lines B, C, and D indicate characteristic features in the DOS, corresponding to the position of Cs, MA, and FA related states, respectively. The extracted positions of VBM and CBM are given by black vertical markers. The procedure for evaluation of the exact positions of the VBM and CBM is elaborated in Figure 2.*

Here, we provide answers to these open questions by determining the absolute energy levels of the complete set of primary Sn and Pb perovskites using optimized material fabrication and consistent data evaluation procedures. The energy levels of these compounds are determined by combining UV and inverse photoelectron spectroscopy (UPS and IPES), as well as absorption measurements. We carefully determine the Ionization Energy (IE) and Electron Affinity (EA) by comparing the experimental and density- functional- theory (DFT) calculated DOSs. Incorporating these values into an intuitive tight-binding model, we are able to give a clear analysis of all trends in IEs and EAs. This study therefore provides a fundamental understanding of the evolution of key electronic energy levels of metal halide perovskites and opens up the possibility for rational materials design for efficient perovskite optoelectronic devices.

**Photoelectron spectroscopy measurements**
Figure 1 shows representative UPS and IPES measurements of all 18 tin- and lead-based $AMX_3$ perovskites. To increase the comparability, all spectra have been shifted along the x-axis in such a way that the high energy cutoff position, marked by line A, is located at the excitation energy of 21.22 eV. This way, the positions of E = 0 eV corresponds to the vacuum level and the onset positions of the valence band and conduction band directly match the IE and EA values. These VBM and CBM positions are indicated by black vertical markers and are extracted by correlating the measured spectra with DOSs obtained from first-principles calculations as elaborated below.

The three additional vertical lines in Figure 1 mark the positions of the Cs $5p_{3/2}$ semicore level (line B at 15.1 eV), a MA molecular level (line C at 13.1 eV) and a FA molecular level (line D at 9.9 eV). Intriguingly, in our experiments we found that these features consistently appear at almost the same positions (within ± 0.2 eV) with respect to the vacuum level, independent on the perovskite structure. This finding is further supported by our DFT-derived semi-core levels of Cs $s$ and $p$ states in $CsMX_3$ perovskites (Supplementary Table S1), where merely minor energetic differences are present (within ± 0.1 eV); only for $CsSnCl_3$, a somewhat larger deviation of ~0.3 eV was found. These key features have been used throughout the experiments as valuable indicators for evaluating proper material preparation procedures. An off-stoichiometry composition usually unintentionally changes the IE of the film[27], which is then also evident by shifts in these atomic and molecular levels. It should be noted that we have extensively optimized the preparation procedure of each compound and have further evaluated the quality of our samples with respect to elemental composition, oxidation state, crystal structure, and morphology; the corresponding measurements can be found in the Supplementary Information, Section 9.

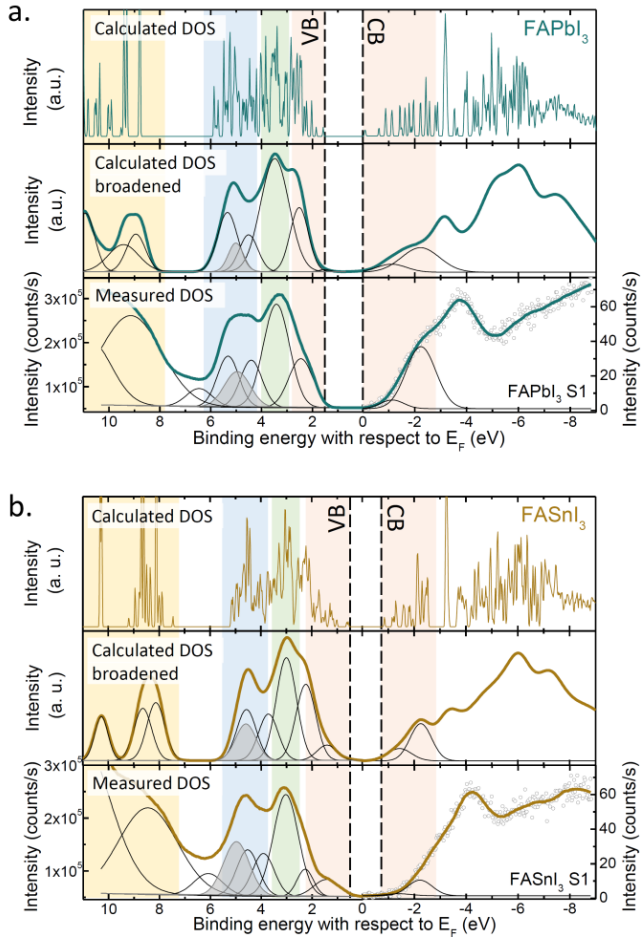

*Figure 2: Comparisons between measured (combined UPS and IPES) and DFT calculated densities of states. a. Top panel: calculated DOS of $FAPbI_3$. Middle panel: same DOS uniformly broadened by a Gaussian function and subsequently, this broadened DOS is fitted with Gaussian peaks. Lower panel: experimental DOS of $FAPbI_3$ fitted with a similar set of Gaussian peaks, so that the DFT and experimental spectra can be aligned. b shows the same as a. but for $FASnI_3$. The different colored vertical bars act as guide for the eye to show how the spectra are lined up; the shaded grey peak corresponds to FA related features. Note that the x-axes of the DFT spectra have been shifted to match the Fermi energies of the measured data sets. Further, the DFT data are plotted (in accordance with the measurement) using positive energies for VB and negative energies for CB bands; in Figs. 3 and 4. we use opposite signs for the energies, in agreement with standard DFT convention.*

As already indicated above, it is challenging to reliably determine the onset of the bands directly from the experimentally measured DOSs as presented in Figure 1. Firstly, gap states, e.g. induced by the imperfect surface or by defects in the film, can introduce additional states at the band edges. Secondly, the typically low DOS of perovskites at the onsets can be missed with common fitting procedures. The latter can be improved upon by fitting the experimental DOSs at the band edges to the corresponding DFT-calculated DOSs as already put forward by Endres et al.[30]. We employ a similar approach in this work, however instead of focusing on band onsets, we align key features of the experimental and DFT DOSs. Fitting over the full DOS region has the advantage that the results are independent of experimental broadening. Furthermore, possible gap states at the band edges are not factored in.

The fitting procedure we employ, is shown in Figure 2, using the materials $FAPbI_3$ and $FASnI_3$ as examples. Data sets concerning the remaining 16 materials can be found in the Supplementary Figure S6. First, the DFT DOS is calculated (Top panels in Figures 2 a and 2 b) and corrected such, that the band gap is equal to the experimentally measured optical gap (UV-vis measurements in the Supplementary Information, Section 9). Next, each point of the DOS is broadened by a Gaussian function such as to match the experimental resolution, resulting in the data sets shown in the middle panels. To be able to match theory and experiment, these broadened DFT spectra are then fitted to a linear combination of Gaussian peaks. These peaks do not have a direct physical meaning, so they do not represent features in the perovskites DOS, but are chosen such that they are able to consistently describe all perovskite compositions. Finally, the same sets of Gaussian peaks are used to fit the experimental DOS, shown in the lower panels of Figures 2 a and b; by this, the experimental and DFT DOSs can be aligned as indicated by the colored regions in Figure 2. The valence and conduction band onsets can be obtained from the VBM and CBM in the original (i.e., not broadened) DFT spectra, as indicated by the dashed lines.

Our procedure proves to be very robust for the VB region and therefore the position of the VBM can be accurately determined. However, in the CB region the agreement between the calculated and experimental DOSs is not as good, which makes it difficult to align them. Currently, it is unclear where the difference comes from, as both the calculations and experiments match previously reported calculated[30,33-35] and measured spectra[30-32,36,37] well; in addition, the sample to sample variation is small (Supplementary Figure S6). We suggest that the inconsistency between the DFT and IPES-derived DOSs may have to do with significant differences in measurement cross section. Due to the uncertainties, we aligned DFT and IPES-derived DOSs using only the first two CB features. This is more error prone than a fit over a wider region, and additional constraints are needed to make the fit more robust. Therefore we included the constraint that the

*Table 1: List of IEs, EAs, and optical band gaps for all 18 tin and lead based perovskites, extracted from data given in Figure 1, as well as data from the Supplementary information in Figure S6 and Section 9. All values are in eV and error bars represent the spread over 3 samples.*

**Pb based perovskites**

|  | I | Br | Cl |  |
| --- | --- | --- | --- | --- |
| $E_{g,opt}$ | 1.72 ± 0.01 | 2.31 ± 0.1 | 2.99 ± 0.02 | **Cs** |
| IE | 6.25 ± 0.1 | 6.53 ± 0.05 | 6.80 ± 0.1 |  |
| EA | 4.47 ± 0.1 | 4.17 ± 0.1 | 3.77 ± 0.1 |  |
| $E_{g,opt}$ | 1.59 | 2.30 ± 0.02 | 3.04 ± 0.01 | **MA** |
| IE | 5.93 ± 0.05 | 6.60 ± 0.05 | 6.92 ± 0.1 |  |
| EA | 4.36 ± 0.1 | 4.25 ± 0.1 | 3.77 ± 0.15 |  |
| $E_{g,opt}$ | 1.51 ± 0.02 | 2.25 ± 0.02 | 3.02 ± 0.05 | **FA** |
| IE | 6.24 ± 0.1 | 6.7 ± 0.1 | 6.94 ± 0.05 |  |
| EA | 4.74 ± 0.15 | 4.51 ± 0.1 | 3.98 ± 0.1 |  |

**Sn based perovskites**

|  | I | Br | Cl |  |
| --- | --- | --- | --- | --- |
| $E_{g,opt}$ | 1.25 ± 0.02 | 1.75 ± 0.05 | 2.88 ± 0.05 | **Cs** |
| IE | 5.69 ± 0.1 | 5.82 ± 0.1 | 6.44 ± 0.05 |  |
| EA | 4.38 ± 0.1 | 4.07 ± 0.1 | 3.47 ± 0.1 |  |
| $E_{g,opt}$ | 1.24 ± 0.02 | 2.13 ± 0.02 | 3.50 ± 0.08 | **MA** |
| IE | 5.39 ± 0.05 | 5.67 ± 0.05 | 6.85 ± 0.15 |  |
| EA | 4.07 ± 0.1 | 3.42 ± 0.1 | 3.36 ± 0.15 |  |
| $E_{g,opt}$ | 1.24 ± 0.1 | 2.63 ± 0.1 | 3.55 ± 0.05 | **FA** |
| IE | 5.34 ± 0.1 | 6.23 ± 0.05 | 7.33 ± 0.1 |  |
| EA | 4.12 ± 0.1 | 3.6 ± 0.1 | 3.83 ± 0.1 |  |

electronic gap has to be close to the optical gap. Differences in electronic and optical band gaps of common three-dimensional metal halide perovskites are typically small (tens of meV) [38-41] and are in the same order of magnitude as standard sample-to-sample variations.

Using the above described fitting routine, we determine the absolute positions of VBM and CBM for all Pb and Sn perovskites, as indicated by the black vertical markers in Figure 1. The extracted values for the IE and the EA are listed in Table 1, together with the optical gap obtained from the UV-vis measurements (Supplementary Information, Section 9). For each material we average these values over three separate samples, the error bars in Table 1 correspond to the sample to sample variation; the individual UPS/IPES spectra and fits are shown in the Supplementary Figure S6. Large differences in both energy level positions and band gaps are found between the different compounds. These trends will be discussed next based on a chemical bonding analysis and a tight-binding model.

**The chemical bonding in AMX$_3$ perovskites**

To rationalize the observed energy differences, we probe the contributions of the different atoms to the calculated DOS (called the partial DOS here). The example of CsPbI$_3$ is shown in Figure 3 a, where for simplicity the position of VBM is set to zero. Consistent with results reported in the literature[21-25], the states at the CB and VB band edges are dominated by Pb and I contributions, whereas Cs-related states are found at much lower and at much higher energies. The sharp feature around -8 eV in the DOS corresponds to a quasi-atomic state of the Cs$^+$ ion, for instance.

Next, we analyze the electronic structure using a Crystal Orbital Hamiltonian Population (COHP) analysis. A COHP analysis gives the covalent bonding and anti-bonding character of the DOS at each energy, and indicates which atomic orbitals are involved. Figures 3 a and b show a direct comparison between the DOS and the COHP of CsPbI$_3$ in the energy range between -13 eV and +5 eV. Cs character is almost absent in the orbital-resolved COHP, implying that Cs does not participate in covalent bonding; clearly, Pb and I contributions dominate. This pattern is observed in all AMX$_3$ compounds studied here. We therefore concentrate on the interactions of metal (M) cation and halide (X) anion, excluding those involving A site cations. Four pairs of bands of bonding/anti-bonding states can be identified in Figure 3 b (for the case of CsPbI$_3$), which result from the hybridization of the *s* and the *p* orbitals of the Pb and I atoms. In particular, the top of the VB can be identified as a Pb,*s* / I,*p* anti-bonding state, whereas the bottom of the CB is a Pb,*p* / I,*s* anti-bonding state. From the COHP we construct the simplified energy level diagram shown in Figure 3 c. Here all bands are interpreted in terms of hybridization of the atomic *s* and *p* states of Pb and I. In the tight-binding analysis, to be described below, we represent the energy levels of those atomic states by $E_{Pb,s}$, $E_{Pb,p}$, $E_{I,s}$, and $E_{I,p}$. The energies of the CBM and the VBM are indicated by $E_1$ and $E_2$ in Figure 3 c. As remarked above, both of these levels correspond to anti-bonding states, while their bonding partners can be found at energies $E_4$ and $E_3$, respectively. These four states $E_{1-4}$ play a central role in our tight-binding analysis.

**Tight-binding analysis**

Our tight-binding analysis focuses on the VBM and the CBM. The analysis becomes more straightforward if one concentrates on cubic symmetry and uses the symmetry analysis presented by Boyer-Richard *et al.* [19]. For cubic perovskites (space group P*m-3m*) the VBM and CBM are situated at the R-point of the Brillouin zone, and one can restrict a tight-binding analysis to states at the R-point.

We apply a nearest neighbor tight-binding model with six parameters: the four effective atomic *s* and *p* energy levels of the M cation and the X anion, see Figure 3 c, and two hybridization strengths, between the M,*s* and X,*p* orbitals, and between the M,*p* and X,*s* orbitals, respectively. Interactions between M,*s* and X,*s* orbitals, and between M,*p* and X,*p* are symmetry forbidden at the R-point of a cubic perovskite, so we do not have to consider the corresponding hybridization strengths[19]. The energy levels of the halide ions, $E_{X,s}$, $E_{X,p}$, can be obtained from the DFT-calculated level spectra at the Γ-point or the R-point, by identifying halide states that are non-bonding in cubic perovskites. The remaining four parameters can then be calculated from the energy levels $E_{1-4}$ in the DFT-calculated level spectrum at the R-point. For details, see the Methods Section. The dominant effect of spin-orbit-coupling (SOC) on the electronic structure stems from the SOC-induced p-level splitting on the M cation. We include this as an atomic parameter $\Delta_{SOC}$, where we use $\Delta_{SOC}$ of 1.65 eV and 0.60 eV for Pb and Sn, respectively[23,42].

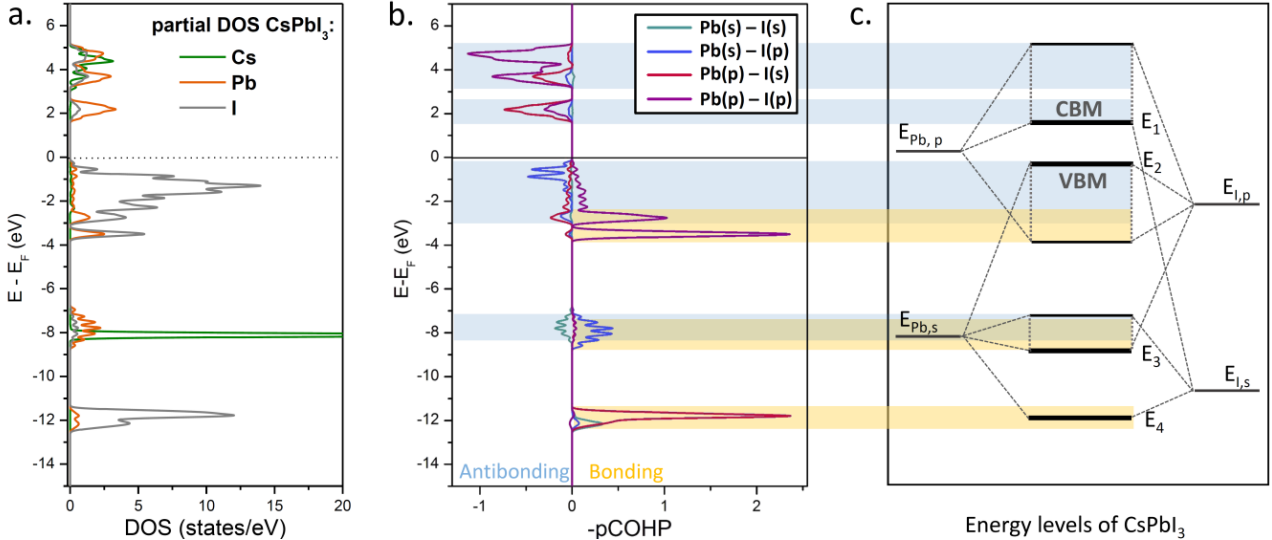

*Figure. 3: Orbital contributions to the energy bands of CsPbI$_3$. a. CsPbI$_3$ DOS projected on the Cs, Pb, and I atoms (partial DOS). b. Orbital resolved COHP; positive (negative) sign indicates bonding (anti-bonding) character. The line colour indicates which atomic orbitals are involved in the bonds (anti-bonds). c. schematic energy level diagram extracted from the COHP analysis. Bonding interactions are shaded in yellow, and anti-bonding interactions in blue.*

A graphical representation of the results of the tight-binding analysis at the R-point is given in Figures 4 a and b; the corresponding values of all relevant energy levels can be found in the Supplementary Table S3. In the following we use these results to analyze the trends in the VBM and CBM in case the halide anions or the metal cations are exchanged. We also discuss the influence of the structural variations in volume and distortions on the trends in these energies levels when the A cations are exchanged (Figure 4 c).

**Varying the X anion**
From Table 1 and Figure 1 it is obvious that exchanging the halide leads to significant changes in both EA and IE. We discuss this finding using CsPbX$_3$ as example; the schematic energy diagram is shown in Figure 4 a. The energy of the CBM is mostly influenced by the position of the Pb,p atomic level which is shifted upward when going from I to Br to Cl. Very likely this is a confinement effect, i.e., as the Pb-X distances decrease going from I to Br to Cl, an electron on a Pb atom is more confined and its energy increases.

The energy of the VBM is influenced by three competing effects. There is a significant downward shift of the X,p level going from I to Br to Cl, which simply reflects increasing electronegativity. This is expected to cause a large downward shift of the VBM. This trend is lessened by two factors working in the opposite direction. Firstly, there is a shift upward of the Pb,s level, which is the same confinement effect as discussed above. Secondly, the Pb,s/X,p hybridization strength increases somewhat going from I to Br to Cl (see Supplementary Table S3). However, the downward shift of the X,p level is still the dominating factor in determining the position of the VBM. A direct consequence of both the upward shift of the CBM and the downward shift of the VBM is the well-known substantial increase in the band gap going from I to Br to Cl.

**Varying the M cation**
Keeping the halide anion fixed and comparing Pb and Sn compounds in Table 1, it is evident that the IEs and EAs of Pb perovskites are larger than those of corresponding Sn ones. We analyze this trend using CsMI$_3$ (M=Pb, Sn) as example, shown in Figure 4 b. Replacing Pb with Sn, the atomic levels shift upwards, which is consistent with the smaller electronegativity of Sn. In absence of large changes in the anion levels, both the VBM and the CBM shift upward, i.e., both the IE and the EA of the Sn compounds are smaller than those of the corresponding Pb compounds. At the same time, we find that the splitting between s and p states in a Sn atom is smaller than in a Pb atom, which means that, going from Pb to Sn, the upward shift of the *s* level is larger than that of the *p* level; the consequence is that the VBM shifts upward more than the CBM. In other words, substituting Pb with Sn gives a strong reduction in the IE and a moderate reduction in the EA. The upward shift of the VBM is further enhanced by a slight increase in M,$s$/X,$p$ hybridization going from Pb to Sn (Supplementary Table S3).

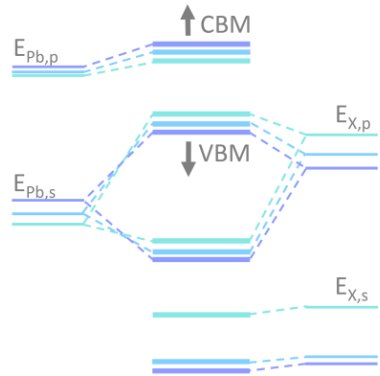 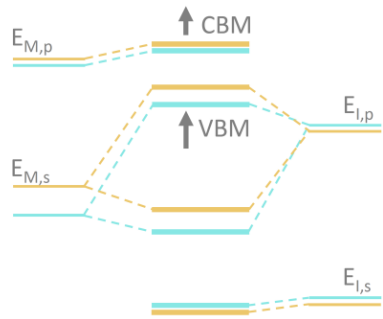 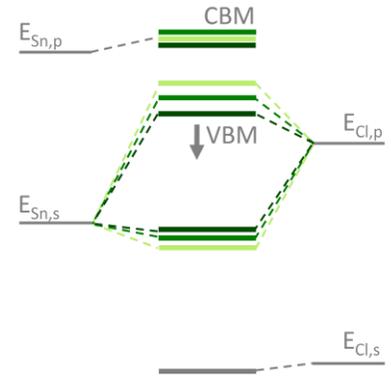

*Figure 4: Schematic energy levels in AMX$_3$ perovskites. a. and b. represent trends in changing the halide anions and the metal cations, respectively, as identified from the tight-binding analysis. c. is an intuitive illustration of energy-level changes based on structural distortions in tin-based perovskites.*

This upward shift in VBM can be nicely observed in the experimental data, i.e. in Figure 1 and Figure 2, where clearly an additional feature appears at the band edge of the VB side for the Sn based perovskites. Overall, the shifts of the VBM and CBM lead to a reduction of the band gaps of Sn compounds compared to their Pb analogues. There are three exceptions to this trend (FASnBr$_3$, MASnCl$_3$, and FASnCl$_3$) where we suggest that secondary effects, such as lattice distortions, play a larger role; these will be discussed next.

**Varying the A cation.**

When changing the A cation, the IEs and EAs do not show a uniform trend. As mentioned earlier, the A site cation does not directly participate in the bonding and only influences the electronic structure indirectly via changing the volume of the AMX$_3$ lattice or by introducing distortion in the ideal perovskite structure.

An indicator for possible distortions of perovskite lattices is the commonly used Goldschmidt's tolerance factor[43] of $TF = \frac{r_A + r_M}{\sqrt{2}(r_M + r_X)}$, where $r_\square$ is the radius of the corresponding ion. It is commonly accepted that 3D perovskite structures form for $TF$ in the range of $0.8 < TF \leq 1$. In the lower part of this range the structures are distorted by tilting of the MX$_6$ octahedra, TF = 1 results a perfect cubic perovskite structure, and for $TF > 1$ or $TF < 0.8$ additional distortion of the octahedra can occur and alternative structures instead of 3D perovskites are possibly formed [44].

Structural deformations, including octahedral tilting and distortion of the octahedra in the MX$_3$ framework, reduce somewhat the hybridization between the M and X states throughout the crystal. This shifts the VBM and CBM downward, whereby the IE is affected most because it is more sensitive to hybridization. Increasing the size of the A cation going from Cs to MA and FA also generally increases the volume (see Supplementary Information, Section 9 for structural information). This lowers the M levels somewhat (due to moderation of the confinement effect, see the discussion above). Again, this increases the IE and EA, but now the EA is affected most, as it is more sensitive to the M levels. In summary, both lattice distortion and volume expansion increase the IE and EA, where the former affects the IE most, and the latter the EA.

The interplay of these factors allows one to rationalize the variations in the IE and the EA of the lead-based perovskites. Here, varying the A cation within one halide class leads to relatively mild changes in volume and structures; hence, the IEs and EAs mostly vary only little. It is notable though that the EA of all FAPbX$_3$ compounds are larger by about 300 meV than their MA and Cs counterparts, which is due to the effect of increased volume described in the previous paragraph. Furthermore, MAPbI$_3$ shows an unusually low IE value which can be explained by its ideal $TF$ close to 1 (see Table S2 in the Supporting Information); it is therefore least affected by lattice distortion and will have the highest degree of hybridization, resulting in an effective upshift of the VB; in contrast, CsPbI$_3$ and FAPbI$_3$ have larger IE due to a reduced hybridization. Indeed, these two compounds are well-known to be distorted and tend to form a 2D yellow phase at room temperature while the black phase is only accessible via a high temperature annealing step [15,45].

In contrast to the subtle changes in Pb perovskites, much larger variations in EAs and IEs are found in Sn perovskites. Likely, the larger influence of A-site substitution on Sn perovskites comes from the smaller ionic radius of Sn compared to Pb leading to the fact that Sn compounds have larger $TF's$ than their Pb counterparts; for Cs containing compounds the *TF* becomes closer to 1, while Sn in combination with MA or FA leads to *TF* values greater than 1 (see Supplementary Table S2). Within the $ASnCl_3$ series in particular, the lattice distortion increases severely going from Cs to MA to FA, which means that the IE and EA increase, as depicted in Figure 4 c. As the IE is affected most, this means that the band gap increases significantly within this series via a downshift of the VB. Notably, also $FASnBr_3$ has a large *TF* and is severely distorted. In fact, the large band gaps found for $FASnBr_3$, $MASnCl_3$, and $FASnCl_3$ go against the general trend we established previously, where we stated that the band gaps of Sn perovskites are generally smaller than those of Pb perovskites.

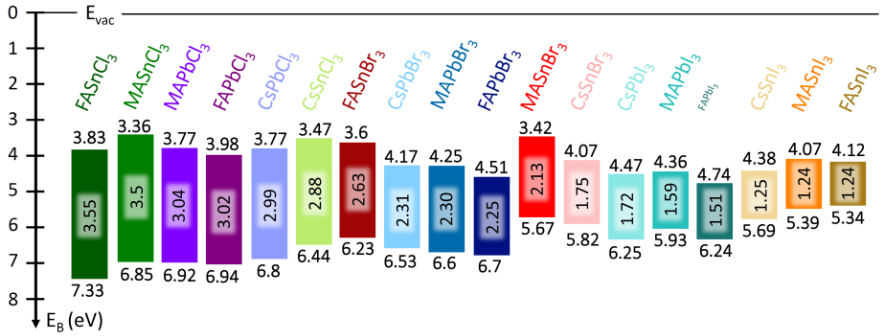

*Figure 5: Schematic energy level diagram of the perovskites under investigation, showing the respective IE and EA values as well as their optical gap (all in eV) as listed in Table 1; compounds are sorted in order of decreasing band gap. Minor deviations between the optical gap and the difference between IE and EA stem from the fact that one is extracted from the UV-vis measurements and the other from the PES investigations, which are furthermore averaged over 3 samples each; the corresponding error bars are given in Table 1.*

**General trends in the $AMX_3$ systems.**
Figure 5 shows a schematic diagram of all extracted energy level values sorted by their optical gaps (see Figure S1 for changes in energy levels sorted by IE's and EA's). Overall, the IE (position of the VBM) varies more strongly than the EA and is determined by the hybridization between the M,*s* and the X,*p* states. We can identify two trends regarding changes in IE: it increases in energy going from I to Br to Cl due to the downshifting of the X,*p* states while it also increases when switching from Sn to Pb, mainly due to the higher lying Sn,*s* states compared to the Pb,*s* states. The EA (energy of the CBM) is determined by the hybridization between the M,*p* and the X,*s* states, whereby the M,*p* position plays the dominant role, as the X,*s* level lies too low to affect the CBM much. The EA decreases when this M,*p* level is shifted upwards, which happens when going from I to Br to Cl due to confinement effects, as well as when changing from Pb to Sn. Combining the trends for the IE and the EA, we find the band gap generally increases going from I to Br to Cl, and going from Sn to Pb. Compounds where the lattice is strongly distorted with respect to the ideal perovskite structure can however break any of the above mentioned trends.

In summary, by combining photoelectron spectroscopy measurements and DFT calculations, we provided a consistent and complete set of absolute energy levels of all primary tin and lead based halide perovskites. We clarified the physical origin of the trends observed in ionization energy, electron affinity, and band gaps by an elaborated analysis based on a simple tight-binding model. Important factors have been identified, which play key roles for determining the absolute energy levels. These are the effective atomic energy levels of metal cations and halide anions, their hybridization strength, as well as structural variations including size and distortion of the crystal lattice. This study therefore provides the basis to optimize interfaces in optoelectronic applications and to further engineer energy levels of more complex (mixed) halide perovskites.

**Methods**
**1. Computational procedures**
**DFT:** All calculations on metal halide perovskites are performed using density-functional theory (DFT) and the projector-augmented wave (PAW) / plane wave techniques, as implemented in the Vienna ab-initio simulation package (VASP)[46,47]. The Perdew, Burke, and Ernzerhof (PBE)[48] functional is used for geometry optimizations. An energy cutoff of 500 eV and a *k*-point scheme of 6×6×4, 6×4×6, 6×6×6 are used for tetragonal, orthorhombic, and cubic structures, respectively, and energy and force convergence parameters are set to 0.1 meV and 2 meV/Å, respectively.

The hybrid functional PBE0[49] is used to calculate the density of states (DOS), with a small smearing parameter of 0.02 eV in a Gaussian smearing scheme. This small smearing parameter allows accurate identification of the onsets of valence and conduction bands in the DOS, as needed for the evaluation of the PES experiments, described in Figure 2. Here, to reduce the computational cost, we use a reduced *k*-point scheme, namely, 6×6×4, 6×4×6, 3×3×3 for tetragonal, orthorhombic, and cubic structures. We have adopted tetragonal cells for all MA and FA perovskites and orthorhombic cells for the Cs perovskites. The underlying reasons are explained in the Supplementary Information Section 3. To support the tight-binding analysis, Cs perovskites with cubic symmetry are used, as the tight-binding analysis becomes considerably easier for cubic symmetry.

**COHP:** To analyse the electronic structure and bonding in halide perovskites, we calculate the density of states (DOS), the partial density of states (PDOS) and the crystal orbital Hamiltonian population (COHP) with the Lobster 2.2.1 code [50-52]. This involves a transformation of the plane wave basis set used by VASP, to a localized basis set of Slater-type orbitals (STO).

The PDOS is defined as:
$$PDOS_i(E) = \sum_n |c_i^n|^2 \delta(E - E_n), \quad (1)$$
where $c_i^n$ are the coefficients associated with the atomic orbitals $\phi_i$ in a molecular orbital $\psi_n = \sum_i c_i^n \phi_i$. The COHP is defined as:
$$-\text{COHP}_{ij}(E) = H_{ij} \sum_n c_i^n c_j^{*n} \delta(E - E_n), \quad (2)$$
where $H_{ij}$ is the Hamiltonian matrix element between the atomic orbitals $\phi_i$ and $\phi_j$. For positive values of $-\text{COHP}_{ij}(E)$ the electronic interaction between the atomic orbitals $i$ and $j$ is bonding, negative values of $-\text{COHP}_{ij}(E)$ symbolize an anti-bonding interaction, while a zero value designates a non-bonding interaction. COHP has proved to be very powerful in the analysis of magnetism[53], phase stability[54], and catalytic reactivities [55,56] of solid state materials.

**Tight binding analysis:** We use a nearest neighbour tight-binding model for the *Pm-3m* cubic structure of perovskites, as described by Boyer-Richard *et al* [42]. The model has six parameters: the on-site energies of the *s* and *p* levels of the X and M atoms, $E_{X,s}$, $E_{X,p}$, $E_{M,s}$, and $E_{M,p}$, and the two hopping parameters (hybridization strengths) $V_{M,p-X,s}$ and $V_{M,s-X,p}$. The VBM and CBM are situated at the R-point of the Brillouin zone, so we restrict the bonding analysis to states at the R-point. The energy levels $E_{1-4}$, see Figure 3 c, are then given by the expressions

$$E_{1,4} = \frac{E_{X,s} + E_{M,p}}{2} \pm \frac{\left[(E_{M,p} - E_{X,s})^2 + 16 V_{M,p-X,s}^2\right]^{\frac{1}{2}}}{2} \quad (3)$$

$$E_{2,3} = \frac{E_{M,s} + E_{X,p}}{2} \pm \frac{\left[(E_{M,s} - E_{X,p})^2 + 48 V_{M,p-X,s}^2\right]^{\frac{1}{2}}}{2} \quad (4)$$

Here, the + signs give the CBM $E_1$ and the VBM $E_2$, while the – signs correspond to levels deep in the valence band $E_3$ and $E_4$. The energies $E_{1,4}$ and $E_{2,3}$ can be identified in the DFT spectrum at the R-point. They correspond to states with $R_4^+$ (3) and $R_1^+$ (1) symmetry, respectively.

The two anion levels, $E_{X,s}$ and $E_{X,p}$, can be identified in the DFT spectrum at Γ-point. They correspond to non-bonding halide states with $\Gamma_3^+$ (2) and $\Gamma_5^-$ (3) symmetry, respectively. Alternatively, we can obtain $E_{X,p}$ from the DFT spectrum at the R-point. In a nearest-neighbour tight binding model, one should find an eight-fold degenerate level of non-bonding halide *p* states. With non-nearest neighbour interactions this degeneracy splits up into a doublet and two triplets with $R_3^+$ (2), $R_4^+$ (3), and $R_5^+$ (3) symmetry, respectively. Indeed, in the DFT spectrum we find a splitting of about 1 eV between the $R_4^+$ (3), and the $R_5^+$ (3) triplet states. If we assume that the next-nearest-neighbour interaction between anion *p*-states is responsible for this splitting, then the $R_4^+$ (3) and $R_5^+$ (3) levels are split by a single hopping-matrix element[57], and the average of the two corresponds to the anion *p* level $E_{X,p}$. Indeed, we find a difference smaller than 0.1 eV between $E_{X,p}$ extracted from the DFT spectra of the R-point and the Γ- point.

The remaining four parameters, $E_{M,s}$, $E_{M,p}$, $V_{M,p-X,s}$, $V_{M,s-X,p}$, can then be extracted from Equations (3) and (4). For simplicity, the symmetry analysis described above is performed in absence of spin-orbit coupling (SOC). SOC is largest for the cation *p* states, where it is easily included by subtracting $\frac{2}{3}\Delta_{SOC}$ from $E_{M,p}$ in Equations (3) and (4), with $\Delta_{SOC}$ the SOC splitting[23,42]. More details can be found in the Supporting Information. It should be noted that the experimental values of IE and EA from Table 1 are used to correct the DFT energy values of $E_2$ (VBM) and $E_1$ (CBM); all other levels extracted from DFT are shifted accordingly. All relevant energy levels and hopping parameters are summarized in the Supplementary Table S3.

## 2. Experimental procedures

**Sample preparation:** Investigating perovskites poses several challenges. It is often reported that variations in processing can lead to sample-to-sample variation, either due to variations in film stoichiometry or partial or complete transition into different crystal structures; such variations can lead to changes in work function (Wf) [58,59] ionization energy [26-28], or the band gap [45,60]. Therefore, in this study great care was taken to ensure the preparation of representative perovskite films. Typically, for each composition dozens of samples were tested using a variety of preparation methods. Most films were prepared by solution processing and variations include the choice of solvent, co-solution vs. sequential deposition, spin speed, antisolvent treatment, and annealing procedure. In some cases thermal co-evaporation was used; this was especially necessary for various Cs containing compounds, since solubility of CsCl, and to some CsBr, provided major challenges. Films were tested and optimized with respect to their absorption properties, film morphology (via Scanning Electron Microscopy, SEM), crystal structure (via X-Ray Diffraction, XRD) and films composition (via X-ray Photoelectron Spectroscopy, XPS). With XPS, the films were also checked for unwanted oxidation states, e.g. the presence of signal originating from $Sn^{4+}$ or $Pb^0$. The corresponding absorption, SEM, XRD, and XPS measurements for representative samples can be found in the Supplementary Information, Section 9.

Samples were prepared on top of PEDOT:PSS (Clevios P VP Al 4083, Heraeus) covered indium tin oxide substrates (ITO from Thin Film Devices). The 40 nm thick PEDOT:PSS layer was employed in order to passivate the ITO surface which is otherwise known to undergo detrimental reactions with the perovskite at the interface[61]. The solution processing was performed under nitrogen atmosphere, always using a 1:1 molar ratio of the precursor salts in either DMF or DMSO. In some cases, an orthogonal solvent was used during the spin coating procedure to induce faster crystallization. Thermal evaporation was used for a set of Cs containing samples ($CsMCl_3$, and $CsMBr_3$). Co-evaporation was done using a molar ratio of the precursors close to 1:1. Some of the samples were annealed in vacuum. A detailed description of the individual preparation procedures, such as concentration, spin speed, and annealing time, is listed in the Supplementary Table S6.

**Photoelectron spectroscopy measurements** were performed in a custom built multi-chamber ultra-high vacuum setup. Thermally evaporated samples were transferred directly into the measurement chamber without breaking the vacuum, while solution processed films were transferred though nitrogen atmosphere; no sample was air exposed at any time and samples were measured within 24h after preparation. For the measurement of the occupied DOS and work function via UV photoelectron spectroscopy, a monochromatic He plasma source (VUV 5k, Scienta Omicron) at an excitation energy of 21.22 eV was used in combination with a hemispherical electron analyser (Phoibos 100, Specs) at an electron pass energy of 2 eV; a sample bias of -8V was applied during measurements to observe the high energy cutoff. The experimental resolution at this low pass energy setting is only determined by thermal broadening and is in the range of 100 meV ($\Delta E = 4k_BT$). For some of the samples, additional Kelvin Probe (KP) measurements (KP6500, McAllister) in vacuum were performed to compare the Wf measured under illumination (UPS) to the one in the dark (KP) since changes in surface dipole have been reported in literature[62]. No significant difference was found between the measurements, except in some samples of $CsPbCl_3$ and $CsSnCl_3$. Here a light dependence change in Wf was observed, with the Wf being lowered by about 0.5 eV due to illumination. After a series of tests with different sample treatments it was found that a moderate annealing in vacuum to 60 - 80°C made the effect vanish (resulting difference between KP and UPS ≤ 60 meV).

Measurements of the unoccupied DOS were performed by inverse photoelectron spectroscopy. Here, a low energy electron gun (ELG-2, Kimball) was used at 2 µA emission current together with a bandpass photon detector ($SrF_2$/NaCl bandpass, IPES 2000, Omnivac). The energy resolution, as determined from the width of an Ag Fermi edge, is approximately 600 meV. Since the electron bombardment during IPES measurements can be harmful to a sample surface, it was always performed after the UPS and XPS measurements were finished. The samples were re-checked via UPS afterwards to ensure no severe change in the DOS was induced by the IPES measurements. Typically, sample containing Cs and FA as cations were very stable, however some of the MA samples ($MASnI_3$, $MASnBr_3$, $MASnCl_3$, and $MAPbI_3$) could not be measured for more than 5-10 minutes by IPES before a change in the DOS occurred.

All other experimental characterization methods, such as SEM, UV-vis, XPS, and XRD, are described in the Supplementary Information.

**References:**
1. Kojima, A., Teshima, K., Shirai, Y. & Miyasaka, T. Organometal halide perovskites as visible-light sensitizers for photovoltaic cells. *J. Am. Chem. Soc*. **131**, 6050 (2009).
2. Burschka, J. *et al.* Sequential deposition as a route to high-performance perovskite-sensitized solar cells. *Nature* **499**, 316 (2013).
3. Lee, M. M., Teuscher, J., Miyasaka, T., Murakami, T. N. & Snaith, H. J. Efficient hybrid solar cells based on meso-superstructured organometal halide perovskites. *Science* **338**, 643 (2012).
4. Saliba, M. *et al.* Cesium-containing Triple Cation Perovskite Solar Cells: Improved Stability, Reproducibility and High Efficiency. *Energy Environ. Sci.* **9**, 1989 (2016).
5. Yang, W. S. *et al.* Iodide management in formamidinium-lead-halide–based perovskite layers for efficient solar cells. *Science* **356**, 1376 (2017).
6. Dou, L. *et al.* Solution-processed hybrid perovskite photodetectors with high detectivity. *Nat. Commun*. **5**, 5404 (2014).
7. Tan, Z.-K. et al. Bright light-emitting diodes based on organometal halide perovskite. *Nat. Nanotechnol*. **9**, 687 (2014).
8. Stranks, S. D. & Snaith, H. J. Metal-halide perovskites for photovoltaic and light-emitting devices. *Nat. Nanotechnol*. **10**, 391 (2015).
9. Cho, H. *et al.* Overcoming the electroluminescence efficiency limitations of perovskite light-emitting diodes. *Science* **350**, 1222 (2015).
10. Xing, G. *et al.* Low-temperature solution-processed wavelength-tunable perovskites for lasing. *Nat. Mater*. **13**, 476 (2014).
11. McMeekin, D. P. *et al.* A mixed-cation lead mixed-halide perovskite absorber for tandem solar cells. *Science* **351**, 151 (2016).
12. Protesescu, L. *et al.* Nanocrystals of Cesium Lead Halide Perovskites (CsPbX3, X = Cl, Br, and I): Novel Optoelectronic Materials Showing Bright Emission with Wide Color Gamut. *Nano Lett*. **15**, 3692 (2015).
13. Noh, J. H., Im, S. H., Heo, J. H., Mandal, T. N. & Seok, S. Il. Chemical management for colorful, efficient, and stable inorganic-organic hybrid nanostructured solar cells. *Nano Lett*. **13**, 1764 (2013).
14. Sargent, E. H. et al. Structural, optical, and electronic studies of wide-bandgap lead halide perovskites. *J. Mater. Chem. C* **3**, 8839 (2015).
15. Stoumpos, C. C., Malliakas, C. D. & Kanatzidis, M. G. Semiconducting tin and lead iodide perovskites with organic cations: phase transitions, high mobilities, and near-infrared photoluminescent properties. *Inorg. Chem*. **52**, 9019 (2013).
16. Buin, A., Comin, R., Xu, J., Ip, A. H. & Sargent, E. H. Halide-Dependent Electronic Structure of Organolead Perovskite Materials. *Chem. Mater*. **27**, 4405 (2015).
17. Butler, K. T., Frost, J. M. & Walsh, A. Band alignment of the hybrid halide perovskites CH3NH3PbCl3, CH3NH3PbBr3 and CH3NH3PbI3. *Mater. Horiz*. **2**, 228 (2015).

18. Walsh, A. Principles of Chemical Bonding and Band Gap Engineering in Hybrid Organic-Inorganic Halide Perovskites. *J. Phys. Chem. C* **119**, 5755 (2015).
19. Berger, R. F. Design Principles for the Atomic and Electronic Structure of Halide Perovskite Photovoltaic Materials: Insights from Computation. *Chem. - A Eur. J.* **24**, 8708 (2018).
20. Zhou, L. *et al.* Density of States Broadening in CH3NH3PbI3 Hybrid Perovskites Understood from ab Initio Molecular Dynamics Simulations. *ACS Energy Lett.* **3**, 787 (2018).
21. Mosconi, E., Umari, P. & De Angelis, F. Electronic and optical properties of MAPbX3 perovskites (X = I, Br, Cl): A unified DFT and GW theoretical analysis. *Phys. Chem. Chem. Phys.* **18**, 27158 (2016).
22. Umari, P., Mosconi, E. & De Angelis, F. Relativistic GW calculations on CH3NH3PbI3 and CH3NH3SnI3 perovskites for solar cell applications. *Sci. Rep.* **4**, 4467 (2014).
23. Tao, S. X., Cao, X. & Bobbert, P. A. Accurate and efficient band gap predictions of metal halide perovskites using the DFT-1/2 method: GW accuracy with DFT expense. *Sci. Rep.* **7**, 14386 (2017).
24. Even, J., Pedesseau, L., Jancu, J.-M. & Katan, C. DFT and k · p modelling of the phase transitions of lead and tin halide perovskites for photovoltaic cells. *Phys. status solidi - Rapid Res. Lett.* **8**, 31 (2014).
25. Goesten, M. G. & Hoffmann, R. Mirrors of Bonding in Metal Halide Perovskites. *J. Am. Chem. Soc.* **140**, 12996 (2018).
26. Kim, T. G., Seo, S. W., Kwon, H., Hahn, J.-H. & Kim, J. W. Influence of halide precursor type and its composition on the electronic properties of vacuum deposited perovskite films. *Phys. Chem. Chem. Phys.* **17**, 24342 (2015).
27. Emara, J. *et al.* Impact of Film Stoichiometry on the Ionization Energy and Electronic Structure of CH3NH3PbI3 Perovskites. *Adv. Mater.* **28**, 553 (2016).
28. Fassl, P. *et al.* Fractional deviations in precursor stoichiometry dictate the properties, performance and stability of perovskite photovoltaic devices. *Energy Environ. Sci.* **11**, 3380 (2018).
29. Olthof, S. Research Update: The electronic structure of hybrid perovskite layers and their energetic alignment in devices. *APL Mater.* **4**, 091502 (2016).
30. Endres, J. *et al.* Valence and Conduction Band Densities of States of Metal Halide Perovskites: A Combined Experimental–Theoretical Study. *J. Phys. Chem. Lett.* **7**, 2722 (2016).
31. Li, C. et al. Halide-Substituted Electronic Properties of Organometal Halide Perovskite Films: Direct and Inverse Photoemission Studies. *ACS Appl. Mater. Interfaces* **8**, 11526 (2016).
32. Schulz, P. *et al.* Interface energetics in organo-metal halide perovskite-based photovoltaic cells. *Energy Environ. Sci.* **7**, 1377 (2014).
33. Mosconi, E., Amat, A., Nazeeruddin, M. K., Grätzel, M. & De Angelis, F. First-Principles Modeling of Mixed Halide Organometal Perovskites for Photovoltaic Applications. *J. Phys. Chem. C* **117**, 13902 (2013).
34. Amat, A. *et al.* Cation-induced band-gap tuning in organohalide perovskites: Interplay of spin-orbit coupling and octahedra tilting. *Nano Lett.* **14**, 3608 (2014).
35. Bernal, C. & Yang, K. First-principles Hybrid Functional Study of the Organic-inorganic Perovskites CH3NH3SnBr3 and CH3NH3SnI3. *J. Phys. Chem. C* **118**, 24383 (2014).
36. Komesu, T. *et al.* Surface Electronic Structure of Hybrid Organo Lead Bromide Perovskite Single Crystals. *J. Phys. Chem. C* **120**, 21710 (2016).
37. Liu, X. et al. Electronic structures at the interface between Au and CH3NH3PbI3. *Phys. Chem. Chem. Phys.* **17**, 896 (2014).
38. Miyata, A. *et al.* Direct measurement of the exciton binding energy and effective masses for charge carriers in organic–inorganic tri-halide perovskites. *Nat. Phys.* **11**, 582 (2015).
39. Yang, Y. *et al.* Comparison of Recombination Dynamics in CH3NH3PbBr3 and CH3NH3PbI3 Perovskite Films: Influence of Exciton Binding *Energy. J. Phys. Chem. Lett.* **6**, 4688 (2015).
40. Saba, M., Quochi, F., Mura, A. & Bongiovanni, G. Excited State Properties of Hybrid Perovskites. *Acc. Chem. Res.* **49**, 166 (2016).
41. Milot, R. L. *et al.* The Effects of Doping Density and Temperature on the Optoelectronic Properties of Formamidinium Tin Triiodide Thin Films. *Adv. Mater.* **30**, 1804506 (2018).
42. Boyer-Richard, S. *et al.* Symmetry-Based Tight Binding Modeling of Halide Perovskite Semiconductors. *J. Phys. Chem. Lett.* **7**, 3833 (2016).
43. Goldschmidt, V. M. Die Gesetze der Krystallochemie. *Naturwissenschaften* **60**, 477 (1926).
44. Saparov, B. & Mitzi, D. B. Organic–Inorganic Perovskites: Structural Versatility for Functional Materials Design. *Chem. Rev.* **116**, 4558 (2016).
45. Eperon, G. E. et al. Inorganic caesium lead iodide perovskite solar cells. *J. Mater. Chem. A* **3**, 19688 (2015).
46. Kresse, G. & Furthmüller, J. Efficiency of ab-initio total energy calculations for metals and semiconductors using a plane-wave basis set. *Comput. Mater. Sci.* **6**, 15 (1996).
47. Kresse, G. & Furthmüller, J. Efficient iterative schemes for ab initio total-energy calculations using a plane-wave basis set. *Phys. Rev. B - Condens. Matter Mater. Phys.* **54**, 11169 (1996).
48. Perdew, J. P., Burke, K. & Ernzerhof, M. Generalized gradient approximation made simple. *Phys. Rev. Lett.* **77**, 3865 (1996).
49. Paier, J., Hirschl, R., Marsman, M. & Kresse, G. The Perdew-Burke-Ernzerhof exchange-correlation functional applied to the G2-1 test set using a plane-wave basis set. *J. Chem. Phys.* **122**, 234102 (2005).
50. Dronskowski, R. & Blöchl, P. E. Crystal orbital hamilton populations (COHP). Energy-resolved visualization of chemical bonding in solids based on density-functional calculations. *J. Phys. Chem*. **97**, 8617 (1993).
51. Deringer, V. L., Tchougréeff, A. L. & Dronskowski, R. Crystal orbital Hamilton population (COHP) analysis as projected from plane-wave basis sets. *J. Phys. Chem. A* **115**, 5461 (2011).
52. Maintz, S., Deringer, V. L., Tchougréeff, A. L. & Dronskowski, R. Analytic projection from plane-wave and PAW wavefunctions and application to chemical-bonding analysis in solids. *J. Comput. Chem*. **34**, 2557 (2013).
53. Landrum, G. A. & Dronskowski, R. The Orbital Origins of Magnetism: From Atoms to Molecules to Ferromagnetic Alloys. *Angew. Chemie Int. Ed.* **39**, 1560 (2000).
54. Wuttig, M. *et al.* The role of vacancies and local distortions in the design of new phase-change materials. *Nat. Mater.* **6**, 122 (2007).
55. Van Santen, R. A., Tranca, I. & Hensen, E. J. M. Theory of surface chemistry and reactivity of reducible oxides. *Catal. Today* **244**, 63 (2015).
56. Van Santen, R. A. Modern Heterogeneous Catalysis: An Introduction. (Wiley-VCH Verlag GmbH & Co. KGaA: Weinheim, Germany, 2017).
57. Mattheiss, L. F. Band structure and fermi surface of ReO3. *Phys. Rev.* **181**, 987 (1969).

58. Wang, Q. et al. Qualifying composition dependent p and n self-doping in CH3NH3PbI3. *Appl. Phys. Lett.* **105**, 163508 (2014).
59. Paul, G., Chatterjee, S., Bhunia, H. & Pal, A. J. Self-Doping in Hybrid Halide Perovskites via Precursor Stoichiometry: To Probe the Type of Conductivity through Scanning Tunneling Spectroscopy. *J. Phys. Chem. C* **122**, 20194 (2018).
60. Brennan, M. C., Draguta, S., Kamat, P. V. & Kuno, M. Light-Induced Anion Phase Segregation in Mixed Halide Perovskites. *ACS Energy Lett*. **3**, 204 (2018).
61. Olthof, S. & Meerholz, K. Substrate-dependent electronic structure and film formation of MAPbI3 perovskites. *Sci. Rep*. **7**, 40267 (2017).
62. Zu, F. et al. Impact of White Light Illumination on the Electronic and Chemical Structures of Mixed Halide and Single Crystal Perovskites. *Adv. Opt. Mater*. **5**, 1700139 (2017).

**Acknowledgement**

This work was in parts supported by the ministry of Science of the state of NRW within the PeroBOOST (EFRE) project. S.O acknowledges funding by the Eleonore Trefftz Programme for Visiting Women Professors at TU Dresden. S.T. and J.J. acknowledge funding by the Computational Sciences for Energy Research (CSER) tenure track program of Shell and NWO (Project number 15CST04-2), the Netherlands. S.T. and I.T. thank the NWO for access to the Dutch national high-performance computing facilities (Cartesius).

Supporting information for
Absolute energy level positions in tin and lead based halide perovskites

*Shuxia Tao [1],*, Ines Schmidt [2], Geert Brocks [1, 3], Junke Jiang [1], Ionut Tranca [4], Klaus Meerholz [2], and Selina Olthof [2],**

*1 Center for Computational Energy Research, Department of Applied Physics, Eindhoven University of Technology, P.O. Box 513, 5600MB, Eindhoven, The Netherlands*

*2 Department of Chemistry, University of Cologne, Luxemburger Straße 116, Cologne 50939, Germany*

*3* Computational Materials Science, *Faculty of Science and Technology and MESA+ Institute for Nanotechnology, University of Twente, P.O. Box 217, 7500 AE Enschede, The Netherlands*

*4 Energy Technology, Department of Mechanical Engineering, Eindhoven University of Technology, 5600 MB Eindhoven, The Netherlands*

Content

1. Comparative plots of changes in energy levels
2. Additional tables
3. Details of DFT calculations
4. Tight-Binding Parameters
5. Broadening and fitting procedure for the DFT calculated DOS
6. Procedures used for the experimentally measured DOS
7. Experimental sample-to-sample variation
8. Description of additional measurement techniques
9. Material measurement sheets (XRD, SEM, XPS, UV-vis)
10. Sample preparation
11. Additional references

1. Comparative plots of changes in energy levels

In the main manuscript the trends in energy level positions are shown sorted by the size of the optical gap; here, in Figure S1, this graph is re-plotted and in addition similar sketches are presented in which the materials are sorted by their IE or EA values, respectively.

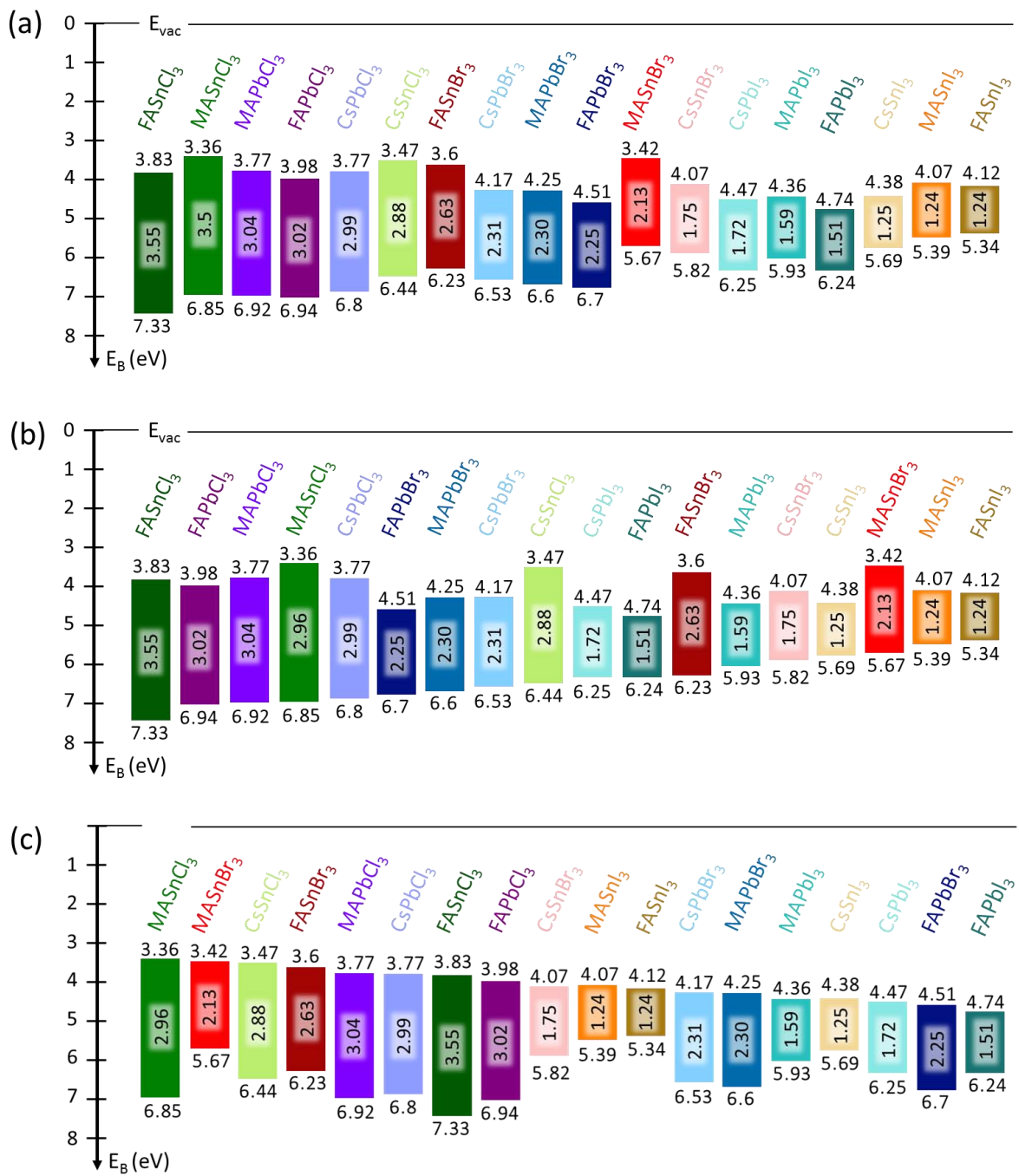

Figure S1. Graphical representation of IE, EA, and band gap data of the perovskites, ordered with respect to (a) decreasing band gap, (b) decreasing IE, and (c) increasing EA value.

2. Additional tables

Semi-core states
As mentioned in the main article and shown there in Figure 1, in the experimental studies we find that features originating from the A cation have approximately the same energetic position with respect to the vacuum level, irrespective of the perovskite composition. This is used, for instance, to identify well-prepared films (in addition to the stoichiometric information obtained by XPS). This result is also found in our DFT analysis of $CsMX_3$, where the Cs 5$s$ and 5$p$ semi-core states for different halides X lie within ± 0.1 eV as shown in Table S1. Only for $CsSnCl_3$, a somewhat larger variation of ~0.3 eV is found; this material shows more deviation in the experimental study as well.

*Table S1: The energies (in eV) of Cs 5s and 5p semi-core states, obtained from DFT calculations on Cs containing perovskites with cubic structure. The values are shifted such that the VBM position matches the position of the experimental IE.*

**Pb**

|      | I      | Br     | Cl     |
|------|--------|--------|--------|
| Cs $s$ | -26.57 | -26.59 | -26.46 |
| Cs $p$ | -13.34 | -13.38 | -13.27 |

**Sn**

|      | I      | Br     | Cl     |
|------|--------|--------|--------|
| Cs $s$ | -26.54 | -26.47 | -26.78 |
| Cs $p$ | -13.32 | -13.27 | -13.63 |

Calculated Goldschmidt's tolerance factors
Goldschmidt's tolerance factor is used as an indicator for the stability of perovskite crystal structures ($AMX_3$), and their tendency to distort; it is defined as: $TF = \frac{r_A + r_M}{\sqrt{2}(r_M + r_X)}$, with $r_\square$ the radius of the ions. Table S2 lists the extracted tolerance factors of the perovskite under investigation here. For the ionic radii we use Shannon's crystal radii. The radii of 1.81, 2.70, 2.79 Å are used for Cs$^+$, MA$^+$, and FA$^+$, respectively; 1.20 and 1.33 Å are used for Sn$^{2+}$ and Pb$^{2+}$, respectively; 1.67, 1.82 and 2.06 Å are used for Cl$^-$, Br$^-$, and I$^-$, respectively. [values taken from Ref. [1]]

*Table S2: Calculated tolerance factors for the different perovskites.*

**Pb**

|    | I    | Br   | Cl   |     |
|----|------|------|------|-----|
| TF | 0.81 | 0.82 | 0.82 | **Cs** |
| TF | 0.99 | 1.02 | 1.03 | **MA** |
| TF | 1.01 | 1.04 | 1.05 | **FA** |

**Sn**

|    | I    | Br   | Cl   |     |
|----|------|------|------|-----|
| TF | 0.84 | 0.85 | 0.86 | **Cs** |
| TF | 1.03 | 1.06 | 1.08 | **MA** |
| TF | 1.05 | 1.08 | 1.10 | **FA** |

3. Details on DFT calculations

To compare the experimental DOSs to the DFT calculated ones, it is important to take into account the structural disorder of metal-halide perovskites at finite temperature in the structural models in DFT calculations. The lattice of the metal halide perovskites is proven to be highly dynamic due to its soft nature. Dynamic disorder of inorganic frameworks and of organic or inorganic cations at finite temperature was observed in previous computational[2,3] and experimental studies[4,5]. In order to model this disorder in a simple way, we adopt a tetragonal cell, consisting of 8 $AMX_3$ units, for organic-cation-containing perovskites (Figure S2 (a)) and an orthorhombic cell, consisting of 4 $AMX_3$ units, for Cs-containing perovskites (Figure S2 (b)). These cells are (i) sufficiently large to model some of the disorder seen in both A cation (MA, FA, and Cs) and the $MX_3$ inorganic framework, and (ii) they are not too large, so that we can afford DFT calculations using a hybrid functional and take spin-orbit-coupling into account. This way, as shown from our comparisons of DFT and experimental DOSs of all 18 perovskites (Figure 2 in main text and Figure S6 here), excellent agreement in the occupied DOSs is found. We note that smaller unit cells are insufficient to describe the DOS realistically due to an unrealistic distribution (in case of Cs) and uniform orientation (in cases of MA and FA) of A cations.

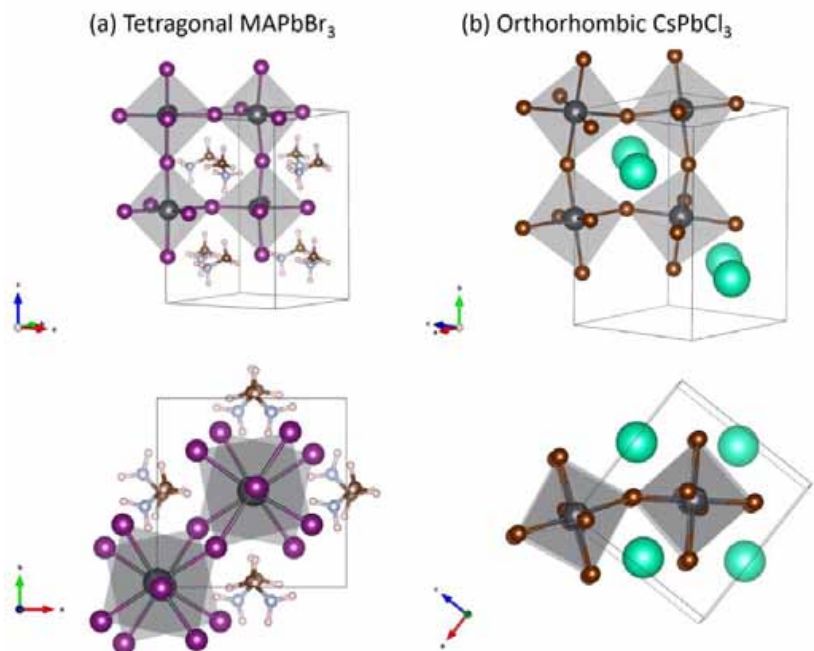

*Figure S2: Structural models used for DOS calculations: tetragonal cell for MAPbBr$_3$ and orthorhombic cell for CsPbCl$_3$.*

To further motivate the choice of crystal structures employed here, Figure S3 shows a comparison between the calculated DOS in a primitive cubic cell vs. the tetragonal cell of Figure S2(a) for MAPbBr$_3$, as well as a primitive cubic cell vs. the orthorhombic cell of Figure S2(b) for CsPbCl$_3$. XRD measurements show that both of these systems form a cubic crystal structure at room temperature (see Section 9 for XRD measurements). However, if a primitive cubic unit cell is used to calculate the DOS, significant deviations from experiments are found. In particular, features at the band edges are missing. Another important discrepancy found in cubic MAPbBr$_3$ is that the second peak deeper in the VB (around 3 to 4 eV, marked by grey arrow) shifts to the left compared to the one found in a tetragonal cell. Again, this shift leads to a significant inconsistency with the experimental DOS. We attribute this to the uniform orientation of MA ions (ferroelectric alignment), enforced when using a primitive cell, which is unlikely to be present at room temperature due to the rotational motions of the MA ions. Highly ordered MA cations create an electrostatic field that induces a shift of the Br and Pb atomic levels, which modifies the band structure. It can, therefore, be concluded that, even though these systems do not show lattice distortion or structural disorder in the XRD measurements, there is nonetheless a dynamic disorder in the MX$_3$ lattice, which renders these systems only cubic on average. In our calculations, we can only represent this dynamic disorder by a static variation in the orientation of the MA ions. Our tetragonal cell containing 8 MA/FA molecules orthorhombic cells with 4 Cs are the smallest one that approximately capture the spread in these orientations.

Based on the above analysis, we have adopted tetragonal structures for all MA and FA perovskites and orthorhombic structures for Cs perovskites. Structural models for MAPbBr$_3$ and CsPbCl$_3$ are found in Figure S2. We use the experimental XRD data of MAPbI$_3$ and CsSnI$_3$ as starting point, and optimize all (tetragonal and orthorhombic) structures of the 18 perovskites using the same DFT settings. It should be noted that slight overestimation or underestimation of lattice parameters in the DFT calculations will add a small uncertainty to the calculated energy levels near band edges and slightly alter the band gaps (due to a change of hybridization strength between M cations and I anions, see the main text). However, varying the size of cells in the DFT calculations by 1-2% only leads to small changes in band gaps in the range of 30 - 60 meV [6]. Another source of uncertainty is introduced by temperature effects, which are not taken into account in these DFT calculations. Experiments show that when the temperature is varied by 150 K, changes in band gap in the range of 30-60 meV are found [7]. Overall, the variations introduced by lattice size and temperature are therefore smaller than the experimental uncertainties and sample-to-sample variations (main text Table 1), so we can disregard them.

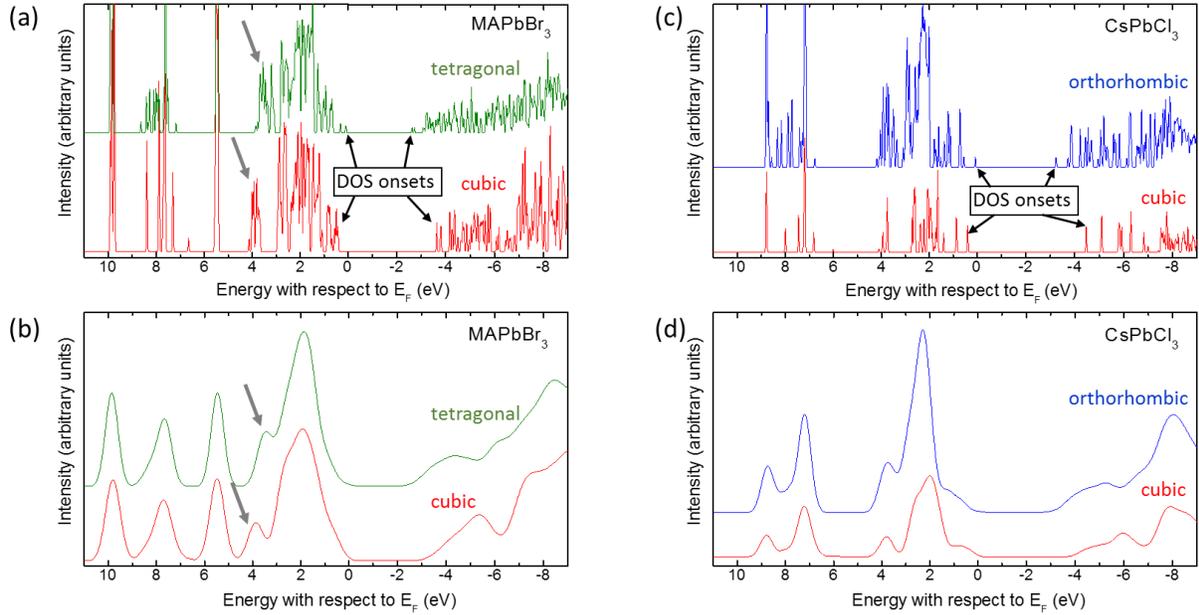

*Figure S3. Comparison between the DFT-calculated DOSs using different structures. (a) shows the calculated DOS and (b) shows the broadened DOS of MAPbBr$_3$, using a primitive cubic unit cell (red curve), and a tetragonal supercell (green curve). The structural motive of the latter is shown in Figure S2. In (c) and (d) the same comparison is made for the DOSs of CsPbCl$_3$, using a primitive cubic vs. an orthorhombic supercell. The procedure for peak broadening is described in Section 5.2. Here, for a direct comparison to experiments, the signs of the energy levels are adjusted to the experimental ones, i.e., positive values for VB energies and negative values for CB energies.*

4. Tight-binding parameters

Table S3 lists the tight-binding parameters of the CsMX$_3$ systems while Figure S4 shows these energy levels for CsPbI$_3$.

*Table S3: Tight-binding parameters (in eV relative to the vacuum level); $E_{X,s}$, $E_{X,p}$, $E_{M,s}$, $E_{M,p}$, are the on-site s and p atomic energy levels of X anions and M cations, respectively; $V_{M,p\text{-}X,s}$, $V_{M,s\text{-}X,p}$ are the hybridization strengths (hopping parameters) between the indicated orbitals. $E_1^{exp}$ and $E_2^{exp}$ are specific energies of the conduction and valence band onsets (CBM and VBM), as extracted from experimental electron affinity and ionization energy, respectively; $E_3$ and $E_4$ are energy levels within the valence band, extracted from the DFT calculations (see main text). $E_{M,p}$ includes the downward shift of $-2/3\ \Delta soc$ due to the spin-orbit coupling, with $2/3\ \Delta soc=1.1$ eV for Pb and $2/3\ \Delta soc=0.4$ eV for Sn.*

|  | CsPbI$_3$ | CsPbBr$_3$ | CsPbCl$_3$ | CsSnI$_3$ | CsSnBr$_3$ | CsSnCl$_3$ |
|---|---|---|---|---|---|---|
| $E_1^{exp}$ | -4.53 | -4.22 | -3.81 | -4.44 | -4.07 | -3.56 |
| $E_2^{exp}$ | -6.25 | -6.53 | -6.80 | -5.69 | -5.85 | -6.44 |
| $E_3$ | -14.51 | -14.88 | -15.11 | -14.04 | -14.39 | -15.24 |
| $E_4$ | -18.71 | -21.47 | -21.94 | -18.87 | -21.56 | -22.52 |
| $E_{X,s}$ | -18.44 | -20.99 | -21.19 | -18.66 | -21.39 | -22.31 |
| $E_{X,p}$ | -8.15 | -9.05 | -9.79 | -8.28 | -9.11 | -10.35 |
| $E_{M,s}$ | -12.61 | -12.36 | -12.12 | -11.45 | -11.13 | -11.33 |
| $E_{M,p}$ | -4.80 | -4.70 | -4.56 | -4.65 | -4.24 | -3.77 |
| $V_{M,p\text{-}X,s}$ | 1.01 | 1.47 | 1.86 | 0.87 | 0.88 | 0.99 |
| $V_{M,s\text{-}X,p}$ | 1.00 | 1.11 | 1.15 | 1.11 | 1.21 | 1.26 |

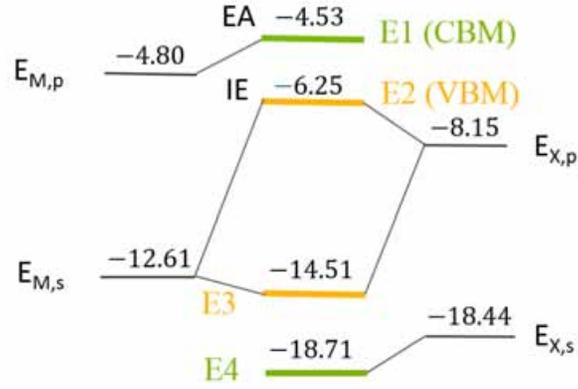

*Figure S4. Scheme of the energy levels involved in the tight-binding analysis, using cubic $CsPbI_3$ as a model system. $E_{X,s}$, $E_{X,p}$ and $E_1$, $E_2$, $E_3$, and $E_4$ are derived from DFT band structure calculations. $E_{M,p}$ and $E_{M,s}$ (listed in Table S3 above) are derived from the tight-binding model. Note, $E_1$ and $E_2$ are set to the experimental values of the electron affinity and the ionization energy; the DFT derived values of $E_{X,s}$, $E_{X,p}$, $E_3$, and $E_4$ are shifted accordingly.*

5. Broadening and fitting procedure for the DFT calculated DOS

5.1. Treatment of unbroadened DFT spectra
The calculated DOSs, shown in Fig. 2 of the main article and in Fig. S6 in this SI, are plotted with respect to Fermi level and for a direct comparison with experiments, the sign of the energy levels are adjusted to the experimental ones, i.e. the positive values for VB and negative values for CB bands are used. For halide perovskites, the standard DFT functional PBE gives a DOS that needs to be stretched to match photoemission data. To partly correct for this, it is common to apply a hybrid functional to calculate the DOS [8], as the Hartree-Fock contribution to such a functional stretches the spectrum. We use the PBE0 hybrid functional, which we find has a similar performance as the HSE functional used in Ref. [8], i.e., the band widths in halide perovskites are a few percent too small. Comparing theory and experiment, we are able to determine the necessary stretching to achieve a best fit. The stretching factors for conduction and valence band regions are always kept the same and are at most 8% (listed in Table S4), which is similar to previous reports on $MAPbI_3$ and $CsPbBr_3$ [8].

*Table S4: Stretching factors used on calculated DFT spectra to achieve best agreement with experiment.*

| material | $CsPbI_3$ | $MAPbI_3$ | $FAPbI_3$ | $CsPbBr_3$ | $MAPbBr_3$ | $FAPbBr_3$ | $CsPbCl_3$ | $MAPbCl_3$ | $FAPbCl_3$ |
|---|---|---|---|---|---|---|---|---|---|
| stretching | 1.06 | 1.08 | 1.08 | 1 | 1.02 | 1.05 | 1.06 | 1.04 | 1.06 |
| material | $CsSnI_3$ | $MASnI_3$ | $FASnI_3$ | $CsSnBr_3$ | $MASnBr_3$ | $FASnBr_3$ | $CsSnCl_3$ | $MASnCl_3$ | $FASnCl_3$ |
| stretching | 1 | 1 | 1.04 | 1 | 1.03 | 1.02 | 1 | 1 | 1 |

Neither pure DFT functionals such as PBE, nor hybrid functionals such as PBE0 or HSE give very good band gaps. PBE0 actually gives band gaps that are too large, whereas PBE and HSE give band gaps that are too small, see Ref. [8]. Since we are mainly interested in a realistic representation of the DOS, we used the PBE0 hybrid functional and correct the DFT gap such that it coincides with the experimental optical gap. This was achieved by shifting the CB DOS while keeping the VB constant. As discussed in the main article, this treatment can induce an error equal to the exciton binding energy. However, as this value is typically small, this will not significantly affect the extracted energy level values.

5.2. Broadening of DFT spectra
Next, we broaden the DFT spectra according to the experimental resolution. This is done separately for the VB and CB region, since the resolutions of UPS (VB) and IPES (CB) differ.

A Gaussian function

$$g(x) = \frac{1}{\sqrt{2\pi\sigma^2}} e^{-(x-\mu)^2/2\sigma^2}$$

($\mu$ is the mean value, and $\sigma$ is the variance) is used on each data point of the spectrum to broaden the features. For the VB region, a value of $\sigma = 0.26$ is used and for the CB region $\sigma = 0.34$. Figure S5a shows examples of Gaussian normal distributions for $\mu = 0$ and various values of $\sigma$. Figure S5 b shows as examples the calculated VB DOS of FAPbI$_3$ (black curve) and the same DOS convoluted with a Gaussian function for different values of $\sigma$.

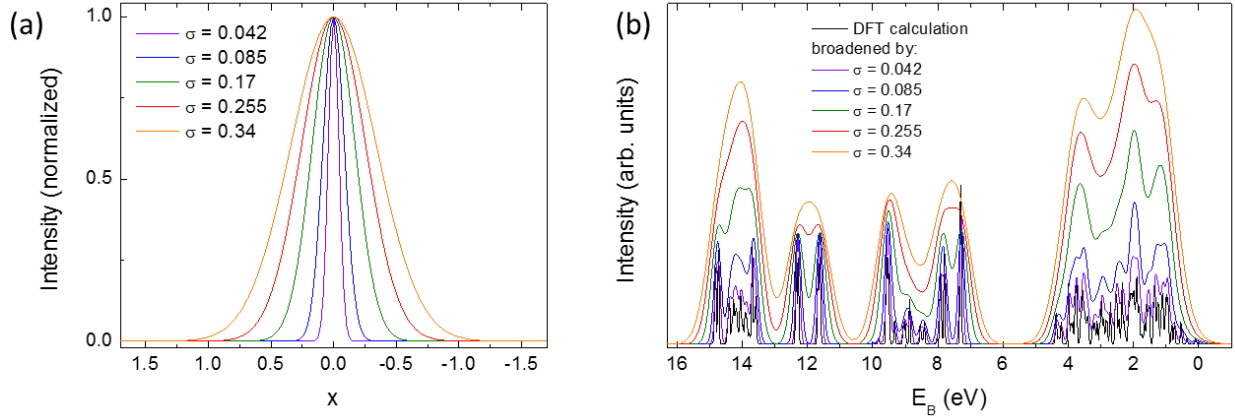

*Fig. S5, a) Gaussian peaks with various broadening. b) example of the theoretical VB DOS of FAPbI$_3$, as calculated (black) and with various broadenings as indicated in the legend.*

5.3 Fitting of the DFT calculated DOS

In order to reliably correlate the calculated to the experimental DOS, as shown for FAPbI$_3$ and FASnI$_3$ in Figure 2 of the main article and for all 18 perovskites in Figure S6 a) - r) below, the calculated DOSs are fitted with a set of Gaussian peaks; these are in principle random, but are chosen in such a way that they describe the different perovskites in a consistent way. While the valence band is fitted over a wide energy range (to approximately 10 eV below the VBM), only the first feature above the CBM is fitted for the CB. This was done since the agreement in the unoccupied DOS between theory and measurement is not sufficient to draw any conclusion from states higher up in the CB.

6. Procedures used for the experimentally measured DOS

6.1 Fitting of the experimentally measured DOS

Next, a similar set of Gaussian peaks is used to fit the experimental DOS in order to be able to align the experimental with the theoretical spectrum as already briefly elaborated in the main text. Fixed distances between the Gaussian peaks, as extracted from the theoretical fit, are employed, while the Full Width Half Max (FWHM) is used as a fitting parameter (this value was then kept constant for different samples of the same material). In the occupied DOS (UPS measurement) good agreement is found between theory and experiment, as shown throughout Figure S6. Here, for each of the 18 perovskites, three separate and representative measurements are used. In analogy to Fig. 2 from the main article, the different colored bars act as guide for the eye to show how different features align and match. Of course intensities of the individual features do not necessarily agree, since they depend on the cross section of the incident light with the material, and therefore on the photon energy.

In the energy region deeper than 6 eV, we find that the theoretical and experimental data deviate. In this energy region we find, for instance, states of the A cation that do not hybridize with the MX$_3$ frame, such as the Cs-related feature in Fig. 3 in the main text. These states are then localized on the A cation. At deeper energies one also finds semicore states of the B cations and X anions, which hybridize little with their environment, and are more localized than the valence states. Standard functionals, either pure density functionals such as PBE, or hybrid functionals such as PBE0, do not give the energy levels of localized states very well, because of the electron self-interaction error associated with these functionals [9]. These localized states are however not relevant for the DOS closer to the VBM. It is important to note that in some cases features at the VB are either broadened compared to theory or there is actually the need to insert an additional peak. This is most notable for $CsPbI_3$, $MASnI_3$, $CsSnBr_3$, $MAPbBr_3$, and $FAPbBr_3$. Such differences can, for example, be introduced by deviations in surface composition, i.e. a specific surface termination. Examples of changes in the shape of the DOS with variations in bulk composition are discussed below in Section 7. In addition, defect states can also induce states at the band edges; one should not forget that UPS is a very surface sensitive technique (probing depth around 1-2 nm) and a surface is always a defect. Such states have been reported to reduce the surface electronic band gap even for cleaved single crystals [10]. This surface effect is less pronounced when higher energy photons are used for excitation since this increases the probing depth [11]. By fitting our measurement to the DFT data, we are however not affected by these states and can distinguish them from the VB and CB DOS. As a further remark to the occupied DOS, the FA cation has a feature that is present only a few eV below the VB onset; its position is indicated as feature D in Figure 1 (main article) and highlighted as a shaded grey peak in the plots in Figure S6. Comparing theory and experiment here, it is obvious that DFT misplaces this feature slightly, which is again due to the more localized nature of this state, which is, of course, localized on the FA cation. In experiment, its position can be placed with high accuracy by comparing the density of states of FA containing species with their MA/Cs counterparts which do not have a contribution by the A-site cation in this region.

Regarding the unoccupied DOS, a similar Gauss fitting procedure is tried for the CB region. As discussed in the main article this turns out to be challenging. We would only like to note here that interestingly, Sn based perovskites always show a much lower intensity at the CBM compared to the Pb analogs. Often the onset is barely visible and hard to distinguish from the background making it absolutely necessary to have the theoretical data to find the correct onset.

6.2 Calculation of IE and EA

The VB and CB onsets (i.e. $E_2^{exp}$ and $E_1^{exp}$) are determined by the position of the first feature in the unbroadened DFT-DOS. Since the spectra have been aligned, as discussed in the previous section, the value can be read out from the dashed line in Figure S6. In practice, the value is extracted by determining the distance between the main peak (for example green shaded area for occupied DOS) and the unbroadened VB onset in the theoretical data set. This offset is then added to the peak values found in the measurement. To calculate the IE and EA, the work function has to be included:

$$IE = Wf + E_2^{exp}, \qquad EA = Wf - E_1^{exp}$$

The values of Wf, IE, and EA for of all samples used for the evaluation (3 samples per material) are listed in Table S6 at the end of this SI.

6.3 All DOS images

The following Figure S6 shows all measurements used to extract IE and EA values listed in the main article; the different materials are divided in the subfigures a) to r). Similar to Fig. 2 in the main article, the theoretical data are shown at the top, while here below three different representative measurements are shown, aligned to the theoretical DOS via the fitting procedure. Sample preparation is varied between these samples to avoid unintentional mistakes in film preparation; additional sample characterization and preparation protocols for each sample are reported at the end of the SI in Section 9 and Table S6, respectively.

(a) CsSnI$_3$

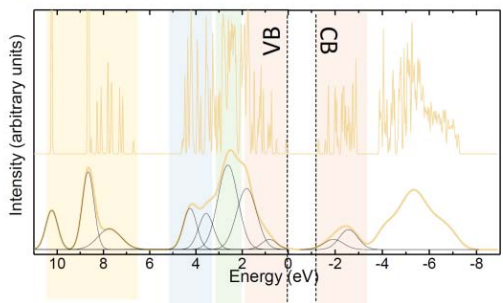

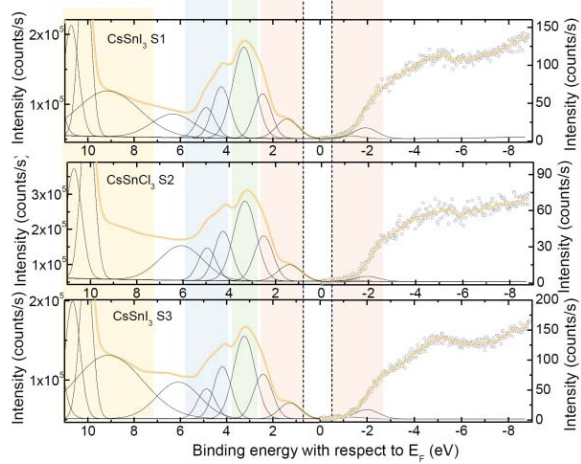

(b) CsPbI$_3$

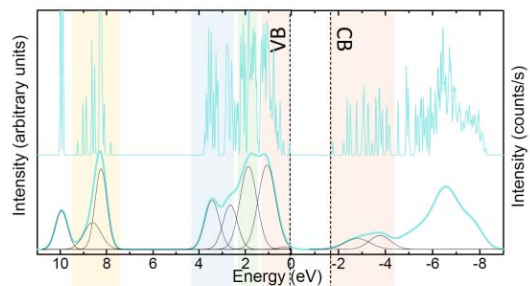

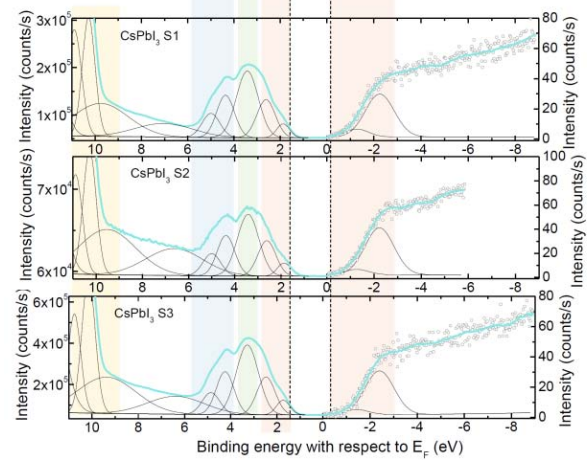

(c) MASnI$_3$

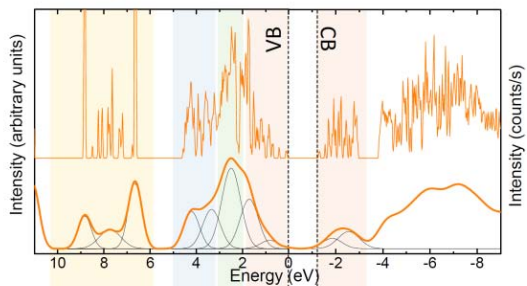

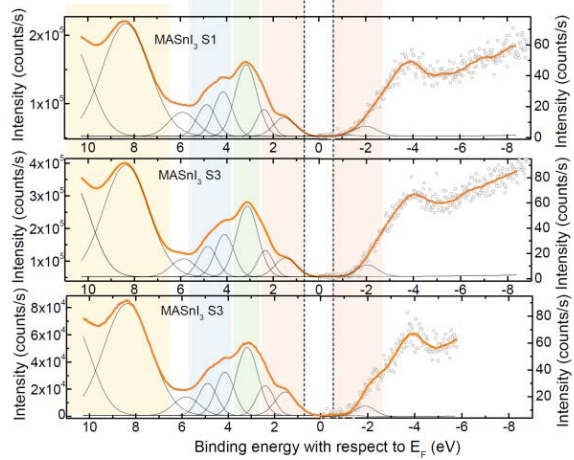

(d) MAPbI$_3$

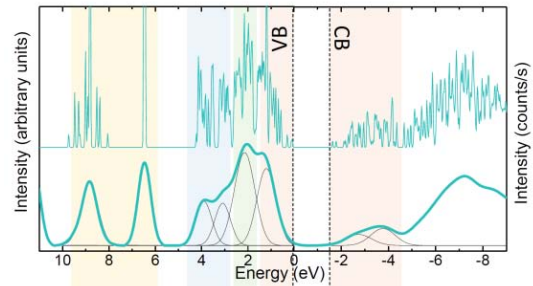

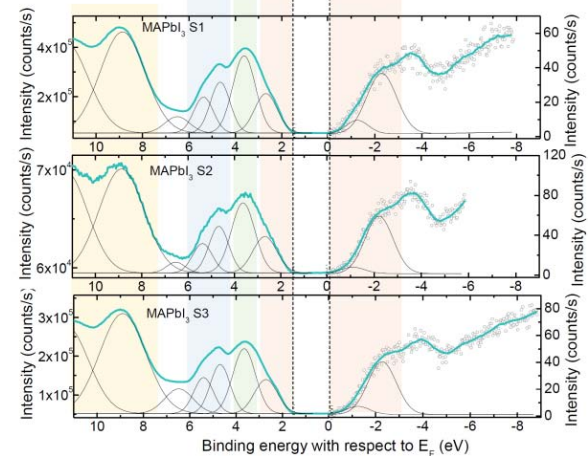

(e) FASnI$_3$

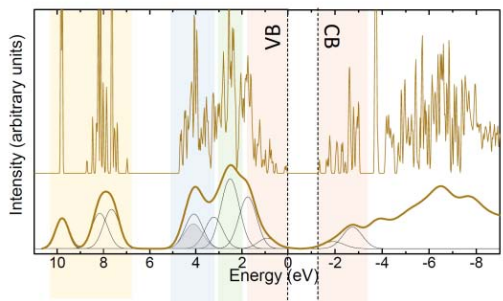

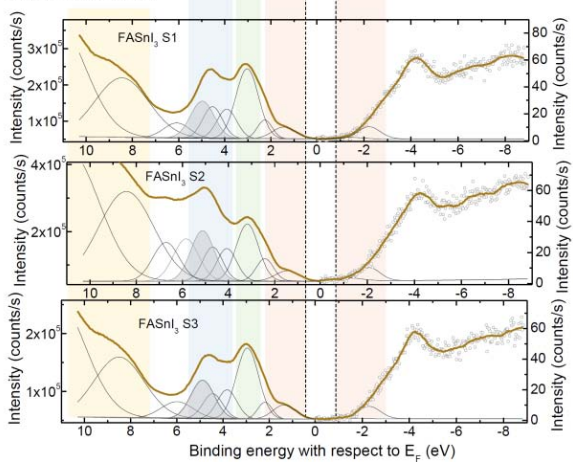

(f) FAPbI$_3$

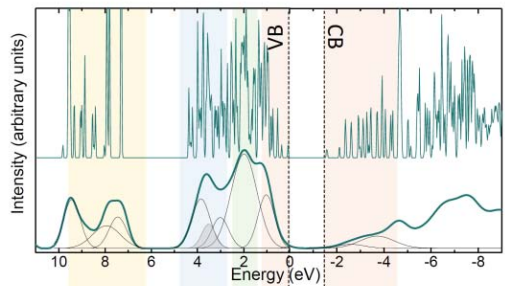

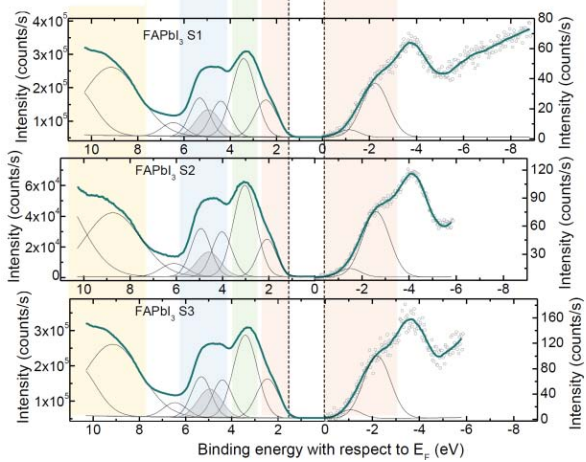

(g) CsSnBr$_3$

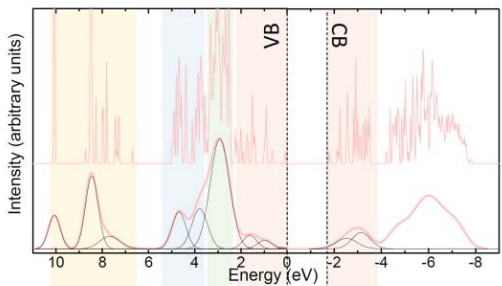

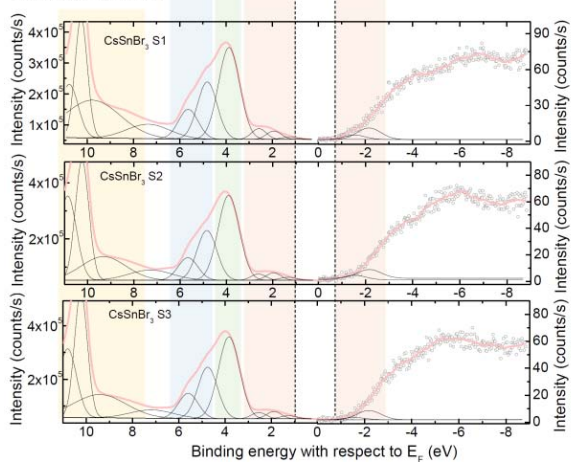

(h) CsPbBr$_3$

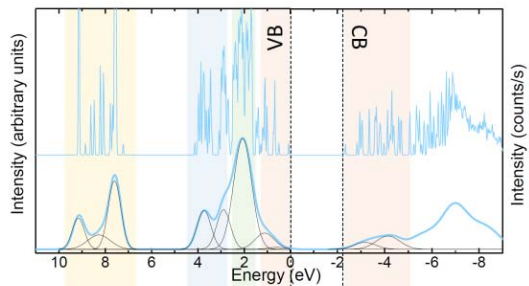

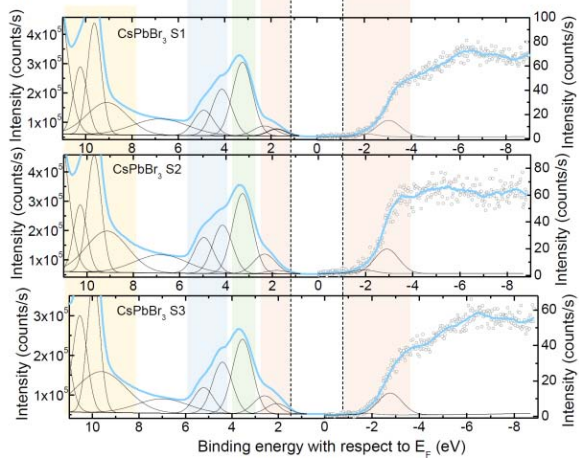

(i) MASnBr$_3$

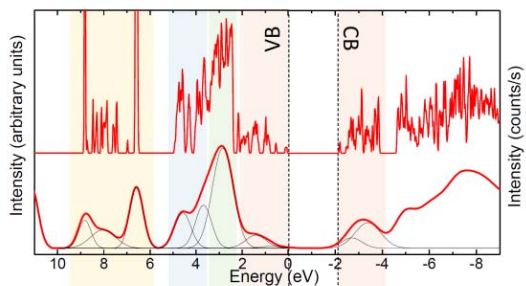

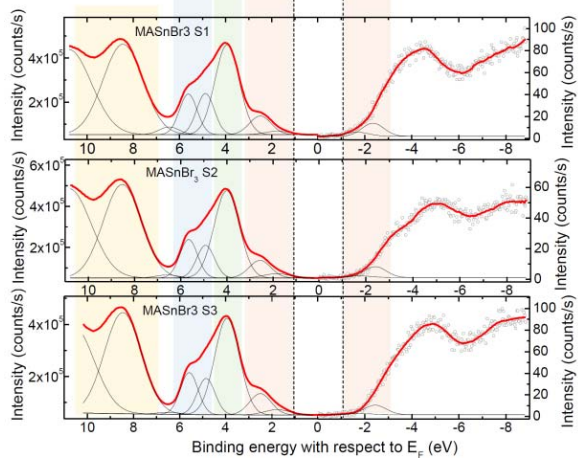

(j) MAPbBr$_3$

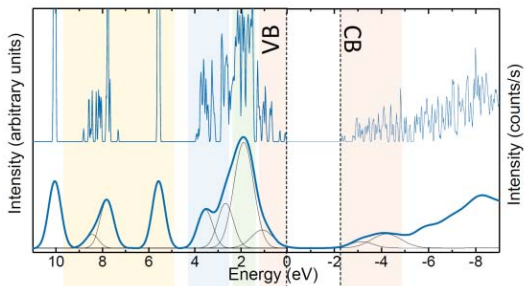

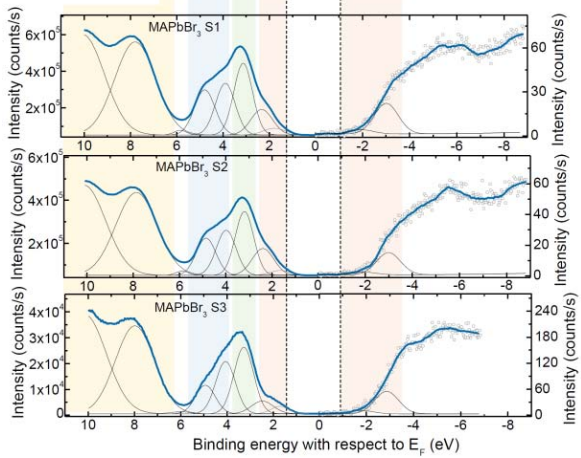

(k) FASnBr$_3$

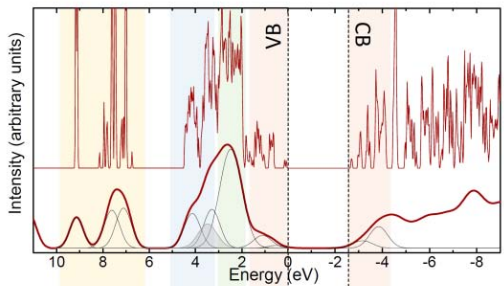

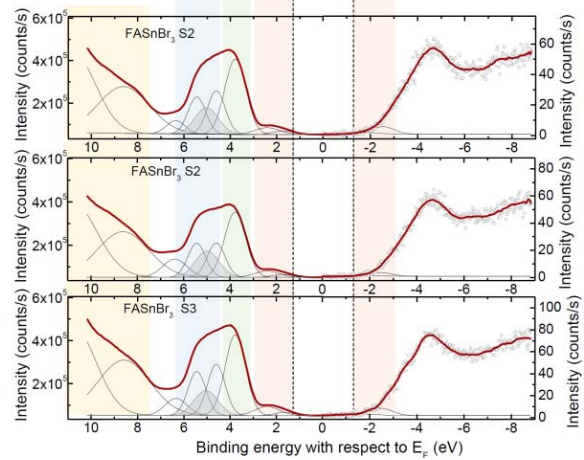

(l) FAPbBr$_3$

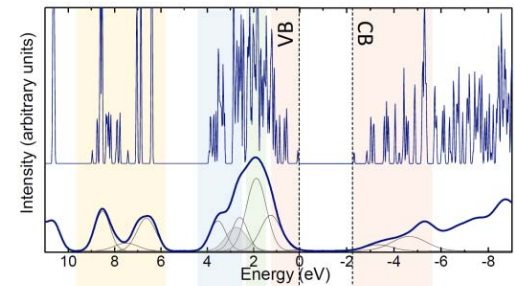

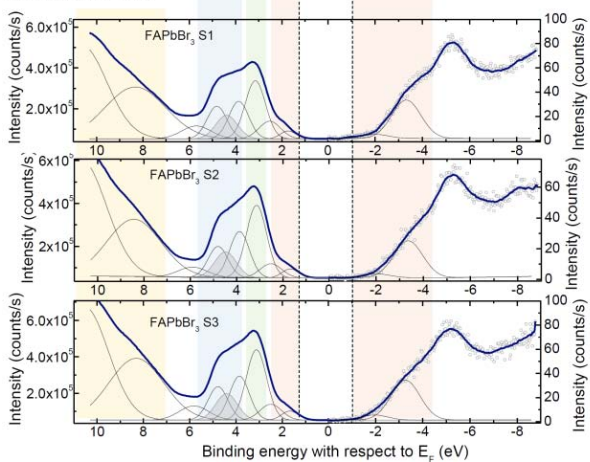

**(m) CsSnCl₃**

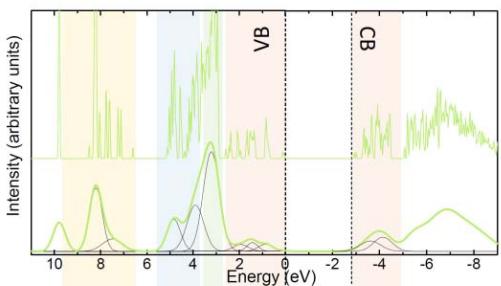
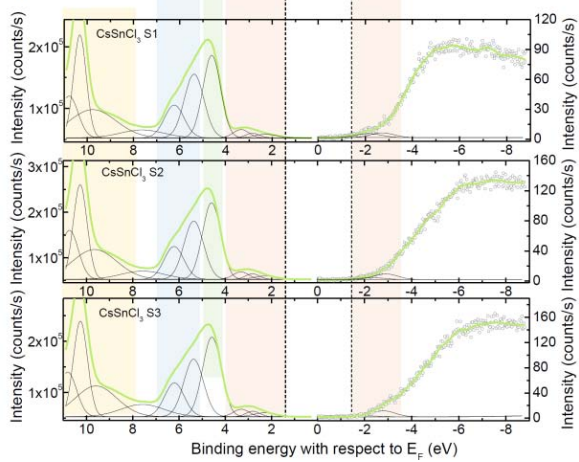

**(n) CsPbCl₃**

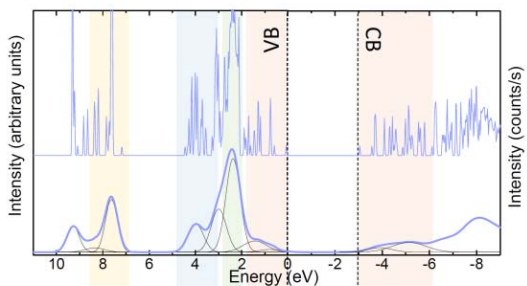
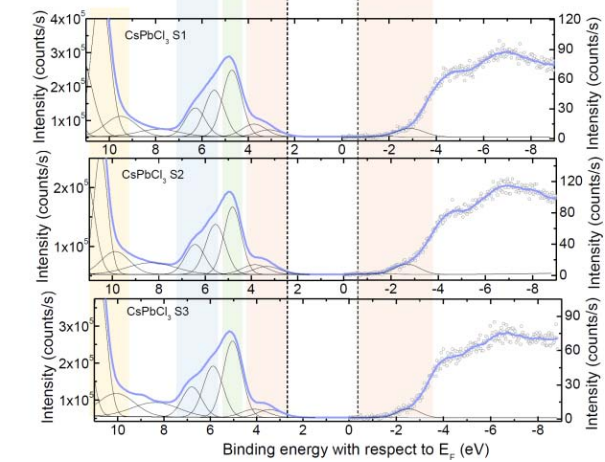

**(o) MASnCl₃**

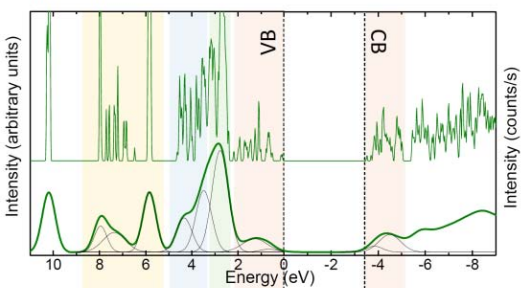
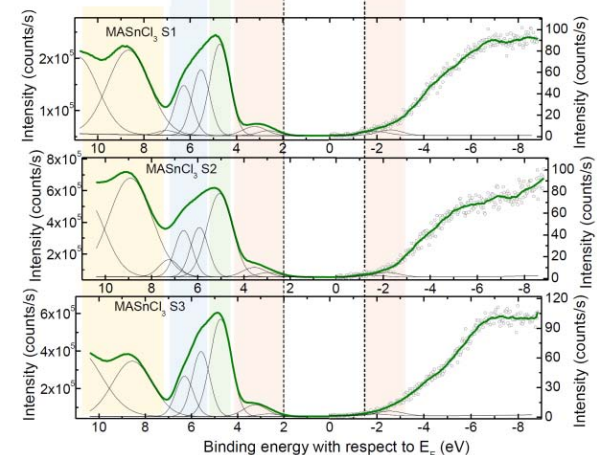

**(p) MAPbCl₃**

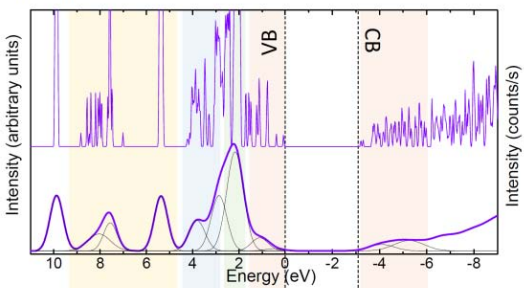
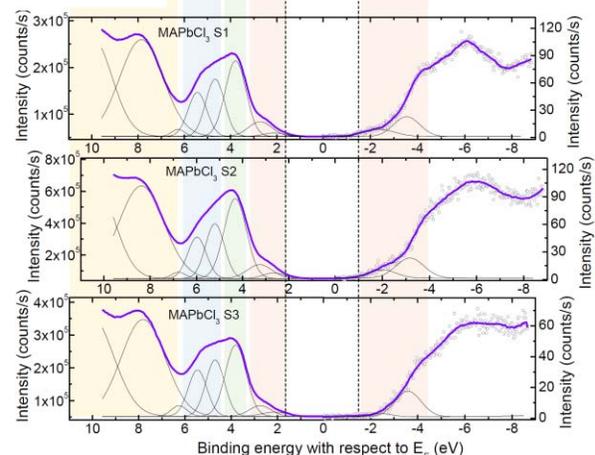

(q) FASnCl$_3$ 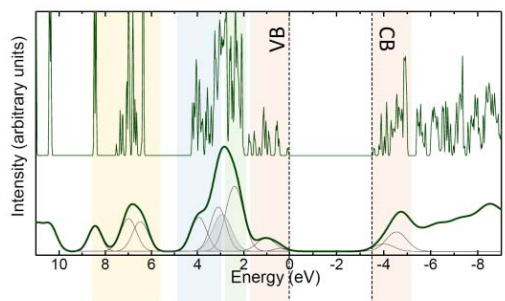
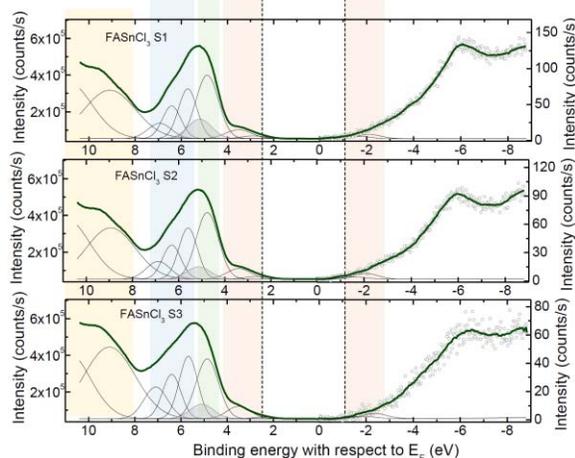

(r) FAPbCl$_3$ 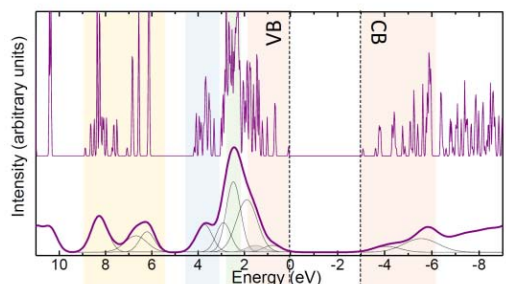
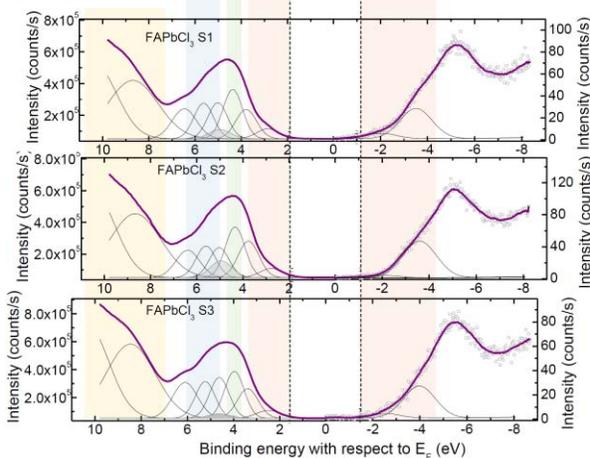

*Figure S6. Comparisons between measured (combined UPS and IPES) and DFT calculated density of states for all 18 perovskite systems, as indicated in the sub-figures a) to r). DFT and experimental spectra are aligned in order to extract the VB and CB onsets. Note that in each top graph, against DFT convention, positive values for VB and negative values for CB are used in analogy to the experimental plot. Shaded peaks show the presence of a FA derived feature, which is situated inside the valence band. In the IPES measurements (negative binding energies) the open circles represent the measurement while the solid line is a smoothed curve through these data point. Sample numbers, such as "S1" etc., indicate the different preparation conditions, which can be found in Table S6 at the end of this SI.*

7. Experimental sample-to-sample variation

The samples for each material shown above in Figure S6 have been selected based on various quality considerations, e.g. XPS stoichiometry, XRD signal, and the correct position of Cs/MA/FA related features (as discussed for Fig. 1 in the main article); these "good" films, therefore, show only little sample-to-sample variation. However, there can be significant variations in the shape of the DOS if the stoichiometry is not met. This will lead to deviations from the calculated DOS due to (i) signals present from additional material phases, or (ii) presence of interstitial or defect states; (iii) in addition, measurements by PES can contain artifacts due to the fact that here only the surface is measured, which will electronically differ to some degree from the bulk and can furthermore be determined by the specific surface termination. While the individual contributions of (i) to (iii) are difficult to quantify here, we would like to show some examples where the stoichiometry (as determined by XPS) deviates from the expected one either due to overall off-stoichiometry or the accumulation of certain phases at the surface.

Three examples are presented, with different degrees of sample-to-sample variation; these "bad" samples are of course not included in the data evaluation throughout this work but help here to identify good film composition.

How much the features at the VBM and CBM were found to be affected by composition strongly depended on the material. For example, CsSnI$_3$, MASnI$_3$, MASnCl$_3$, CsPbBr$_3$, and MASnBr$_3$ were quite sensitive, while MAPbI$_3$, FAPbI$_3$, CsPbI$_3$, FASnI$_3$, CsSnCl$_3$, or FASnBr$_3$ were much less affected in the band onset region by preparation conditions.

### 7.1 Example of CsSnI$_3$

In order to compare the different perovskite spectra here and in the following two examples, the measurements have been shifted such that the VB features overlap; this way variations in Wf are eliminated and the position of $E_F$ is arbitrary. The first example, presented in Figure S7, is CsSnI$_3$ which showes one of the most extreme variations in both VB onset region and unoccupied DOS. Samples with additional CsI (as observed by XPS) show additional DOS in the region between 4 and 8 eV. Comparing this with pure CsI spectrum (red curve) it is clear that this is simply the DOS of CsI appearing. At the same time the high binding energy cutoff shifts to the left, meaning that the Wf decreases and accordingly (since spectra are aligned at VB edge) the IE decreases. With excess SnI$_2$ on the other hand, a more pronounced DOS at the VB onset is observed. This most likely originates from the strong VB feature between 1 and 2eV of an additional SnI$_2$ phase, as seen in the bottom spectrum (green). The surface of CsSnI$_3$ seems to be extremely sensitive, switching easily from CsI rich to SnI$_2$ rich due to minor changes in composition. For the unoccupied DOS, measured by IPES, we find a strong influence as well. Probably the feature between -2 and -4 eV originates from the SnI$_2$ as well and vanishes for the CsI rich films.

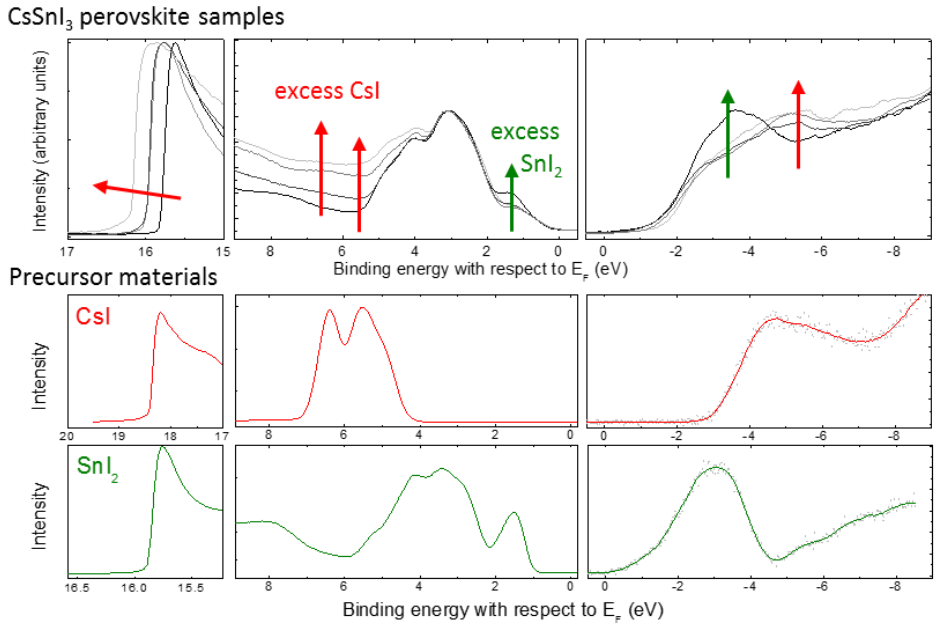

*Figure S7. Examples of sample-to-sample variation in UPS/IPES measured DOSs for CsSnI$_3$, where also samples with "bad" stoichiometry are included (upper panel). The two bottom graphs show measurements of the pure precursor materials (CsI and SnI$_2$) used for this perovskite preparation.*

### 7.2 Example of MAPbI$_3$

Sample-to-sample variations for MAPbI$_3$ are shown in Figure S8. Here, changes close to the band onsets are not very pronounced even though again the Wf changes, meaning that the IE changes as well (as previously reported by us in Ref. [12]). The overall similarity of the "good" and "bad" stoichiometry films could be due to the fact that MAPbI$_3$, MAI, and PbI$_2$ all have somewhat similar shapes of the VB region and also in the CB region MAPbI$_3$ and PbI$_2$ are difficult to distinguish. More pronounced changes are found deeper in the band, around 4 to 7 eV in the VB region and -4 to -7 in the CB region. Here, with excess MAI the DOS increases which can be correlated to features in the pure MAI spectrum; these indicate therefore that excess MAI is present.

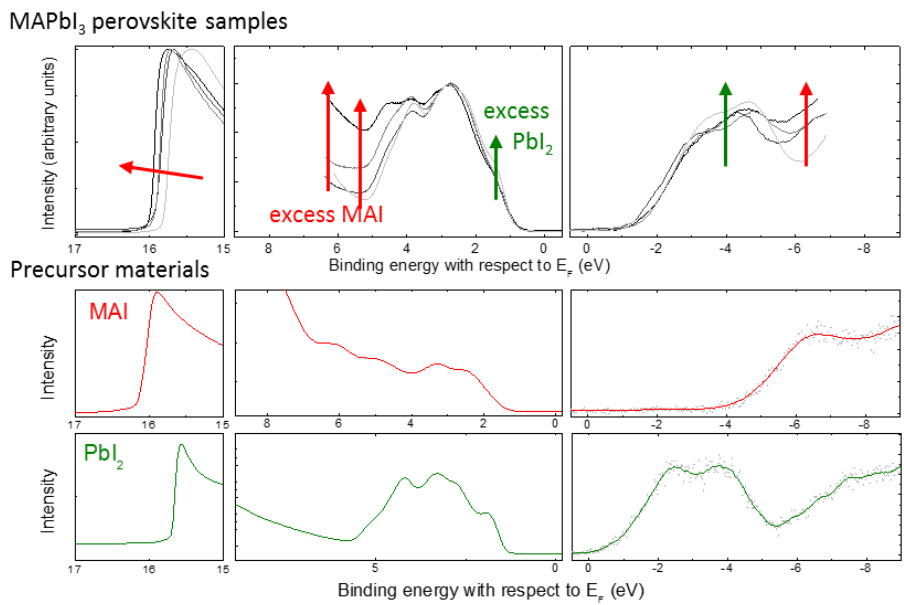

Figure S8. *Examples of sample-to-sample variation in UPS/IPES measured DOSs for MAPbI$_3$, where also samples with "bad" stoichiometry are included (upper panel). The two bottom graphs show measurements of the pure precursor materials (MAI and PbI$_2$) used for this perovskite preparation.*

7.3 Example of MAPbCl$_3$

Finally, in Figure S9 MAPbCl$_3$ is shown. Since the solubility of PbCl$_2$ is limited, no sample ever showed an excess in PbCl$_2$ content. However, we had a large number of samples with different degrees of MACl excess which are shown here. Strong variations around 4eV are observed in the occupied DOS, which must originate from excess MACl material that has a pronounced feature here. The DOS of PbCl$_2$ shows characteristic features deeper in the bands, around 5 to 6 eV and -6 to -8 eV, where variations can also be found. Similar to the previous two examples, with increasing amount of MACl the high binding energy cutoff shifts, resulting in a decrease of the Wf and therefore a decrease in IE.

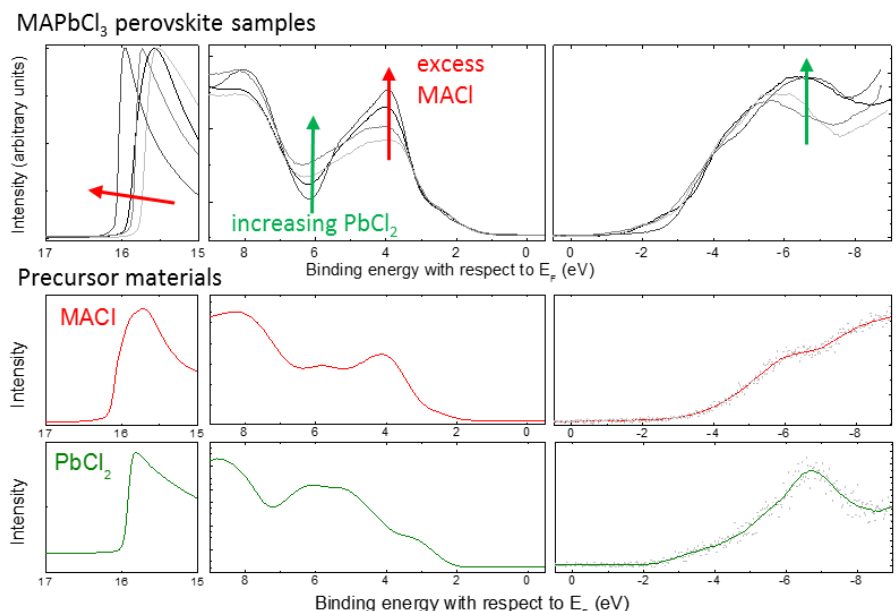

*Figure S9: Examples of sample-to-sample variation in UPS/IPES measured DOSs for MAPbCl$_3$, where also samples with "bad" stoichiometry are included (upper panel). The two bottom graphs show measurements of the pure precursor materials (MACl and PbCl$_2$) used for this perovskite preparation.*

## 8. Description of additional measurement techniques

Various additional measurement techniques were used to characterize the perovskite films, especially during the initial optimization steps of material preparation. The resulting measurements are shown and shortly discussed in Section 9 below.

### Absorption measurements (UV-vis)
Absorption measurements were performed in air using a Cary 50 UV-Vis spectrometer from Varian in ranges from 250 - 1100 nm with a scan rate of 60 nm/min. The spectrum of a reference sample, corresponding to the substrate (glass/ ITO / PEDOT:PSS), was subtracted from each measurement.

### X-ray photoelectron spectroscopy (XPS)
XPS measurements were performed in the same setup as UPS, using also the hemispherical electron energy analyzer (Phoibos 100, Specs). For excitation, a non-monochromatic x-ray source with Mg anode was used, having a photon energy of hν = 1253.6 eV. The survey scans are taken with a pass energy of 70 eV and the detailed peaks with a pass energy of 10 eV; here, the energy resolution is approximately 800 meV. Peak fits of the XPS peaks were done using the program *XPS Peak Fit 4.1* and film stoichiometry was determined by evaluating the individual peak areas and correcting those with the relative sensitivity factors (RSF). These RSF factors were calibrated for this system relative to carbon; the values used are RSF(N 1s) = 1.8, RSF(Cs $3d_{5/2}$) = 40.3, RSF(I $3d_{5/2}$) = 32.8, RSF(Br $3d_{5/2}$) = 2.13, RSF(Cl $2p_{3/2}$) = 1.8, RSF(Pb $4f_{7/2}$) = 16.5, and RSF(Sn $3d_{5/2}$) = 24.

### X-ray diffraction (XRD)
XRD spectra were recorded using a Panalytical Empyrean system with an excitation via a Cu Kα anode (λ = 1.54056Å); the angle 2Θ was scanned between 10 and 40° with a step width of 0.0131°. For most of the measurements Glass/PEDOT:PSS was used as substrate, while in some cases Glass/ITO/PEDOT:PSS was used. The latter ones produce additional reflexes from the ITO layer, which are indicated in the measurements in Section 9. The measurements are performed in air. Here, to prevent degradation, the sensitive perovskite layers (e.g. tin-based ones or $CsPbI_3$/$FAPbI_3$) were coated with PMMA (30 mg/ml in Toluene, spin coating at 3000 rpm for 30 s) to form a protective barrier layer on top.

### Scanning electron microscopy (SEM)
SEM measurements were performed in a Zeiss Neon-40 system, images shown here are taken with an In-lens detector. Images of different magnifications are shown, since samples showed structures on rather different length scales.

## 9. Material measurement sheets
**CsSnI₃ data sheet**

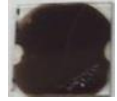

XRD measurement:

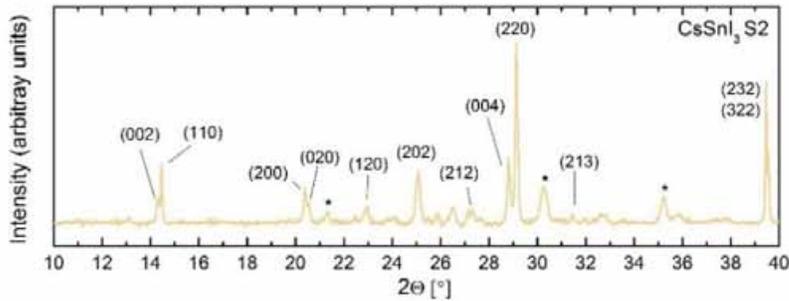

Orthorhombic crystal structure

a = 8.71 Å
b = 8.65 Å
c = 12.38 Å

*XRD measurement of CsSnI₃, sample 2; reflexes of the orthorhombic structures are marked and extracted lattice constants are displayed on the right. Reflexes marked with * come from the ITO substrate.*

SEM measurement:                                  Absorption measurement:

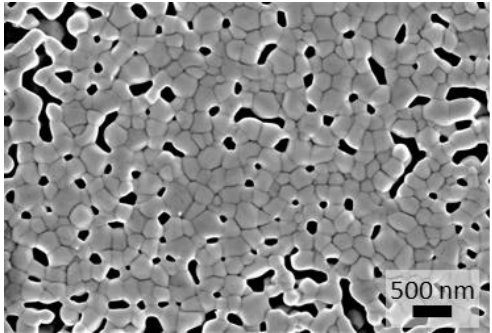 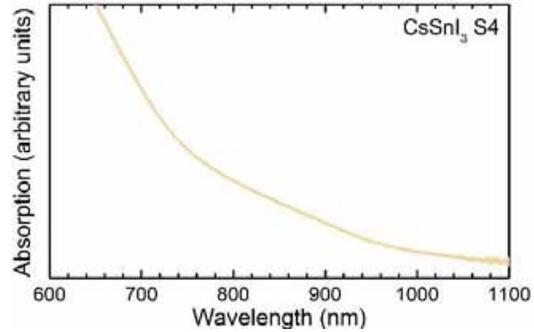

*SEM image of CsSnI₃, sample 2.*        *UV-vis measurement of CsSnI₃, sample 4.*

XPS analysis:

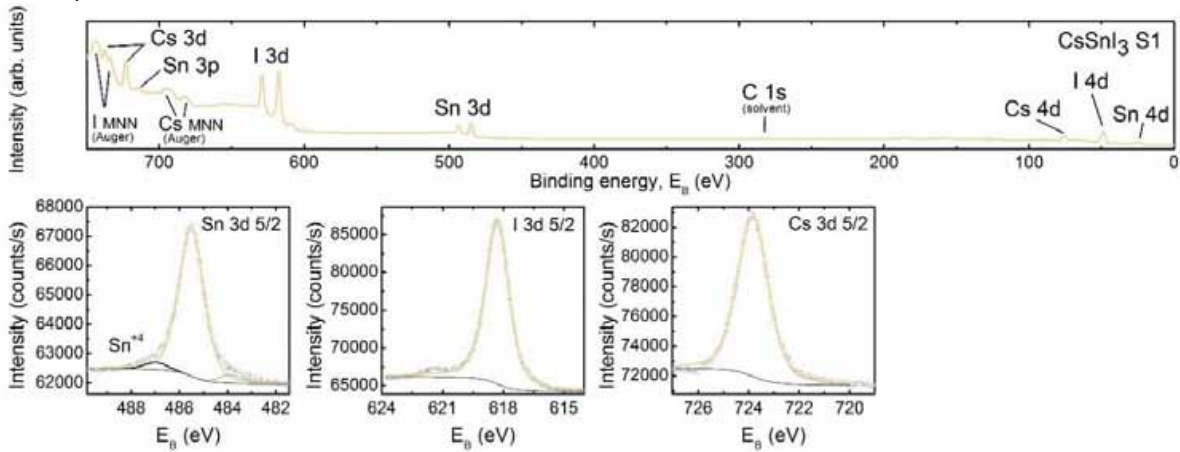

*XPS measurement of CsSnI₃, sample 1, showing a survey spectrum on top as well as detailed scans of perovskite specific core level signals at the bottom; the solid lines are fits to the measurement (open circles).*

Short discussion: In XRD, we find a large number of peaks, indicating an unordered film growth, which can be clearly associated with an orthorhombic crystal structure. Notable in the SEM image, it is usually difficult to form densely packed films for this material, probably due to limited solubility of CsI. XPS shows all expected features and oxidation states, but also a small shoulder for Sn at higher binding energies, likely originating from $Sn^{+4}$ (approx. 6%). Note that the feature in the Sn signal at lower binding energy comes from an asymmetric peak shape and is not a metallic Sn feature (it is equally present in all Sn samples). UV-vis measurements yield an optical gap of 1.25 eV.

**MASnI₃ data sheet**

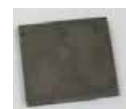

XRD measurement:

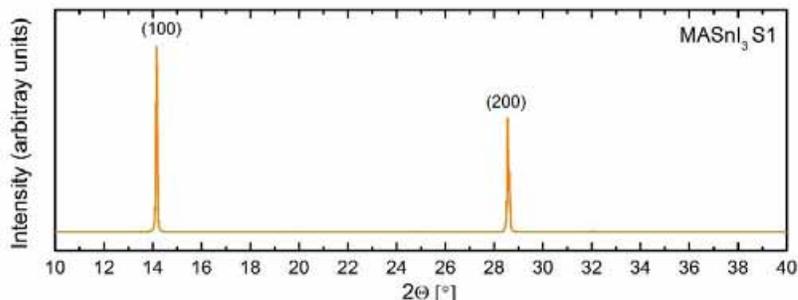

Cubic crystal structure

a = 6.25 Å

*XRD measurement of MASnI₃, sample 1; reflexes of the cubic structure are marked and the extracted lattice constant is displayed on the right.*

SEM: 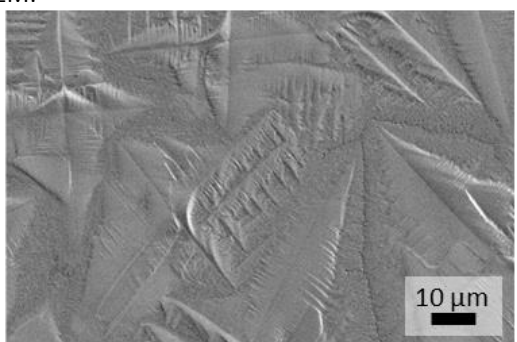  Absorption: 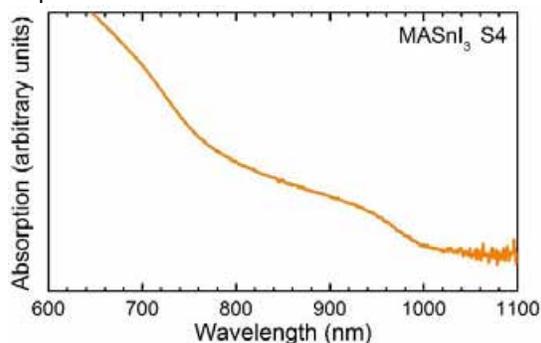

*SEM image of MASnI₃, sample 1.*     *UV-vis measurement of MASnI₃, sample 4.*

XPS analysis:

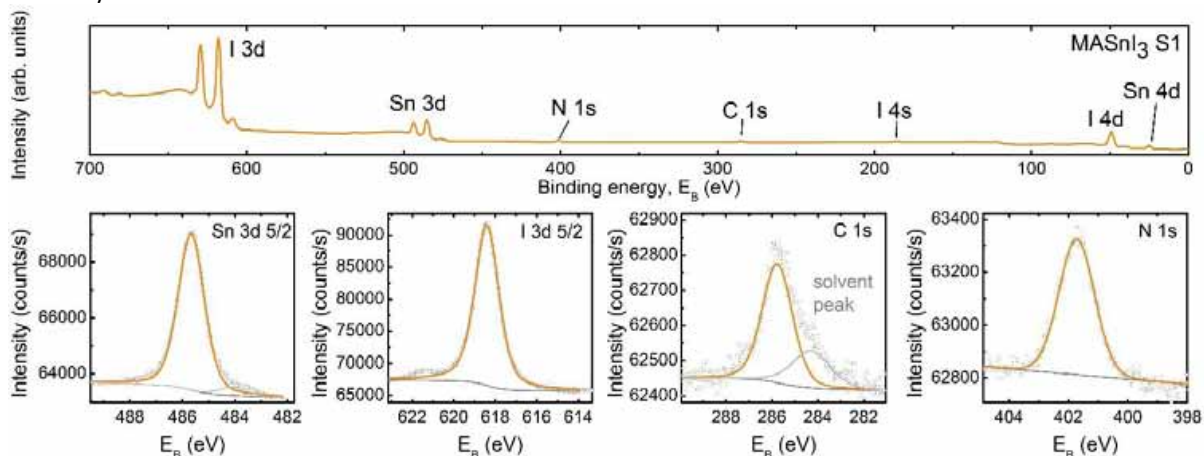

*XPS measurement of MASnI₃, sample 1, showing a survey spectrum on top as well as detailed scans of perovskite specific core level signals at the bottom; the solid lines are fits to the measurement (open circles).*

Short discussion: XRD of MASnI₃ consistently shows highly ordered films, with only (h00) reflexes, indicating a preferred crystallite orientation. The low number of reflexes does not allow to unambiguously tell the crystal structure, but from powder diffraction studies in literature it is known to be cubic [13]. This material forms large intercalated crystals, usually in a "leaf" like structure as seen in the SEM image. XPS shows all expected features and oxidation states with some additional more neutral carbon species (likely solvent remaining in the film). Note that the feature in the Sn signal at lower binding energy comes from an asymmetric peak shape and is not a metallic Sn feature (it is equally present in all Sn samples). UV-vis measurements yield an optical gap of 1.24 eV.

**FASnI$_3$ data sheet**

XRD measurement:

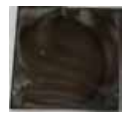

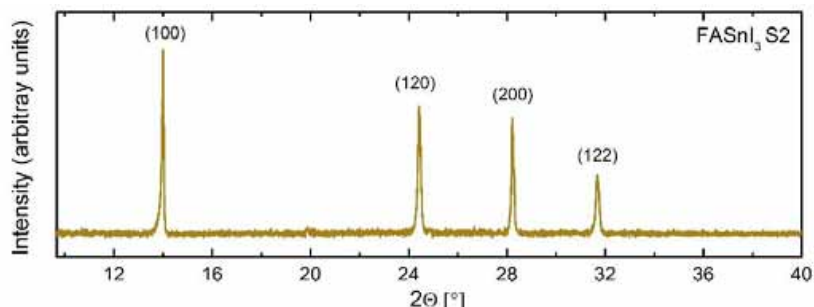

Tetragonal crystal structure

a = 6.32 Å
b = c = 8.93 Å

*XRD measurement of FASnI$_3$, sample S2; reflexes of the tetragonal structure are marked and extracted lattice constants are displayed on the right.*

SEM:                                             Absorption:

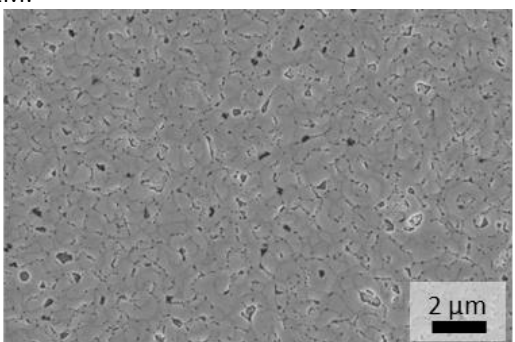     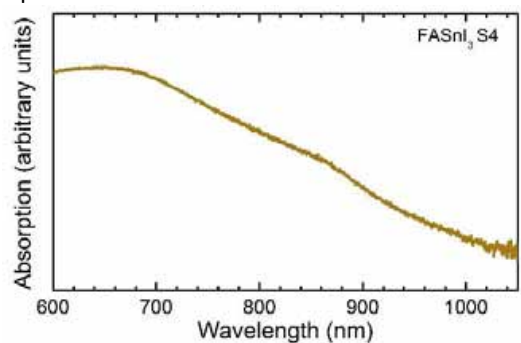

*SEM image of FASnI$_3$, sample 1.*           *UV-vis measurement of FASnI$_3$, sample 4.*

XPS analysis:

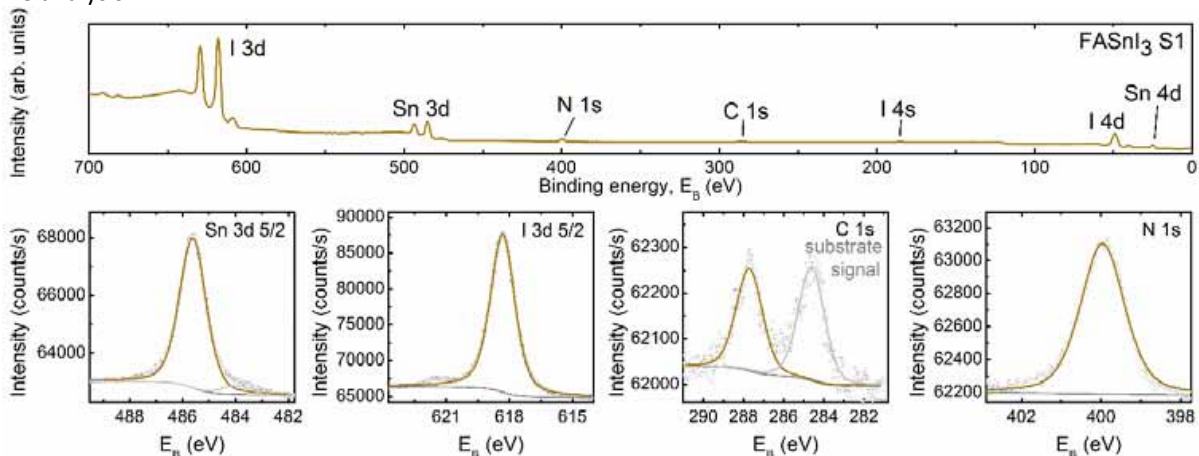

*XPS measurement of FASnI$_3$, sample 1, showing a survey spectrum on top as well as detailed scans of perovskite specific core level signals at the bottom; the solid lines are fits to the measurement (open circles).*

Short discussion: XRD of FASnI$_3$ consistently shows all reflexes associated with a cubic structure indicating that no preferred crystal orientation is present. It is challenging to form dense films, usually "leaf" like structures similar to the MASnI$_3$ case are observed (not shown) or smoother layers with some pinholes, as shown in the SEM image. XPS shows all expected features and oxidation states, the additional carbon peak likely originates from the substrate that is visible through pin holes. Note that the feature in the Sn signal at lower binding energy comes from an asymmetric peak shape and is not a metallic Sn feature (it is equally present in all Sn samples). UV-vis measurements yield an optical gap of 1.24 eV, the onset is weak since the detector is not working efficiently any more in this energy range.

**CsSnBr₃ data sheet**

XRD measurement:

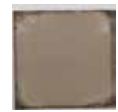

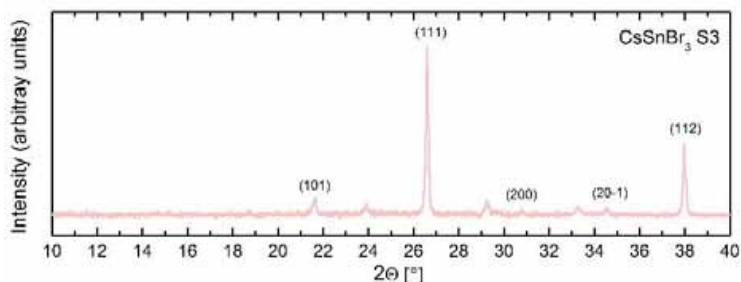

Cubic crystal structure

a = 5.81 Å

*XRD measurement of CsSnBr₃, sample 3; reflexes of the cubic structure are marked and the extracted lattice constant is displayed on the right.*

SEM:                                                Absorption:

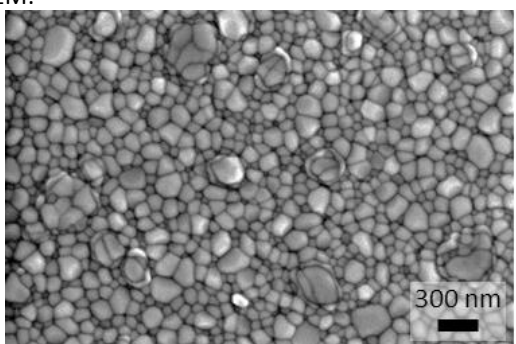 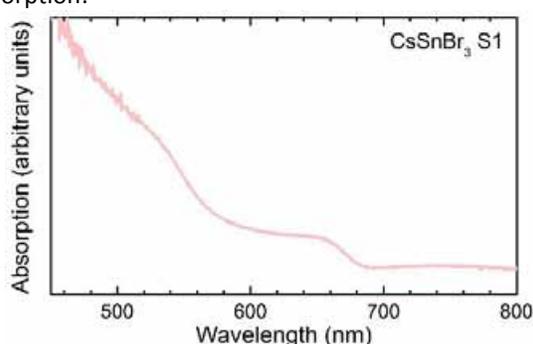

*SEM image of CsSnBr₃, sample 1.*      *UV-vis measurement of CsSnBr₃, sample 1.*

XPS analysis:

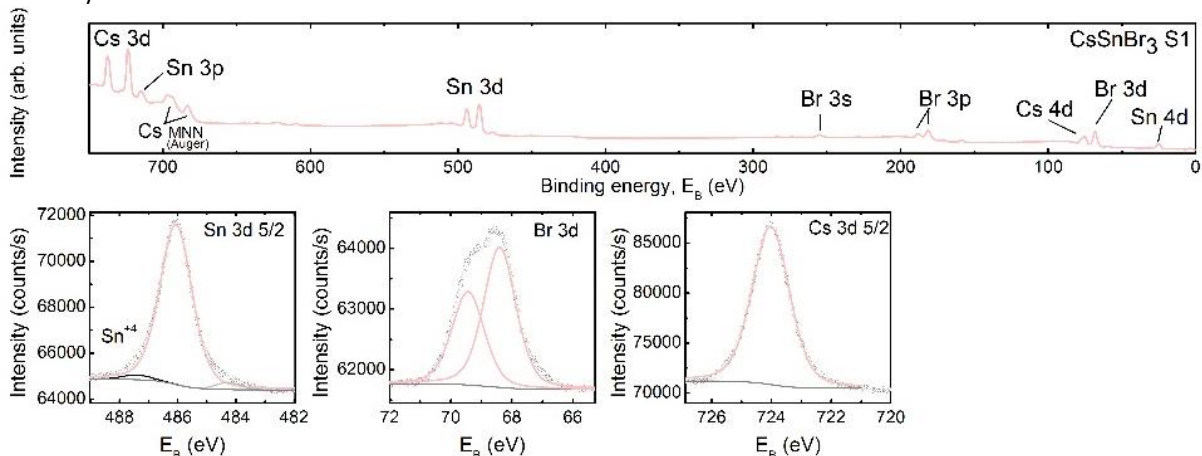

*XPS measurement of CsSnBr₃, sample 1, showing a survey spectrum on top as well as detailed scans of perovskite specific core level signals at the bottom; the solid lines are fits to the measurement (open circles).*

Short discussion: In XRD measurements of evaporated CsSnBr₃, such as the one shown above, the (h00) peak is missing, indicating preferred film grown along the (111) direction for this cubic system, which is in contrast to other perovskite materials. Samples prepared by solution processing usually show no preferred orientation, but always a residual CsBr feature appears due to issues with solubility; no dense films can be formed. Therefore, only evaporated layers are presented in this paper, which form dense crystalline films, as seen in the SEM image. XPS shows all expected features and oxidation states, but also a small shoulder at higher binding energies for Sn, likely originating from $Sn^{+4}$ (less than 3% signal). Note that the feature in the Sn signal at lower binding energy comes from an asymmetric peak shape and is not a metallic Sn feature (it is equally present in all Sn samples). UV-vis yields an optical gap of 1.81 eV.

**MASnBr₃ data sheet**

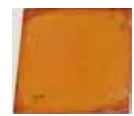

XRD measurement:

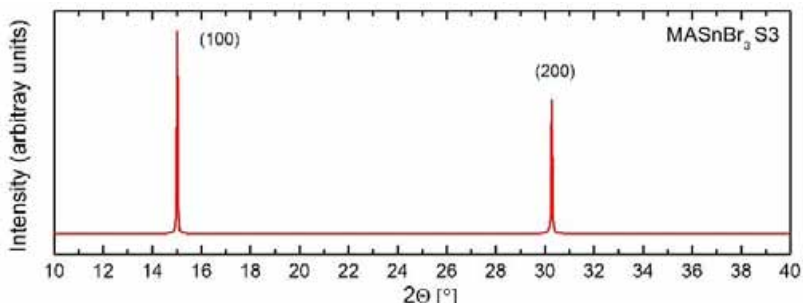

Cubic crystal structure

a = 5.90 Å

*XRD measurement of MASnBr₃, sample 3; reflexes of the cubic structure are marked and the extracted lattice constant is displayed on the right.*

SEM: 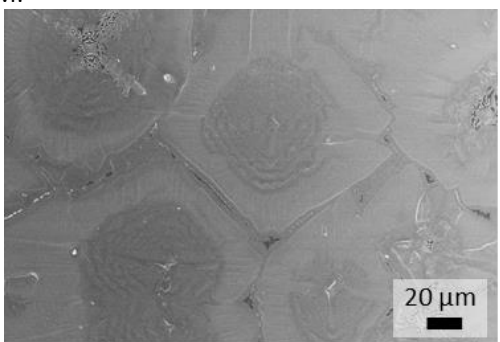   Absorption: 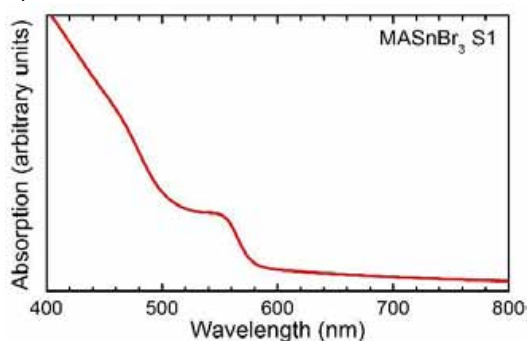

*SEM image of MASnBr₃, sample 3.*   *UV-vis measurement of MASnBr₃, sample 1.*

XPS analysis:

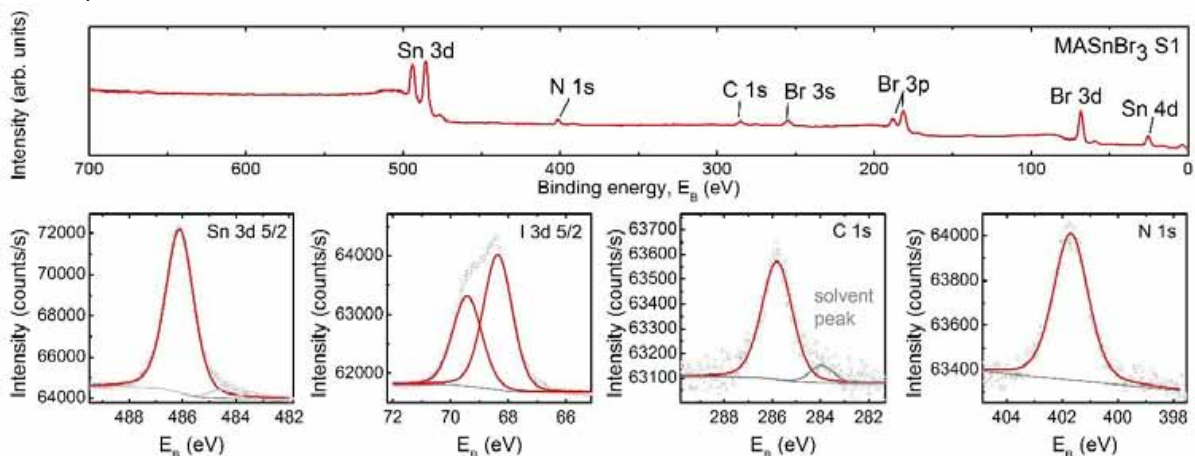

*XPS measurement of MASnBr₃, sample 1, showing a survey spectrum on top as well as detailed scans of perovskite specific core level signals at the bottom; the solid lines are fits to the measurement (open circles).*

Short discussion: XRD of MASnBr₃ consistently shows highly ordered films, with only (h00) reflexes, indicating a preferred crystal orientation. The low number of reflexes does not allow to unambiguously identify the crystal structure, but from powder diffraction studies in literature it is known to be cubic [14]. Like many of the tin based systems, this material forms large domains as seen in the SEM image. XPS shows all expected features and oxidation states with some additional more neutral carbon species (likely solvent remaining in the film). Note that the feature in the Sn signal at lower binding energy comes from an asymmetric peak shape and is not a metallic Sn feature (it is equally present in all Sn samples). UV-vis yields an optical gap of 2.15 eV.

**FASnBr₃ data sheet**

XRD measurement:

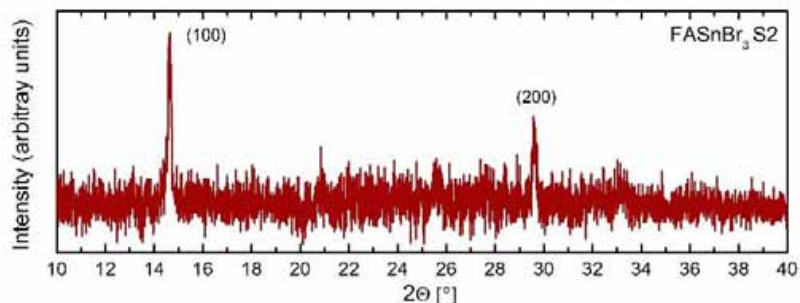
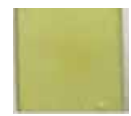

Cubic crystal structure

a = 6.05 Å

*XRD measurement of FASnBr₃, sample 2; reflexes of the cubic structure are marked and the extracted lattice constant is displayed on the right.*

SEM: 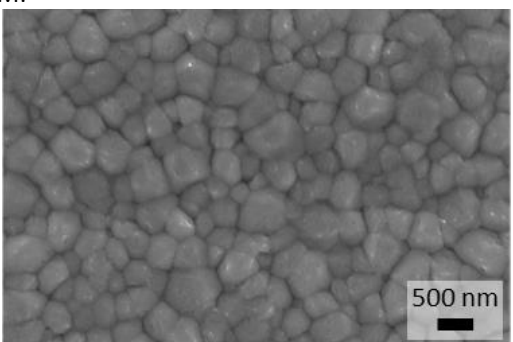  Absorption: 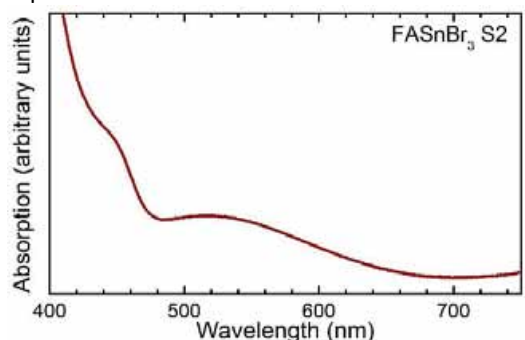

*SEM image of FASnBr₃, sample 3.*   *UV-vis measurement of FASnBr₃, sample 2.*

XPS analysis:

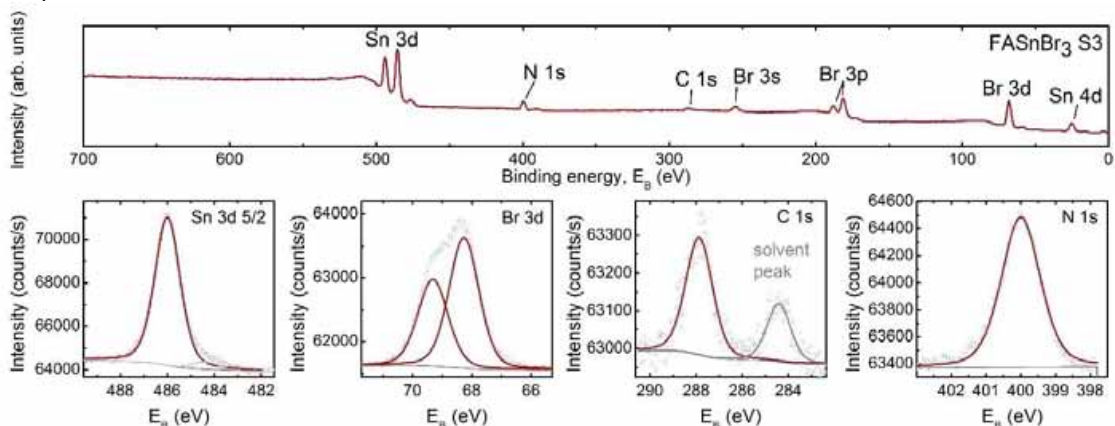

*XPS measurement of FASnBr₃, sample 3, showing a survey spectrum on top as well as detailed scans of perovskite specific core level signals at the bottom; the solid lines are fits to the measurement (open circles).*

Short discussion: In XRD we are only able to record rather noisy spectra, indicating that the crystallinity is not very good, even though SEM measurements show nicely formed films with typically close-packed crystallites. There seems to be a preferred orientation, but from the few reflexes it is not possible to unambiguously determine the crystal structure; from literature we can infer that it should be cubic [15]. This low crystallinity is consistent with the discussion in the main article, where we find that lattice must be very distorted due to the size difference between the small Sn/Br and the large FA, leading to the destabilization of the valance band. XPS shows all expected features and oxidation states with some additional more neutral carbon species (likely solvent remaining in the film). Note that the feature in the Sn signal at lower binding energy comes from an asymmetric peak shape and is not a metallic Sn feature (it is equally present in all Sn samples). UV-vis yields an optical gap of 2.63 eV.

**CsSnCl₃ data sheet**

XRD measurement:

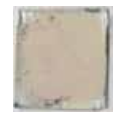

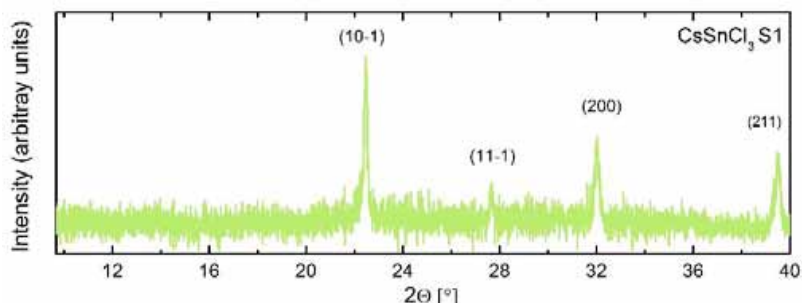

Cubic crystal structure

a = 5.59 Å

*XRD measurement of CsSnCl₃, sample 1; reflexes of the cubic structure are marked and the extracted lattice constant is displayed on the right.*

SEM:  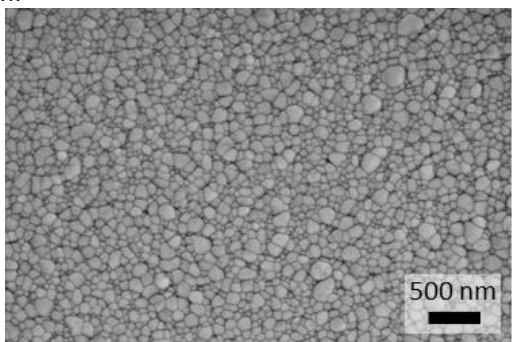

Absorption: 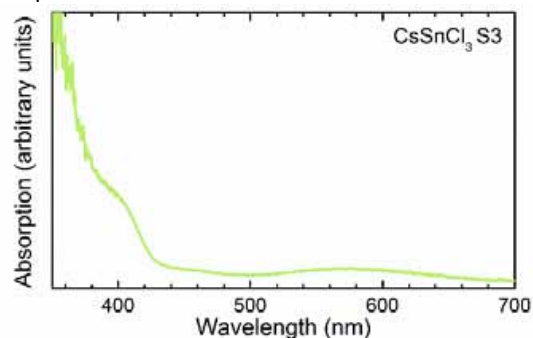

*SEM image of CsSnCl₃, sample 3.*  *UV-vis measurement of CsSnCl₃, sample 3.*

XPS analysis:

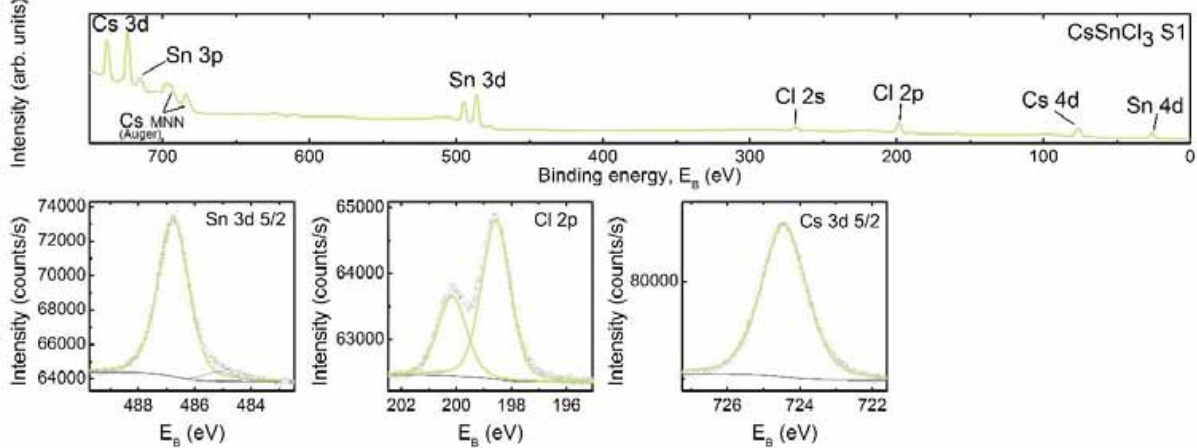

*XPS measurement of CsSnI₃, sample 1, showing a survey spectrum on top as well as detailed scans of perovskite specific core level signals at the bottom; the solid lines are fits to the measurement (open circles).*

Short discussion: All samples were evaporated, since by solution processing no perovskite could be formed (limited solubility of CsCl). Generally, this material does not show good crystallinity in XRD and similar to the case of CsSnBr₃ the (100) reflex is missing, indicating a preferred orientation. The few reflexes indicate a cubic structure, which is in agreement with high temperature measurements reported in literature [16]. SEM images show densely packed crystallites, which is typical for vapor deposition. XPS shows all expected features and oxidation states. Note that the feature in the Sn signal at lower binding energy comes from an asymmetric peak shape and is not a metallic Sn feature (it is equally present in all Sn samples). UV-vis measurements yield an optical gap of 2.88 eV.

**MASnCl₃ data sheet**

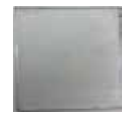

XRD measurement:

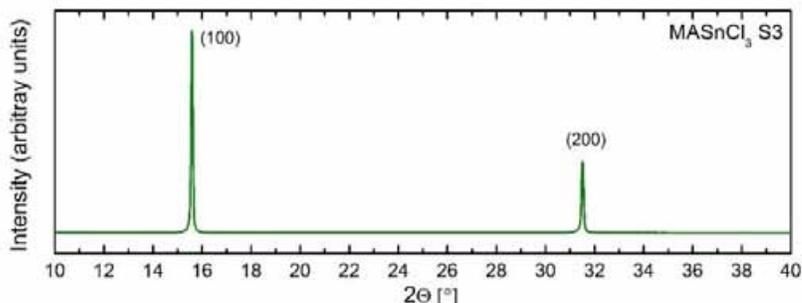

Monoclinic crystal structure

*lattice constants cannot be extracted from this data set.*

*XRD measurement of MASnCl₃, sample 3; detectable reflexes are marked.*

SEM: 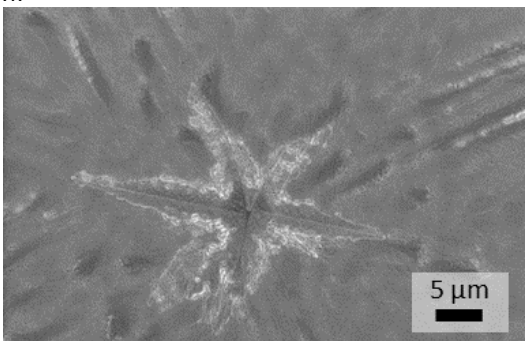

Absorption: 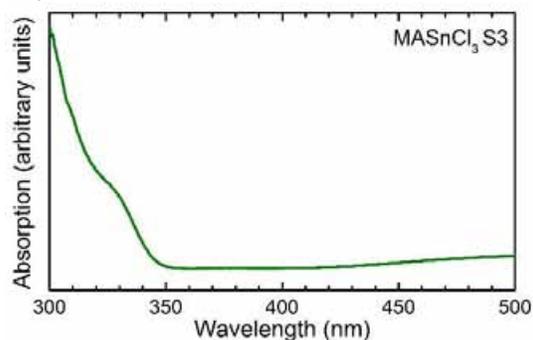

*SEM image of MASnCl₃, sample 3.*   *UV-vis measurement of MASnCl₃, sample 3.*

XPS analysis:

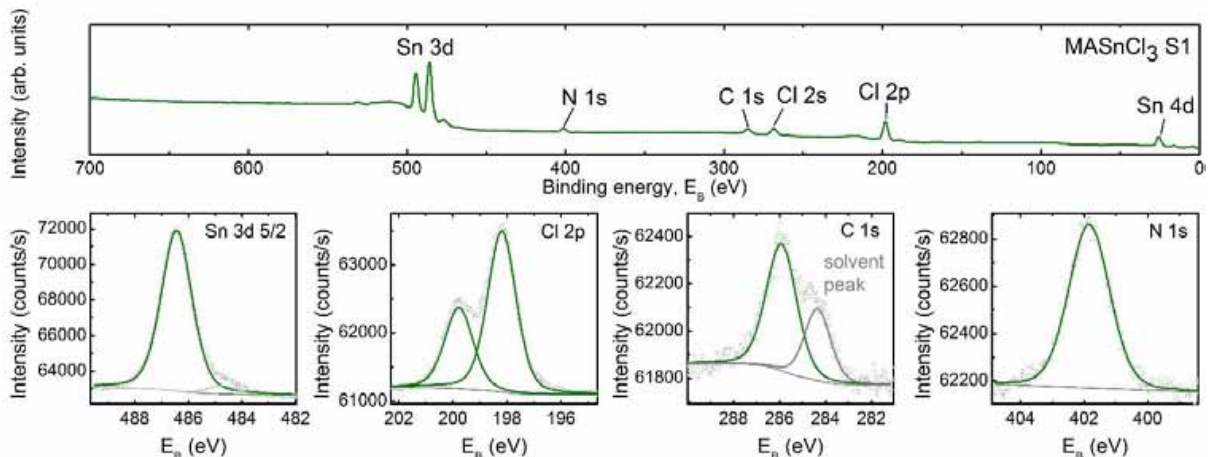

*XPS measurement of MASnCl₃, sample 1, showing a survey spectrum on top as well as detailed scans of perovskite specific core level signals at the bottom; the solid lines are fits to the measurement (open circles).*

Short discussion: The XRD spectrum indicates a cubic structure however one cannot be certain due to the low number of diffraction peaks. From literature it is known that this material is monocline at room temperature [17]; from our data set the exact lattice constants are therefore not extractable as we only see two reflexes. The before mentioned paper reports a =5.69 Å, b = 8.231 Å, c = 7.94 Å, β = 93.2° and their positions of the (100) and (200) reflexes are in good agreement with the current work. SEM measurements show large domains with "leaf" like structure and dense film formation. XPS shows all expected features and oxidation states with some additional more neutral carbon species (likely solvent remaining in the film). Note that the feature in the Sn signal at lower binding energy comes from an asymmetric peak shape and is not a metallic Sn feature (it is equally present in all Sn samples). UV-vis yields an optical gap of 3.58 eV.

**FASnCl$_3$ data sheet**

XRD measurement:

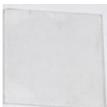

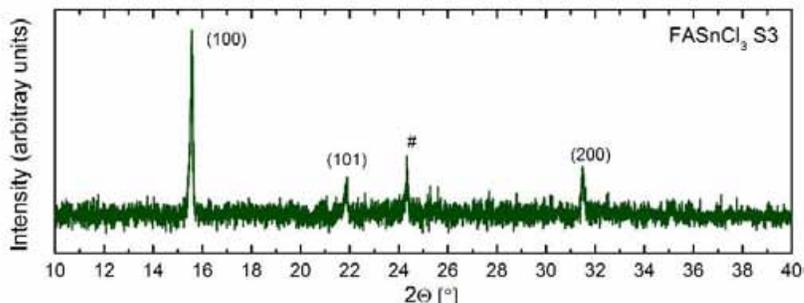

Cubic crystal structure (assumed)

a = 5.69 Å

*XRD measurement of FASnCl$_3$, sample S3; reflexes of the cubic structure are marked and the extracted lattice constant is displayed on the right.*

SEM: 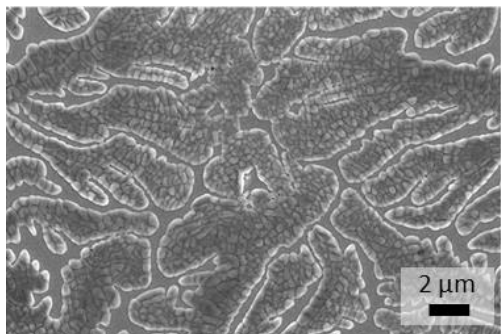

Absorption: 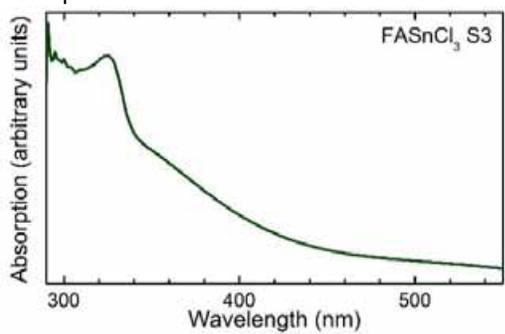

*SEM image of FASnCl$_3$, sample 2.*          *UV-vis measurement of FASnCl$_3$, sample 3.*

XPS analysis:

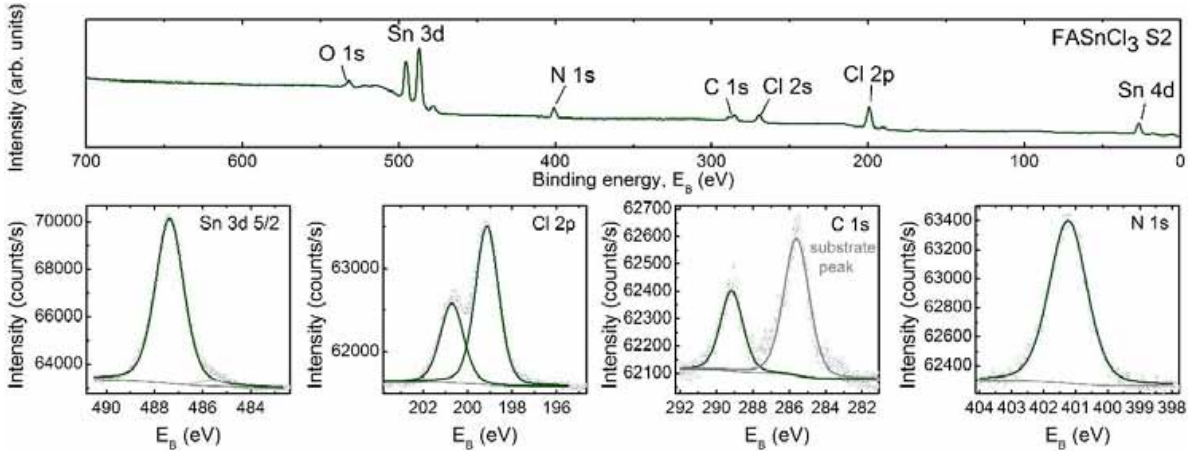

*XPS measurement of FASnCl$_3$, sample 2, showing a survey spectrum on top as well as detailed scans of perovskite specific core level signals at the bottom; the solid lines are fits to the measurement (open circles).*

Short discussion: The XRD spectrum only shows a low number of diffraction peaks indicating high order, however, it is unclear where the reflection marked as (#) originates from. We assume the crystal structure to be cubic, even though this cannot be unambiguously determined from a thin film measurement; in literature no thin film or powder XRD data have been reported. SEM images show densely packed crystallites even though the substrate is not fully covered. This is notable in the XPS measurements as well, where a significant additional carbon peak is dominated by a signal from the underlying PEDOT:PSS. All other elements show the expected oxidation states. Note that the feature in the Sn signal at lower binding energy comes from an asymmetric peak shape and is not a metallic Sn feature (it is equally present in all Sn samples). UV-vis yields an optical gap of 3.51 eV.

**CsPbI₃ data sheet**

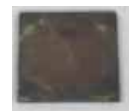

XRD measurement:

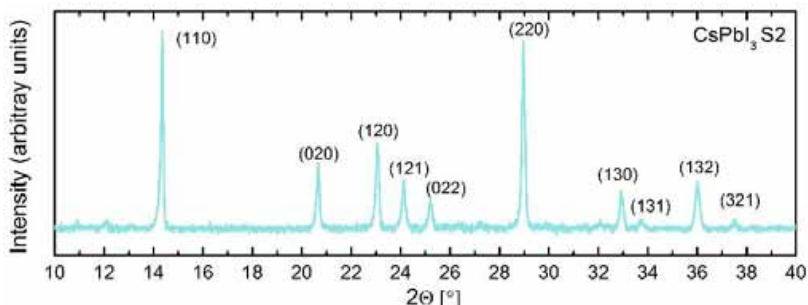

Orthorhombic crystal structure

a = 8.77 Å
b = 8.59 Å
c = 12.41 Å

*XRD measurement of CsPbI₃, sample S2; reflexes of the orthorhombic structure are marked and the extracted lattice constants are displayed on the right.*

SEM:                                                                 Absorption:

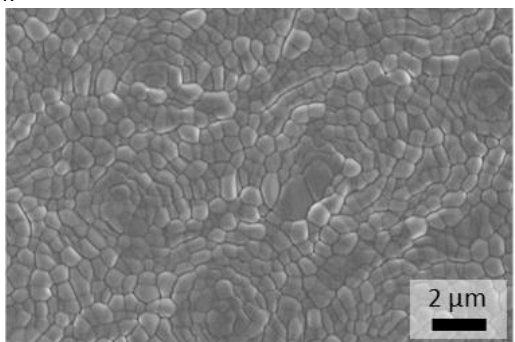
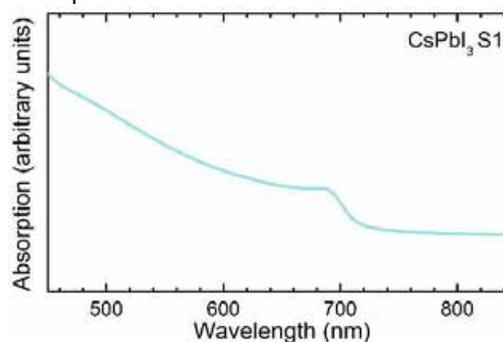

*SEM image of CsPbI₃, sample 1.*           *UV-vis measurement of CsPbI₃, sample 2.*

XPS analysis:

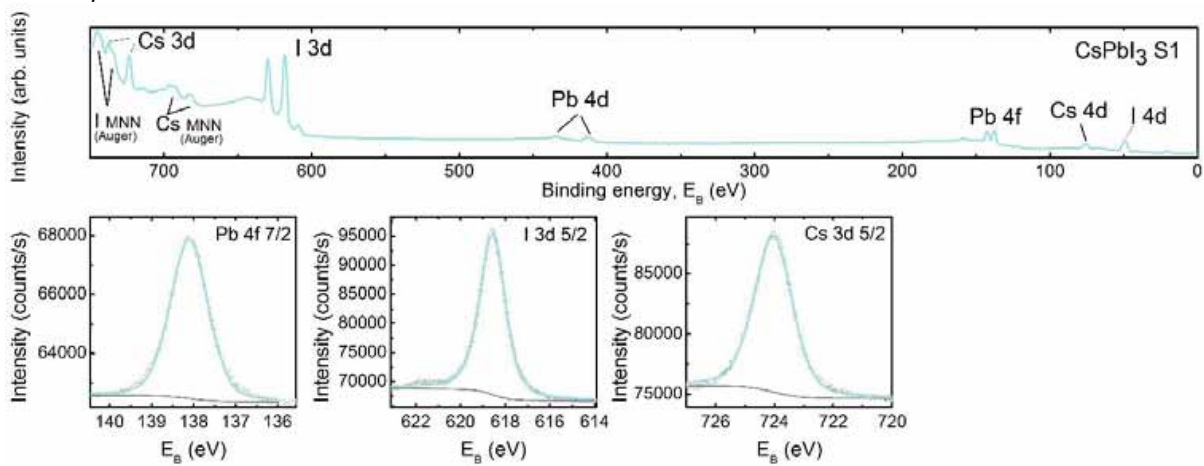

*XPS measurement of CsPbI₃, sample 1, showing a survey spectrum on top as well as detailed scans of perovskite specific core level signals at the bottom; the solid lines are fits to the measurement (open circles).*

Short discussion: CsPbI₃ is commonly reported to be cubic at room temperature, which however does not agree with our spectra. Recently, Sutton et al [18] published an orthorhombic structure, which is in excellent agreement with the XRD spectrum shown above and was used to identify the reflexes here. SEM images show densely packed films, quite similar to what is found for MAPbI₃. XPS peak analysis shows only the expected features and oxidation states. UV-vis yields an optical gap of 1.73 eV.

**MAPbI₃ data sheet**

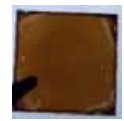

XRD measurement:

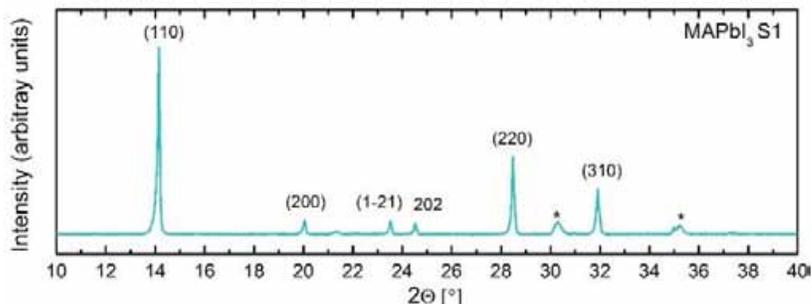

Tetragonal crystal structure

a = b = 8.88 Å
c = 12.68 Å

*XRD measurement of MAPbI₃, sample 1; reflexes of the tetragonal structure are marked and extracted lattice constants are displayed on the right. Reflexes marked with * come from underlying ITO substrate.*

SEM:                                                                 Absorption:

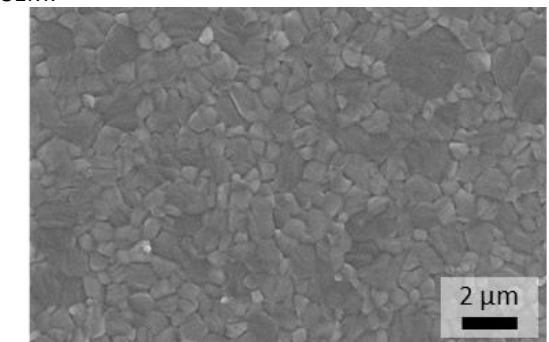
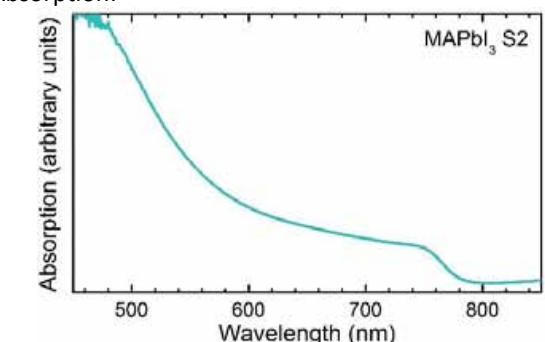

*SEM image of MAPbI₃, sample 2.*                *UV-vis measurement of MAPbI₃, sample 2.*

XPS analysis:

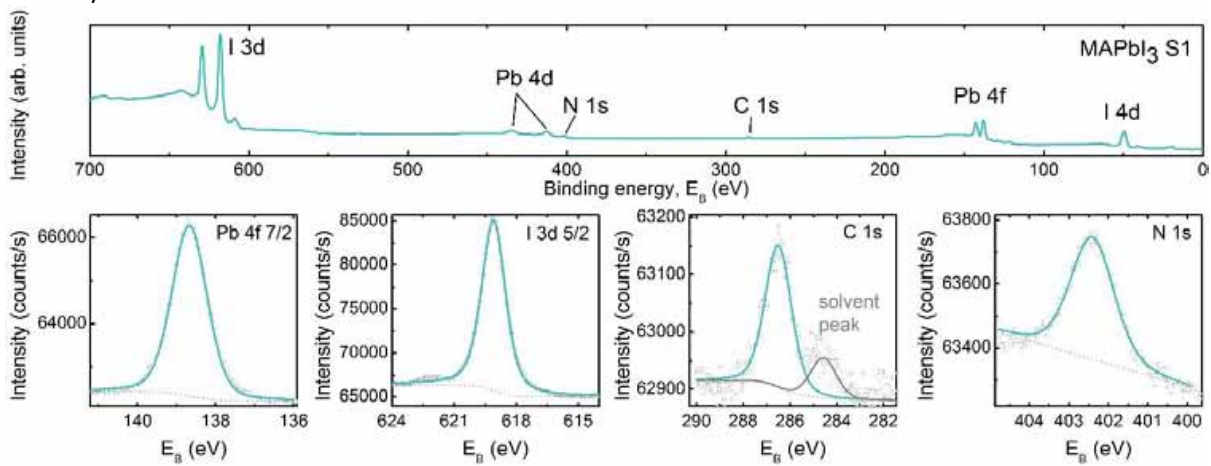

*XPS measurement of MAPbI₃, sample 1, showing a survey spectrum on top as well as detailed scans of perovskite specific core level signals at the bottom; the solid lines are fits to the measurement (open circles).*

Short discussion: XRD measurements show the well-known tetragonal crystal structure, usually with a preferential growth in the (110) direction. Peaks with asterisk (*) originate from underlying ITO substrate. SEM measurements show the typical close packed crystallites and XPS show all expected peaks and oxidation states with some additional more neutral carbon species (likely solvent remaining in the film). UV-vis yields an optical gap of 1.59 eV.

**FAPbI₃ data sheet**

XRD measurement:

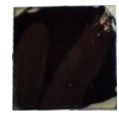

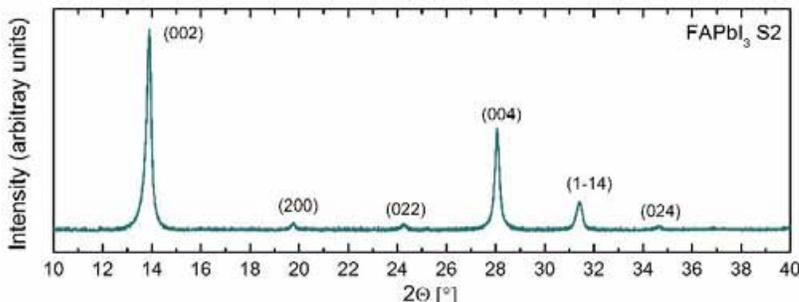

Tetragonal crystal structure

a = b = 8.99 Å
c = 12.75 Å

*XRD measurement of FAPbI₃, sample S2; reflexes of the tetragonal structure are marked and extracted lattice constants are displayed on the right.*

SEM:                                                Absorption:

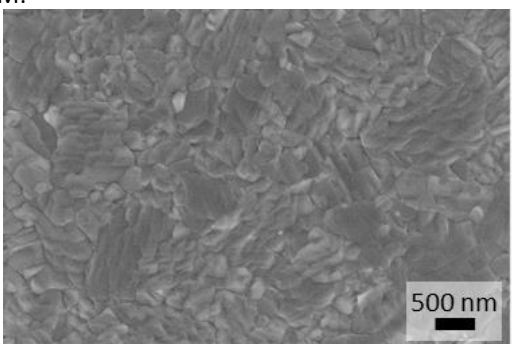 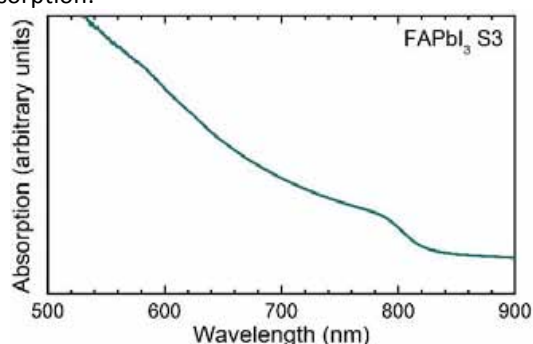

*SEM image of FAPbI₃, sample 1.*         *UV-vis measurement of FAPbI₃, sample 3.*

XPS analysis:

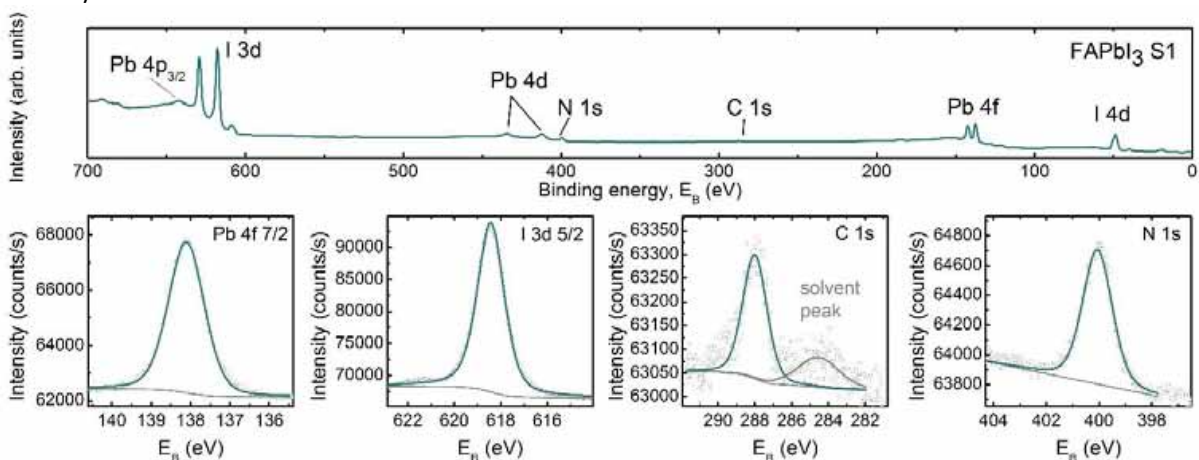

*XPS measurement of FAPbI₃, sample 1, showing a survey spectrum on top as well as detailed scans of perovskite specific core level signals at the bottom; the solid lines are fits to the measurement (open circles).*

Short discussion: XRD shows the well-known tetragonal crystal structure of this black FAPbI₃ phase. SEM measurements show close packed crystallites similar in morphology to MAPbI₃. XPS shows all expected oxidation states for the elements with some additional more neutral carbon species (likely solvent remaining in the film). UV-vis yields an optical gap of 1.51 eV.

**CsPbBr₃ data sheet**

XRD measurement:

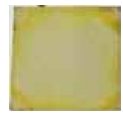

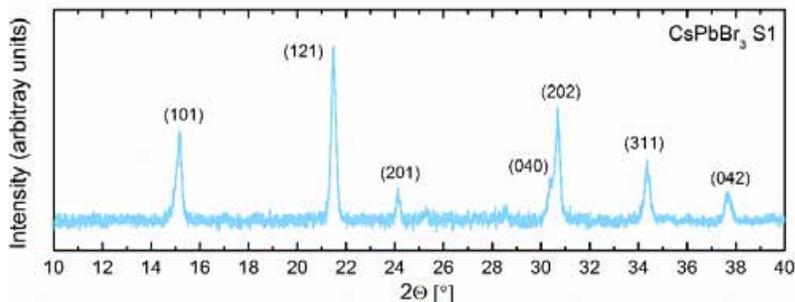

Orthorhombic crystal structure

a = 8.31 Å
b = 11.79 Å
c = 8.21 Å

*XRD measurement of CsPbBr$_3$, sample 1; reflexes of the orthorhombic structure are marked and extracted lattice constants are displayed on the right.*

SEM:                                           Absorption:

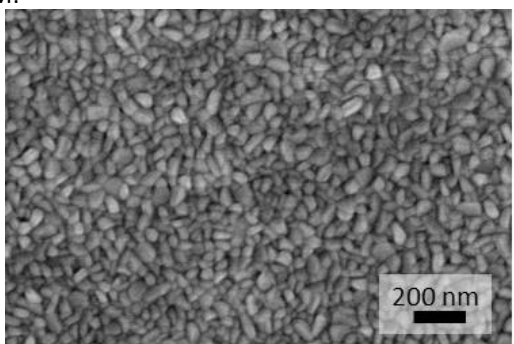    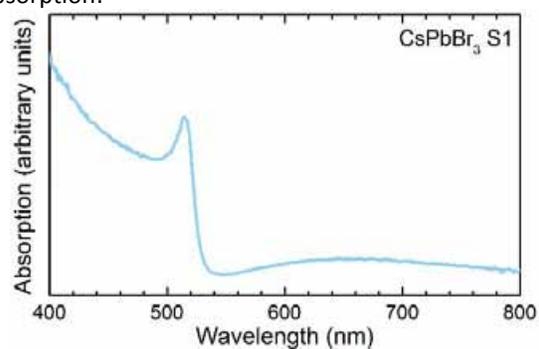

*SEM image of CsPbBr$_3$, sample 1.*    *UV-vis measurement of CsPbBr$_3$, sample 1.*

XPS analysis:

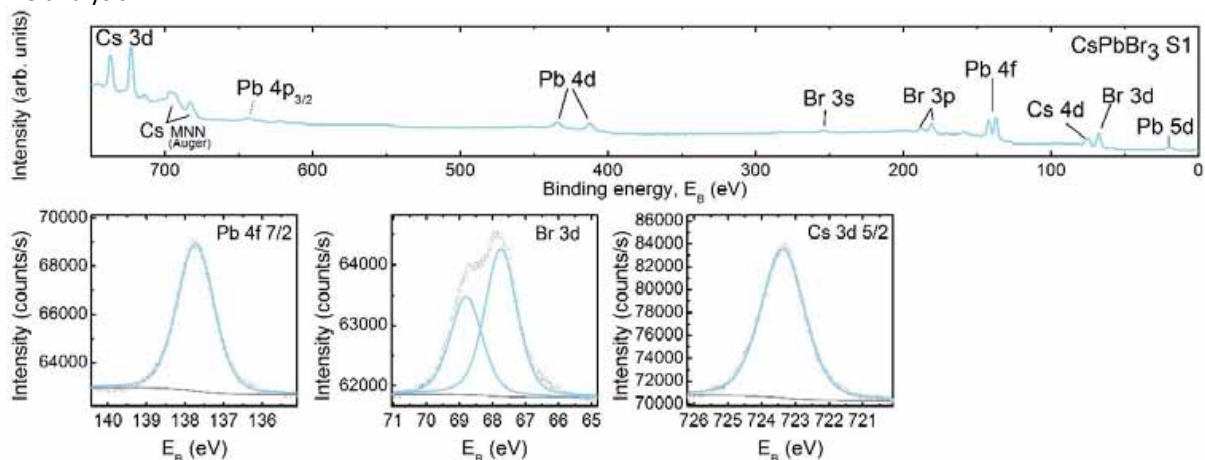

*XPS measurement of CsPbBr$_3$, sample 1, showing a survey spectrum on top as well as detailed scans of perovskite specific core level signals at the bottom; the solid lines are fits to the measurement (open circles).*

Short discussion: In XRD we find that solution processed films have a preferred orientation ( (h00) reflexes missing), while evaporated films are more randomly orientated (not shown). Overall, an orthorhombic crystal structure is found. SEM measurements show close packed crystallites. XPS peak analysis shows the expected oxidations states while UV-vis yields an optical gap of 2.34 eV.

**MAPbBr₃ data sheet**

XRD measurements:

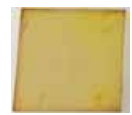

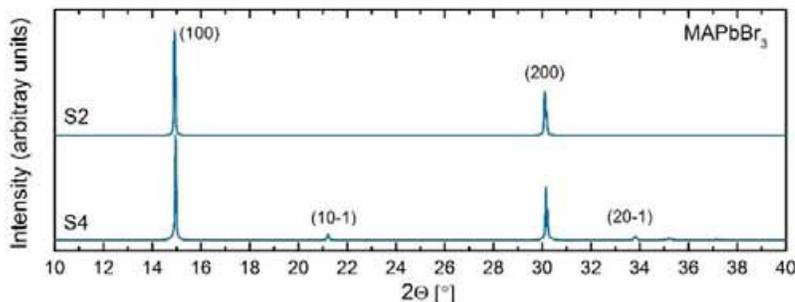

Cubic crystal structure

a = 5.92 Å

*XRD measurements of CsPbBr₃, samples 2 and 4; reflexes of the cubic structure are marked and the extracted lattice constant is displayed on the right.*

SEM: 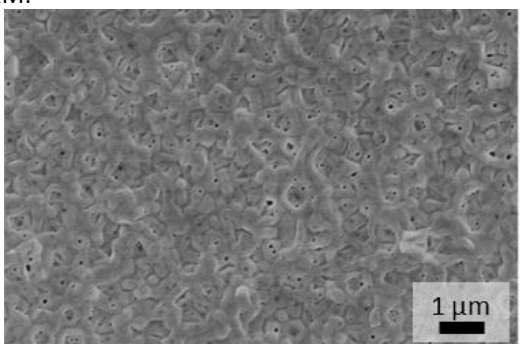   Absorption: 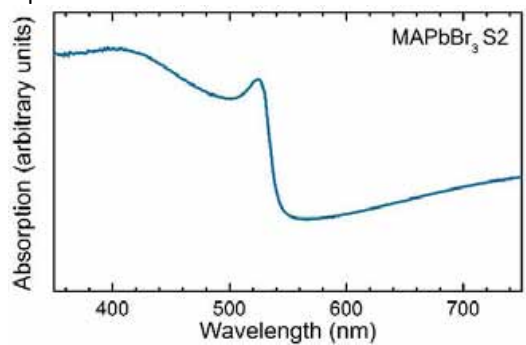

*SEM image of MAPbBr₃, sample 1.*   *UV-vis measurement of MAPbBr₃, sample 2.*

XPS analysis:

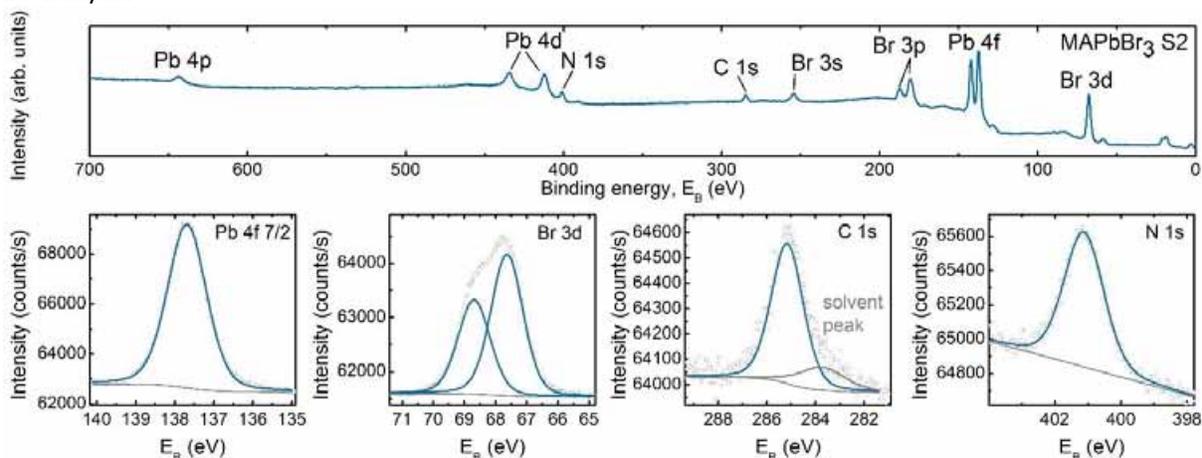

*XPS measurement of MAPbBr₃, sample 2, showing a survey spectrum on top as well as detailed scans of perovskite specific core level signals at the bottom; the solid lines are fits to the measurement (open circles).*

Short discussion: XRD shows the well-known cubic crystal structure, usually with a high degree of order so that only the (100) and (200) reflexes show. A second sample, shown in the image above, also has some weak features from additional reflexes. SEM measurements reveal densely packed films even though the crystallites are not as pronounces as in the case of the *A*PbI₃ compounds. XPS only shows the expected oxidation states with a small additional more neutral carbon species (likely solvent remaining in the film). UV-vis yields an optical gap of 2.28 eV.

**FAPbBr₃ data sheet**

XRD measurement:

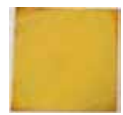

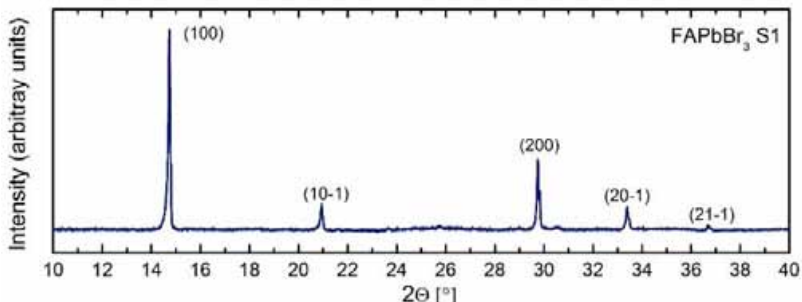

Cubic crystal structure

a = 6.00 Å

*XRD measurement of FAPbBr₃, sample 1; reflexes of the cubic structure are marked and the extracted lattice constant is displayed on the right.*

SEM: 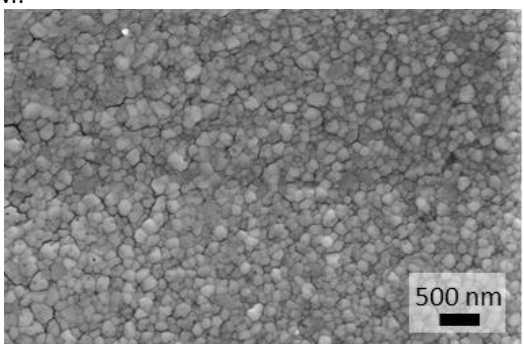  Absorption: 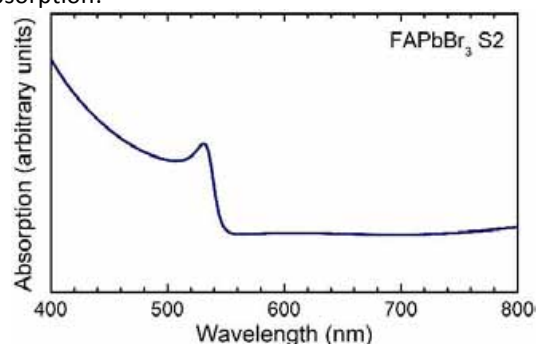

*SEM image of FAPbBr₃, sample 2.*   *UV-vis measurement of FAPbBr₃, sample 2.*

XPS analysis:

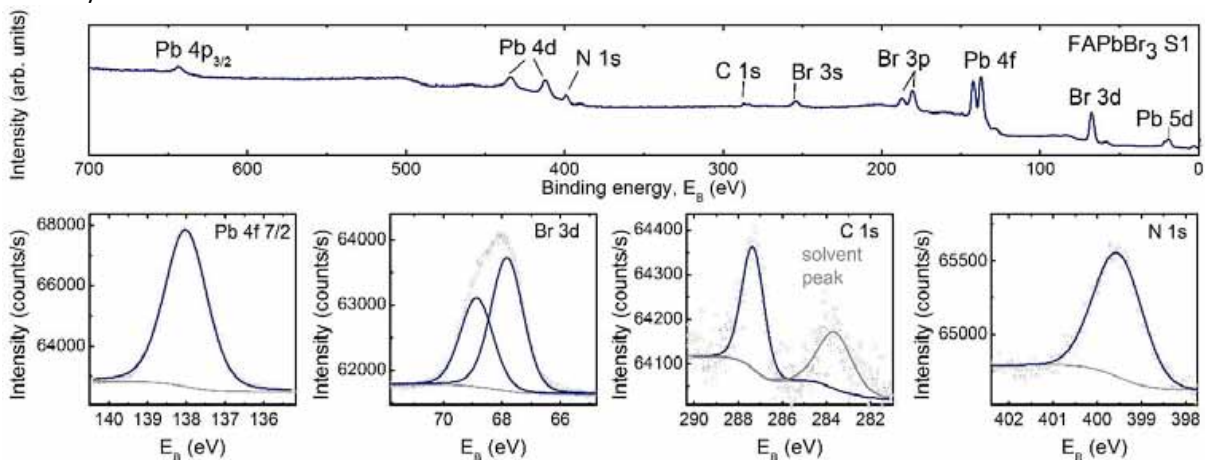

*XPS measurement of FAPbBr₃, sample 1, showing a survey spectrum on top as well as detailed scans of perovskite specific core level signals at the bottom; the solid lines are fits to the measurement (open circles).*

Short discussion: XRD measurements indicate a cubic crystal structure, which has also been observed in powder diffraction measurements in literature [19]. SEM measurements show densely packed films with small crystallites. XPS shows all expected oxidation states for the elements with some additional more neutral carbon species (likely solvent remaining in the film). UV-vis yields an optical gap of 2.26 eV.

**CsPbCl$_3$ data sheet**

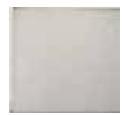

XRD measurement:

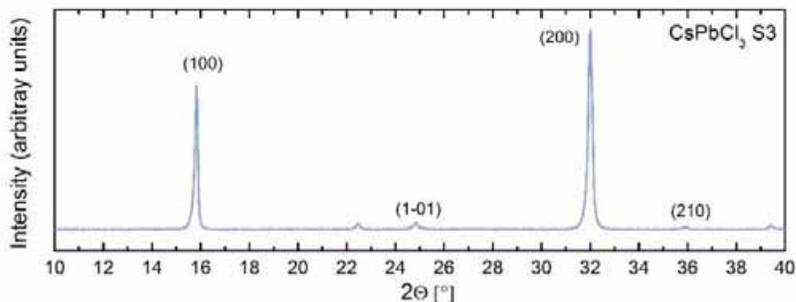

Cubic crystal structure

a = 5.60 Å

*XRD measurement of CsPbCl$_3$, sample 3; reflexes of the cubic structure are marked and the extracted lattice constant is displayed on the right.*

SEM:                                                                      Absorption:

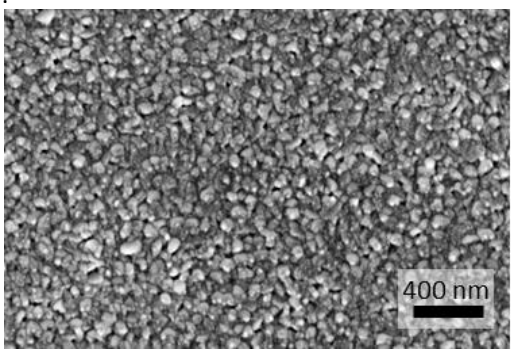
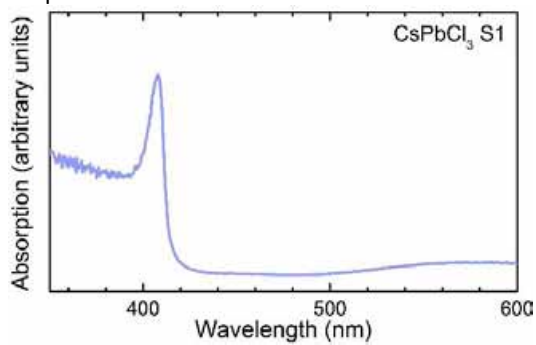

*SEM image of CsPbCl$_3$, sample 1.*                 *UV-vis measurement of CsPbCl$_3$, sample 1.*

XPS analysis:

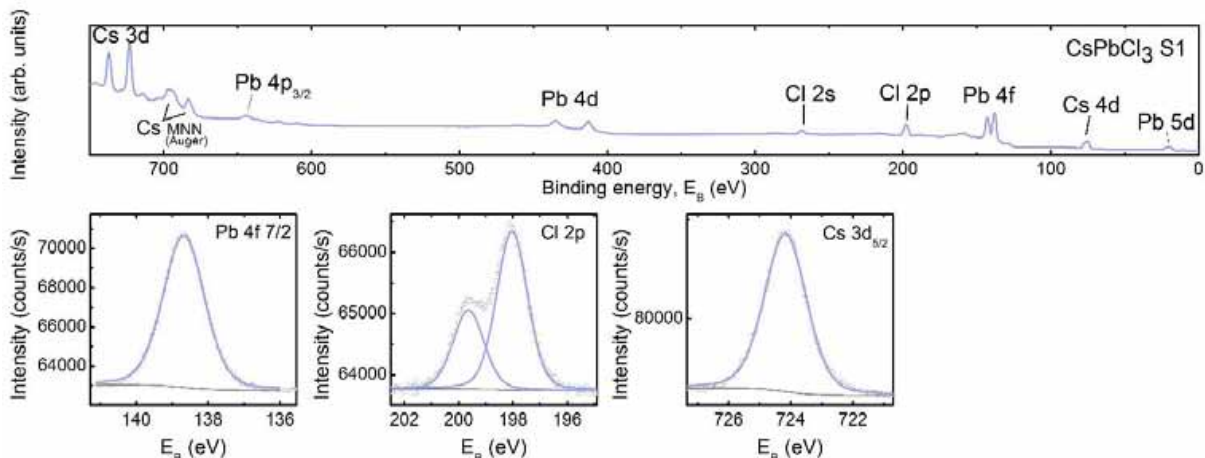

*XPS measurement of CsPbCl$_3$, sample 1, showing a survey spectrum on top as well as detailed scans of perovskite specific core level signals at the bottom; the solid lines are fits to the measurement (open circles).*

Short discussion: All samples had to be prepared by vacuum evaporation, since CsCl is not soluble enough in commonly used solvents. XRD measurements indicate a cubic structure, which is in agreement with published powder measurements [20]. SEM measurements show close packed films, with relatively small crystallites, as often observed for vapor deposition. XPS shows the expected oxidation states for all elements. UV-vis yields an optical gap of 2.98 eV.

**MAPbCl₃ data sheet**

XRD measurements:

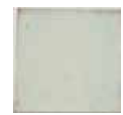

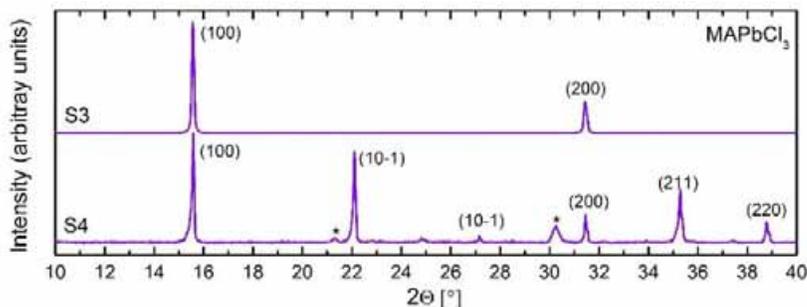

Cubic crystal structure

a = 5.69 Å

*XRD measurements of MAPbCl₃, samples 3 and 4; reflexes of the cubic structure are marked and the extracted lattice constant is displayed on the right. Peaks marked by * in sample 4 originate from the ITO substrate.*

SEM:                                                        Absorption:

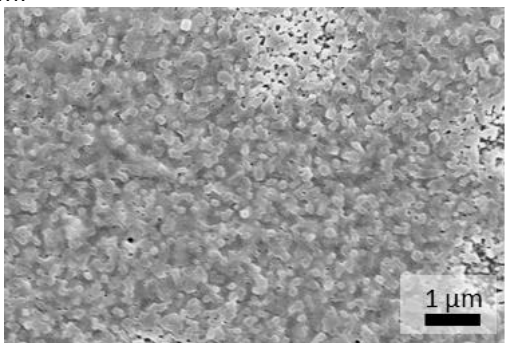 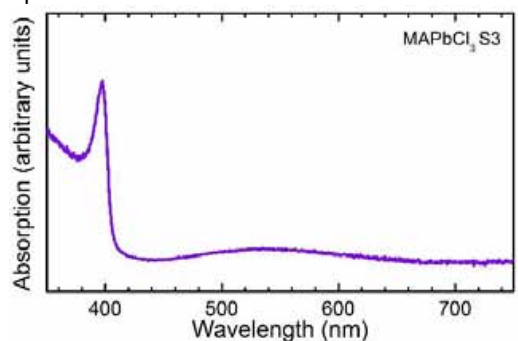

*SEM image of MAPbCl₃, sample 1.*          *UV-vis measurement of MAPbCl₃, sample 3.*

XPS analysis:

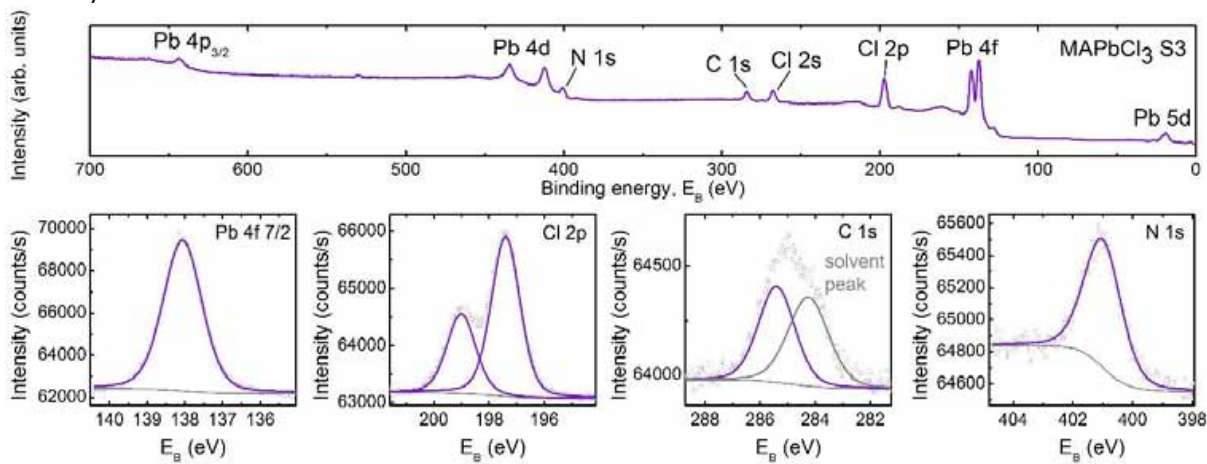

*XPS measurement of MAPbCl₃, sample 3, showing a survey spectrum on top as well as detailed scans of perovskite specific core level signals at the bottom; here the solid lines are fits to the measurement (open circles).*

Short discussion: XRD measurements indicate a cubic crystal structure, which is in agreement to published powder data [21]. SEM measurements show a densely packed film with a few small pinholes and irregularly shaped crystallites, quite similar to MAPbBr₃. XPS shows the expected oxidation states, however with a significant additional contribution for C1s. Here, it is likely not only solvent remaining in the film, but in addition the underlying PEDOT:PSS layer is showing through the pinholes. UV-vis yields an optical gap of 3.05 eV.

**FAPbCl₃ data sheet**

XRD measurement:

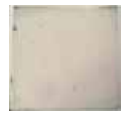

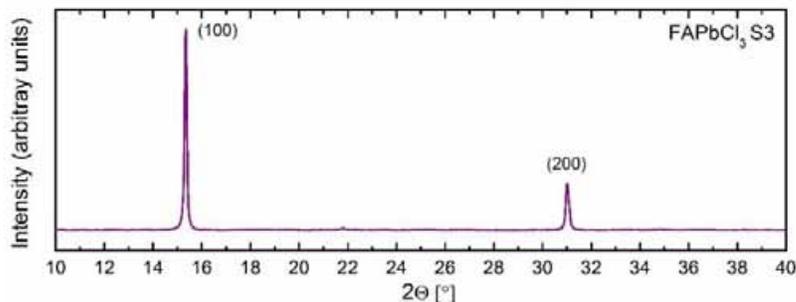

Cubic crystal structure

a = 5.77 Å

*XRD measurement of FAPbCl₃, sample 3; reflexes of the cubic structure are marked and the extracted lattice constant is displayed on the right.*

SEM:                                                                  Absorption:

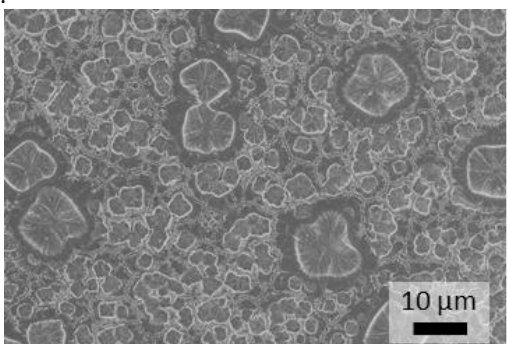 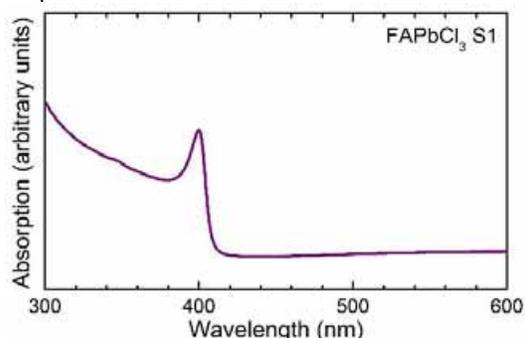

*SEM image of FAPbCl₃, sample 1.*                        *Fig b UV-vis FAPbCl₃ S1.*

XPS analysis:

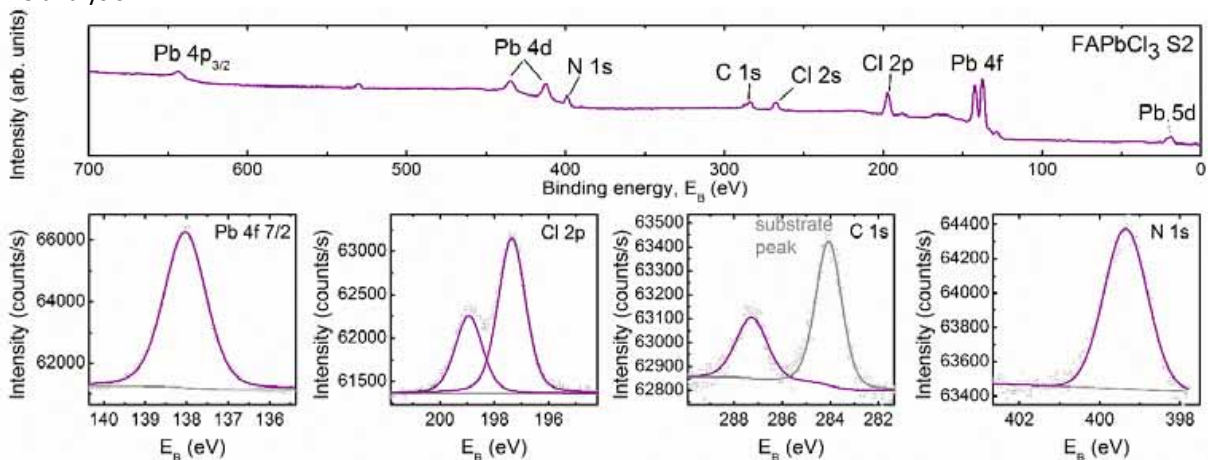

*XPS measurement of FAPbCl₃, sample 2, showing a survey spectrum on top as well as detailed scans of perovskite specific core level signals at the bottom; here the solid lines are fits to the measurement (open circles).*

Short discussion: XRD indicates a cubic crystal structure, which has been reported before for single crystals at RT [22]. SEM images reveal that film formation is not as good as for other lead based systems. Rather large "flower like" features show on the surface, leading to significant parts of the surface not being covered. This is obvious as well in XPS, where a strong contribution of C 1s at lower binding energy is observed, which originates from the underlying PEDOT:PSS layer. The perovskite related peaks show the expected oxidation states. UV-vis yields an optical gap of 3.02 eV.

10. Sample preparation

All films throughout this work are prepared under $N_2$ atmosphere (for solution processing) or in ultra-high vacuum (for thermal evaporation). Samples were given different names (i.e. MAPbI$_3$ S1, CsPbBr$_3$ S2, ...) depending on the preparation protocol used, as listed in Table S6. It should, however, be noted that samples called e.g. S1 throughout the article are not necessarily identical samples, it only means they were similarly prepared (usually separate samples for PES, XRD, and UV-vis were fabricated). Exceptions are the UPS/IPES/XPS measurements that are always done together on the same sample.

For UPS/XPS and SEM investigations, PEDOT:PSS covered ITO substrates were used while mostly glass substrates covered with PEDOT:PSS were employed for UV-Vis and XRD measurements. The filtered PEDOT:PSS dispersion (Clevios P VP Al 4083 from Heraeus) was deposited on top by a two-step spin coating procedure, spinning first for 25 s at 4000 rpm and then 5 s at 4000 rpm to yield about 40 nm thick layers. The PEDOT:PSS films were annealed at 150 °C for 10 min. After that, the substrates were transferred into the nitrogen atmosphere.

The vendors for the solvents and materials used for perovskite preparation are listed in Table S5 while the processing procedure are listed together with extracted values of work function (Wf), ionization energy (IE), electron affinity (EA), and XPS determined film composition in Table S6.

*Table S5: List of used solvents and precursors and their suppliers.*

| Formula | Chemical | supplier | purity |
|---|---|---|---|
| DMF | N N-Dimethylformamide, HPLC grade | Sigma-Aldrich | ≥ 99.9% |
| DMSO | Dimethyl Sulfoxide, analytical reagent grade | Fisher Scientific | ≥ 99.9% |
| CB | Chlorobenzene for HPLC | Sigma-Aldrich | 99.9% |
| Tol | Toluene for HPLC | Sigma-Aldrich | ≥ 99.9% |
| ODCB | 1,2-Dichlorobenzene | Sigma-Aldrich | 99% |
| IPA | 2-Propanol CHROMASOLV LC-MS | Honeywell | 99.9% |
| CsI | Cesium iodide | Sigma-Aldrich | 99.999% |
| CsBr | Cesium Bromide | TCI | >99.0% |
| CsCl | Cesium Chloride | TCI | >99.0% |
| PbI$_2$ | Lead(II) iodide trace metals basis | Sigma-Aldrich | 99.999% |
| PbBr$_2$ | Lead(II) bromide, trace metals basis | Sigma-Aldrich | 99.999% |
| PbCl$_2$ | Lead(II) chloride, anhydrous | Sigma-Aldrich | 99.999% |
| SnI$_2$ | Tin(II) iodide, trace metals basis | Sigma-Aldrich | 99.99% |
| SnBr$_2$ | Tin(II) bromide | Aldrich | |
| SnCl$_2$ | Tin(II) chloride, trace metals basis | Sigma-Aldrich | ≥ 99.99% |
| NH$_4$Cl | Ammonium Chloride, Suprapur | EMD Millipore | 99.995% |
| MA I/Br/Cl FA I/Br/Cl | Methylammonium and formamidinium salts were prepared in house using a standard synthesis procedure | | |

*Table S6: Sample preparation and extracted values of work function (Wf), ionization energy (IE), electron affinity (EA), and XPS determined film composition.*

| Sample | Preparation (precursor, concentration, solvent, processing, annealing) | | | | | Wf [eV] | IE [eV] | EA [eV] | XPS stoichiometry |
|---|---|---|---|---|---|---|---|---|---|
| CsSnI$_3$ S1 | CsI and SnI$_2$ | 0.5 M | DMF | Spin coating: 30s at 5000 rpm | 10 min at 100 °C | 5.02 | 5.70 | 4.44 | Cs(1.2)Sn(0.9)I(3) |
| CsSnI$_3$ S2 | CsI and SnI$_2$ with additive NH$_4$Cl (ratio 3:1) | 1 M | DMF | Spin coating: 60s at 2000 rpm After 5s ODCB as antisolvent | 10 min at 80 °C | 5.02 | 5.76 | 4.42 | Cs(1.2)Sn(0.9)I(3) |
| CsSnI$_3$ S3 | CsI and SnI$_2$ | 0.5 M | DMF | Spin coating: 5s at 500 rpm, 40s at 1000 rpm and 50s at 5000 rpm | 10 min at 100 °C | 4.94 | 5.60 | 4.29 | Cs(1.3)Sn(0.8)I(3) |
| CsSnI$_3$ S4 (UV-vis only) | CsI and SnI$_2$ | 0.5 M | DMSO | Spin coating: 5s at 500 rpm, 40s at 1000 rpm and 50 s at 5000 rpm | 10 min at 80 °C | *Not measured* | | | |
| MASnI$_3$ S1 | MAI and SnI$_2$ | 1 M | DMSO | Spin coating: 30s at 3000 rpm After 20s ODCB as antisolvent | 10 min at 80 °C | 4.70 | 5.38 | 4.04 | C(1.2)N(0.9)Sn(0.8)I(3) |
| MASnI$_3$ S2 | MAI and SnI$_2$ | 1 M | DMSO | Spin coating: 30s at 3000 rpm After 20s ODCB as antisolvent | 30 s at 100 °C | 4.72 | 5.36 | 4.00 | C(1.3)N(1)Sn(0.8)I(3) |
| MASnI$_3$ S3 | MAI and SnI$_2$ with additive NH$_4$Cl (1:0.84) | 0.4 M | DMF | Spin coating: 30s at 3000 rpm | Not annealed | 4.73 | 5.44 | 4.16 | *Not measured* |
| MASnI$_3$ S4 (UV-vis only) | MAI and SnI$_2$ | 1 M | DMSO | Spin coating: 50 s at 5000 rpm using ODCB antisolvent after 10 s. | 10 min at 100 °C | *Not measured* | | | |
| FASnI$_3$ S1 | FAI and SnI$_2$ | 1 M | DMSO | Spin coating: 30s at 3000 rpm After 20s ODCB as antisolvent | 30 s at 100 °C | 4.90 | 5.38 | 4.15 | N(2.1)C(1)Sn(0.8)I(3) |
| FASnI$_3$ S2 | FAI and SnI$_2$ | 1 M | DMSO | Spin coating: 5s at 500 rpm, 40s at 1000 rpm, and 50s at 5000 rpm After 33s of last step ODCB as antisolvent | 10 min at 100 °C | 4.77 | 5.41 | 4.15 | N(2)C(1)Sn(0.9)I(3) |
| FASnI$_3$ S3 | FAI and SnI$_2$ | 1 M | DMSO | Spin coating: 30s at 3000 rpm | 10 min at 100 °C | 4.88 | 5.24 | 4.06 | N(2.5)C(1.3)Sn(0.65)I(3) |
| FASnI$_3$ S4 (UV-vis only) | FAI and SnI$_2$ | 1 M | DMSO | Spin coating: 5s at 500 rpm, 40s at 1000 rpm, and 50s at 5000 rpm After 20 s of last step Toluene as antisolvent | 10 min at 80 °C | *Not measured* | | | |
| CsSnBr$_3$ S1 | CsBr and SnBr$_2$ | - | - | Co-evaporation at molar ratio 1 : 1 | Not annealed | 4.88 | 5.87 | 4.15 | Cs(1.1)Sn(0.9)Br(3) |
| CsSnBr$_3$ S2 | CsBr and SnBr$_2$ | - | - | Co-evaporation at molar ratio 1.1 : 1 (CsBr:SnBr$_2$) | Not annealed | 4.70 | 5.75 | 3.96 | Cs(1.3)Sn(0.9)Br(3) |
| CsSnBr$_3$ S3 | CsBr and SnBr$_2$ | - | - | Co-evaporation at molar ratio 1 : 1 | 1 h at 60°C in vacuum | 4.88 | 5.83 | 4.11 | Cs(1.2)Sn(0.9)Br(3) |

| Sample | Preparation (precursor, concentration, solvent, processing, annealing) | | | | | Wf [eV] | IE [eV] | EA [eV] | XPS stoichiometry |
|---|---|---|---|---|---|---|---|---|---|
| MASnBr$_3$ S1 | MABr and SnBr$_2$ | 1 M | DMSO | Spin coating: 30s at 3000 rpm | 10 min at 100 °C | 4.60 | 5.71 | 3.40 | C(1.7)N(1.3)Sn(1.1)Br(3) |
| MASnBr$_3$ S2 | MABr and SnBr$_2$ | 1 M | DMSO | Spin coating: 30s at 3000 rpm | 10 min at 100 °C | 4.47 | 5.64 | 3.40 | Not measured |
| MASnBr$_3$ S3 | MABr and SnBr$_2$ | 1 M | DMSO | Spin coating: 30s at 3000 rpm | 10 min at 100 °C | 4.56 | 5.67 | 3.45 | C(1.8)N(1.4)Sn(1)Br(3) |
| FASnBr$_3$ S1 | FABr and SnBr$_2$ | 1 M | DMSO | Spin coating: 5s at 500 rpm, 40s at 1000 rpm, and 50s at 5000 rpm. After 33s of the last step ODCB as antisolvent | 10 min at 100 °C | 4.91 | 6.25 | 3.59 | Not measured |
| FASnBr$_3$ S2 | FABr and SnBr$_2$ | 0.5 M | DMSO | Spin coating: 5s at 500 rpm, 40s at 1000 rpm, and 50s at 5000 rpm. After 15s of last step ODCB as antisolvent | 10 min at 100 °C | 4.92 | 6.25 | 3.65 | Not measured |
| FASnBr$_3$ S3 | FABr and SnBr$_2$ | 1 M | DMSO | Spin coating: 5s at 500 rpm, 40s at 1000 rpm, and 50s at 5000 rpm. After 33s of last step ODCB as antisolvent | 10 min at 100 °C | 4.84 | 6.18 | 3.57 | C(1.1)N(2.2)Sn(1)Br(3) |
| CsSnCl$_3$ S1 | CsCl and SnCl$_2$ | - | - | Co-evaporation at molar ratio 1 : 1 | 1 h at 80 °C in vacuum | 4.97 | 6.42 | 3.52 | Cs(1.1)Sn(1)Cl(3) |
| CsSnCl$_3$ S2 | CsCl and SnCl$_2$ | - | - | Co-evaporation at molar ratio 1 : 1.1 (CsI:SnCl$_2$) | Not annealed | 5.06 | 6.48 | 3.43 | Cs(1)Sn(1)Cl(3) |
| CsSnCl$_3$ S3 | CsCl and SnCl$_2$ | - | - | Co-evaporation at molar ratio 1 : 1 | 1 h at 60 °C in vacuum | 5.00 | 6.41 | 3.47 | Cs(1)Sn(1.1)Cl(3) |
| MASnCl$_3$ S1 | MACl and SnCl$_2$ | 1.92 M | DMSO | Spin coating: 30s at 4000 rpm After 5s CB as antisolvent | 5 min at 110 °C | 4.87 | 6.84 | 3.31 | C(1.5)N(1)Sn(1)Cl(3) |
| MASnCl$_3$ S2 | MACl and SnCl$_2$ with additive NH$_4$Cl (5:1) | 1.5 M | DMF | Spin coating: 60s at 2000 rpm | 10 min at 80 °C | 4.67 | 6.98 | 3.53 | C(1.7)N(1.4)Sn(1.1)Cl(3) |
| MASnCl$_3$ S3 | MACl and SnCl$_2$ | 0.5 M | DMSO | Spin coating: 15s at 250 rpm and 25s at 1500 rpm | 10 min at 100 °C | 4.75 | 6.74 | 3.32 | C(1.4)N(1)Sn(1.1)Cl(3) |
| FASnCl$_3$ S1 | FACl and SnCl$_2$ | 0.5 M | DMSO | Soaking of 15 µl solution on sample for 30s, then spin coating 30s at 3000 rpm | 10 min at 100 °C | 4.86 | 7.36 | 3.78 | C(1)N(2)Sn(1)Cl(3) |
| FASnCl$_3$ S2 | FACl and SnCl$_2$ | 0.5 M | DMSO | Soaking of 15 µl solution on sample for 10s, then spin coating 30s at 3000 rpm | 10 min at 100 °C | 4.92 | 7.3 | 3.88 | C(1)N(2)Sn(1)Cl(3) |
| FASnCl$_3$ S3 | FACl and SnCl$_2$ | 0.5 M | DMSO | Spin coating: 15s at 250 rpm and 25s at 1500 rpm | 10 min at 100 °C | 4.57 | 7.05 | 3.24 | Not measured |
| CsPbI$_3$ S1 | CsI and PbI$_2$ | 0.5 M | DMF | Spin coating: 30s at 3000 rpm | 10 min at 350 °C | 4.66 | 6.25 | 4.47 | Cs(1.2)Pb(0.8)I(3) |
| CsPbI$_3$ S2 | CsI and PbI$_2$ | 1 M | DMF | Spin coating: 30s at 5000 rpm, After 5s CB as antisolvent | 5 min 350 °C | 4.72 | 6.32 | 4.55 | Cs(1.2)Pb(0.9)I(3) |
| CsPbI$_3$ S3 | CsI and PbI$_2$ | 1 M | DMF | Spin coating: 30s at 5000 rpm, After 5s CB as antisolvent | 5 min 350 °C | 4.73 | 6.19 | 4.39 | Not measured |

| Sample | Preparation (precursor, concentration, solvent, processing, annealing) | | | | | Wf [eV] | IE [eV] | EA [eV] | XPS stoichiometry |
|---|---|---|---|---|---|---|---|---|---|
| MAPbI$_3$ S1 | MAI and PbI$_2$ | 1 M | DMF | Spin coating: 30s at 3000 rpm After 3s ODCB as antisolvent | 1 h at 100 °C | 4.41 | 5.90 | 4.31 | C(1.1)N(1)Pb(0.9)I(3) |
| MAPbI$_3$ S2 | MAI and PbI$_2$ | 1 M | DMF | Spin coating: 30s at 5000 rpm, After 5s CB as antisolvent | 1 h at 100 °C | 4.45 | 5.95 | 4.43 | C(1)N(1)Pb(0.8)I(3) |
| MAPbI$_3$ S3 | MAI and PbI$_2$ | 1 M | DMF | Spin coating: 30s at 5000 rpm, After 3s CB as antisolvent | 1 h at 100 °C | 4.41 | 5.94 | 4.33 | C(1)N(0.9)Pb(1)I(3) |
| FAPbI$_3$ S1 | FAI and PbI$_2$ | 1 M | DMF | Spin coating: 30s at 3000 rpm, After 4s CB as antisolvent | 10 min at 140 °C | 4.84 | 6.29 | 4.80 | C(1)N(1.7)Pb(0.9)I(3) |
| FAPbI$_3$ S2 | FAI and PbI$_2$ | 1 M | DMF | Spin coating: 30s at 5000 rpm, After 4s CB as antisolvent | 20 min at 140 °C | 5.23 | 6.30 | 4.83 | *Not measured* |
| FAPbI$_3$ S3 | FAI and PbI$_2$ | 1 M | DMF | Spin coating: 30s at 5000 rpm, After 4s as antisolvent | 5 min at 140 °C | 4.58 | 6.13 | 4.58 | C(0.8)N(1.5)Pb(0.8)I(3) |
| CsPbBr$_3$ S1 | CsBr and PbBr$_2$ | - | - | Co-evaporation at molar ratio 1 : 1 | Not annealed | 5.32 | 6.5 | 4.16 | Cs(1.1)Pb(0.9)Br(3) |
| CsPbBr$_3$ S2 | CsBr and PbBr$_2$ | - | - | Co-evaporation at molar ratio 1.1 : 1 (CsBr:PbBr$_2$) | Not annealed | 5.20 | 6.44 | 4.17 | Cs(1)Pb(1.1)Br(3) |
| CsPbBr$_3$ S3 | CsBr and PbBr$_2$ | 0.2 M | DMSO | Spin coating: 5s at 500 rpm, 40s at 1000 rpm, 50s at 5000 rpm. After 20s of last step Toluene as antisolvent | 10 min at 100 °C | 5.06 | 6.56 | 4.17 | Cs(1.1)Pb(0.7)Br(3) |
| MAPbBr$_3$ S1 | MABr and PbBr$_2$ | 1 M | DMSO | Spin coating: 5s at 500 rpm, 40s at 1000 rpm, 50s at 5000 rpm. After 6s of last step ODCB as antisolvent | 10 min at 100 °C | 5.33 | 6.59 | 4.26 | *Not measured* |
| MAPbBr$_3$ S2 | MABr and PbBr$_2$ | 1 M | DMSO | Spin coating: 5s at 500 rpm, 40s at 1000 rpm, 50s at 5000 rpm. After 6s of last step ODCB as antisolvent | 10 min at 100 °C | 5.27 | 6.60 | 4.21 | C(1.7)N(1.3)Pb(1)Br(3) |
| MAPbBr$_3$ S3 | PbBr$_2$ and MABr | 0.5 M and 0.13 M | DMF and ml in IPA | First PbBr$_2$ spin coated 30s at 2000 rpm and pre-annealing at 80 °C, next 50 µl MABr solution on top and spin coated 30s at 2000 rpm | 10 min at 80 °C | 5.22 | 6.61 | 4.29 | C(1.7)N(1.4)Pb(0.9)Br(3) |
| MAPbBr$_3$ S4 (XRD only) | MABr and PbBr$_2$ | 1 M | DMSO | Spin coating: at 3000 rpm for 30s | 10min at 100C | | *Not measured* | | |
| FAPbBr$_3$ S1 | FABr and PbBr$_2$ | 1 M | DMSO | Spin coating: 5s at 500 rpm, 40s at 1000 rpm, 50s at 5000 rpm. After 33s of last step ODCB as antisolvent | 10 min at 100 °C | 5.48 | 6.77 | 4.56 | C(0.7)N(1.3)Pb(1)Br(3) |
| FAPbBr$_3$ S2 | FABr and PbBr$_2$ | 0.5 M | DMSO | Spin coating: 5s at 500 rpm, 40s at 1000 rpm and 50s at 5000 rpm. After 10s of last step ODCB as antisolvent | 10 min at 100 °C | 5.39 | 6.67 | 4.43 | C(1.1)N(1.7)Pb(1)Br(3) |

| Sample | Preparation (precursor, concentration, solvent, processing, annealing) | | | | | Wf [eV] | IE [eV] | EA [eV] | XPS stoichiometry |
|---|---|---|---|---|---|---|---|---|---|
| FAPbBr$_3$ S3 | FABr and PbBr$_2$ | 0.5 M | DMSO | Spin coating: 5s at 500 rpm, 40s at 1000 rpm, 50s at 5000 rpm. After 13 s of last step Toluene as antisolvent | 10 min at 130 °C | 5.41 | 6.67 | 4.54 | C(1.3)N(2.1)Pb(1)Br(3 |
| CsPbCl$_3$ S1 | CsCl and PbCl$_2$ | - | - | Co-evaporation at molar ratio 1 : 1 | Not annealed | 4.58 | 6.91 | 3.85 | Cs(1)Pb(1)Cl(3) |
| CsPbCl$_3$ S2 | CsCl and PbCl$_2$ | - | - | Co-evaporation at molar ratio 1 : 1 | 1 h at 80 °C | 4.3 | 6.76 | 3.77 | Cs(1)Pb(0.9)Cl(3) |
| CsPbCl$_3$ S3 | CsCl and PbCl$_2$ | - | - | Co-evaporation at molar ratio 1 : 1 | 1 h at 80 °C | 4.15 | 6.81 | 3.77 | Cs(0.9)Pb(0.9)Cl(3) |
| MAPbCl$_3$ S1 | MACl and PbCl$_2$ | 1.2 M | DMSO | Spin coating: 30s at 4000 rpm, after 20s CB was put on the spinning substrate | 5 min at 110 °C | 5.34 | 6.98 | 3.91 | N(0.9)C(1.3)Pb(0.8)Cl(3) |
| MAPbCl$_3$ S2 | PbCl$_2$ and MACl | 1.4 M and 0.2 M | DMSO and IPA | First PbCl$_2$ spin coated 60s at 2000 rpm, next MABr solution on top, soaked for 10s then spun off at 2000 rpm for 60s | 10 min at 100 °C | 4.85 | 7.02 | 3.79 | N(1.3)C(1.4)Pb(0.6)Cl(3) |
| MAPbCl$_3$ S3 | MACl and PbCl$_2$ | 0.5 M | DMSO | Spin coating: 5s at 500 rpm, 40s at 1000 rpm, 50 s at 5000 rpm. After 13 s of last step Toluene as antisolvent | 10 min at 100 °C | 5.19 | 6.85 | 3.71 | N(1)C(1.1)Pb(0.8)Cl(3) |
| MAPbCl$_3$ S4 (XRD only) | MACl and PbCl$_2$ | 0.5 M | DMSO | Spin coating: 60s at 2000 rpm | 10 min at 100 °C | Not measured | | | |
| FAPbCl$_3$ S1 | FACl and PbCl$_2$ | 0.5 M | DMSO | Spin coating: 5s at 500 rpm, 40s at 1000 rpm, and 50s at 5000 rpm. After 13s of last step Toluene as antisolvent | 10 min at 100 °C | 5.03 | 6.97 | 4.05 | Not measured |
| FAPbCl$_3$ S2 | FACl and PbCl$_2$ | 0.5 M | DMSO | Spin coating: 5s at 500 rpm, 40s at 1000 rpm, 50s at 5000 rpm. After 15 s of last step ODCB as antisolvent | 10 min at 100 °C | 4.99 | 6.90 | 3.93 | N(1.6)C(1)Pb(0.9)Cl(3) |
| FAPbCl$_3$ S3 | FACl and PbCl$_2$ | 0.5 M | DMSO | Spin coating: 5s at 500 rpm, 40s at 100 rpm, 50s at 5000 rpm. After 15 s of last step ODCB as antisolvent | 10 min at 100 °C | 5.42 | 6.96 | 3.97 | Not measured |

11. Additional References

1. Amat, A. *et al.* Cation-induced band-gap tuning in organohalide perovskites: Interplay of spin-orbit coupling and octahedra tilting. *Nano Lett.* **14,** 3608–3616 (2014).
2. Kang, J. & Wang, L. W. Dynamic Disorder and Potential Fluctuation in Two-Dimensional Perovskite. *J. Phys. Chem. Lett.* **8,** 3875–3880 (2017).
3. Ma, J. & Wang, L. W. Nanoscale charge localization induced by random orientations of organic molecules in hybrid perovskite CH3NH3PbI3. *Nano Lett.* **15,** 248–253 (2015).
4. Zhu, H. *et al.* Organic Cations Might Not Be Essential to the Remarkable Properties of Band Edge Carriers in Lead Halide Perovskites. *Adv. Mater.* **29,** 1–6 (2017).
5. Yaffe, O. *et al.* Local Polar Fluctuations in Lead Halide Perovskite Crystals. *Phys. Rev. Lett.* **118,** 136001 (2017).
6. Tao, S. X., Cao, X. & Bobbert, P. A. Accurate and efficient band gap predictions of metal halide perovskites using the DFT-1/2 method: GW accuracy with DFT expense. *Sci. Rep.* **7,** 14386 (2017).
7. Davies, C. L. *et al.* Bimolecular recombination in methylammonium lead triiodide perovskite is an inverse absorption process. *Nat. Commun.* **9,** 293 (2018).
8. Endres, J. *et al.* Valence and Conduction Band Densities of States of Metal Halide Perovskites: A Combined Experimental–Theoretical Study. *J. Phys. Chem. Lett.* **7,** 2722–2729 (2016).
9. Bao, J. L., Gagliardi, L. & Truhlar, D. G. Self-Interaction Error in Density Functional Theory: An Appraisal. *J. Phys. Chem. Lett.* **9,** 2353–2358 (2018).
10. Komesu, T. *et al.* Surface Electronic Structure of Hybrid Organo Lead Bromide Perovskite Single Crystals. *J. Phys. Chem. C* **120,** 21710–21715 (2016).
11. Philippe, B. *et al.* Valence Level Character in a Mixed Perovskite Material and Determination of the Valence Band Maximum from Photoelectron Spectroscopy: Variation with Photon Energy. *J. Phys. Chem. C* **121,** 26655–26666 (2017).
12. Emara, J. *et al.* Impact of Film Stoichiometry on the Ionization Energy and Electronic Structure of CH 3 NH 3 PbI 3 Perovskites. *Adv. Mater.* **28,** 553–559 (2016).
13. Dang, Y. *et al.* Formation of Hybrid Perovskite Tin Iodide Single Crystals by Top-Seeded Solution Growth. *Angew. Chemie - Int. Ed.* **55,** 3447–3450 (2016).
14. Chiarella, F. *et al.* Preparation and transport properties of hybrid organic–inorganic CH3NH3SnBr3 films. *Appl. Phys. A* **86,** 89–93 (2006).
15. Ferrara, C. *et al.* Wide band-gap tuning in Sn-based hybrid perovskites through cation replacement: The FA1-:XMAxSnBr3mixed system. *J. Mater. Chem. A* **5,** 9391–9395 (2017).
16. Scaife, D. E., Weller, P. F. & Fisher, W. G. Crystal preparation and properties of cesium tin(II) trihalides. *J. Solid State Chem.* **9,** 308–314 (1974).
17. Chiarella, F. *et al.* Combined experimental and theoretical investigation of optical, structural, and electronic properties of CH3NH3SnX3 thin films (X=Cl,Br). *Phys. Rev. B - Condens. Matter Mater. Phys.* **77,** 045129 (2008).
18. Sutton, R. J. *et al.* Cubic or Orthorhombic? Revealing the Crystal Structure of Metastable Black-Phase CsPbI3by Theory and Experiment. *ACS Energy Lett.* **3,** 1787–1794 (2018).
19. Zhumekenov, A. A. *et al.* Formamidinium Lead Halide Perovskite Crystals with Unprecedented Long Carrier Dynamics and Diffusion Length. *ACS Energy Lett.* **1,** 32–37 (2016).
20. Sebastian, M. *et al.* Excitonic emissions and above-band-gap luminescence in the single-crystal perovskite semiconductors CsPbB r3 and CsPbC l3. *Phys. Rev. B - Condens. Matter Mater. Phys.* **92,** 1–9 (2015).
21. Baikie, T. *et al.* A combined single crystal neutron/X-ray diffraction and solid-state nuclear magnetic resonance study of the hybrid perovskites CH3NH3PbX3 (X = I, Br and Cl). *J. Mater. Chem. A* **3,** 9298–9307 (2015).
22. Govinda, S. *et al.* Critical Comparison of FAPbX3and MAPbX3(X = Br and Cl): How Do They Differ? *J. Phys. Chem. C* **122,** 13758–13766 (2018).